\pdfoutput = 1

\documentclass[12pt,a4paper]{article}

\usepackage{amsmath}
\usepackage{amssymb}
\usepackage{cite}
\usepackage{hyperref}
\usepackage{graphicx}
\usepackage{booktabs}
\usepackage[caption=false]{subfig}
\usepackage[export]{adjustbox}

\newcommand{\lsim}{\mathrel{\hbox{\rlap{\hbox{\lower4pt\hbox{$\sim$}}}\hbox{$<$}}}}

\newcommand{\myexp}[1]{\ensuremath{\stackrel{\text{#1}}{=}}}

\usepackage{xspace}

\newcommand{\B}{\ensuremath{B}\xspace}
\newcommand{\Bbar}{\kern 0.18em\overline{\kern -0.18em B}{}\xspace}

\newcommand{\Bd }{\ensuremath{\B^0_d}\xspace}

\newcommand{\uBd}{\ensuremath{\B_d^{\vphantom{+}}}\xspace}

\newcommand{\Kbar}{\kern 0.18em\overline{\kern -0.18em K}{}\xspace}

\newcommand{\Bdtopipi}{\mbox{\ensuremath{\Bd\to \pi^-\pi^+}}\xspace}

\usepackage{tikz}
\usetikzlibrary{shapes.geometric, arrows}
\tikzstyle{input2} = [rectangle, rounded corners, text width=2.5cm, minimum height=.5cm, text centered, draw=black, fill=black!35!green!30]
\tikzstyle{input3} = [rectangle, rounded corners, text width=3.5cm, minimum height=.5cm, text centered, draw=black, fill=black!35!green!30]
\tikzstyle{highlight} = [rectangle, thick, rounded corners, text width=2cm, minimum height=.75cm, text centered, draw=black, fill=yellow!10]
\tikzstyle{highlight2} = [rectangle, rounded corners, text width=1.5cm, minimum height=.5cm, text centered, draw=black, fill=gray!20]
\tikzstyle{observable} = [rectangle, rounded corners, text width=3.5cm, minimum height=.5cm, text centered, draw=black, fill=black!35!red!30]
\tikzstyle{bigobservable} = [rectangle, rounded corners, text width=5cm, minimum height=.5cm, text centered, draw=black, fill=black!35!red!30]
\tikzstyle{smallobservable} = [rectangle, rounded corners, text width=2.5cm, minimum height=.5cm, text centered, draw=black, fill=black!35!red!30]
\tikzstyle{parameter} = [rectangle, rounded corners, text width=2.5cm, minimum height=.5cm, text centered, draw=black, fill=black!35!blue!30]
\tikzstyle{parametermid} = [rectangle, rounded corners, text width=3.5cm, minimum height=.5cm, text centered, draw=black, fill=black!35!blue!30]\tikzstyle{parametermid2} = [rectangle, rounded corners, text width=4.5cm, minimum height=.5cm, text centered, draw=black, fill=black!35!blue!30]
\tikzstyle{bigparameter} = [rectangle, rounded corners, text width=6.5cm, minimum height=.5cm, text centered, draw=black, fill=black!35!blue!30]
\tikzstyle{input} = [rectangle, rounded corners, text width=3.5cm, minimum height=.5cm, text centered, draw=black, fill=black!35!green!30]
\tikzstyle{arrow} = [thick,->,>=stealth]
\tikzstyle{leadsto} = [dashed,thick,->,>=stealth]
\tikzstyle{line} = [draw, -latex']
\begin{document}


\begin{titlepage}

\vspace*{-0.0truecm}

\begin{flushright}
Nikhef-2018-021\\
SI-HEP-2018-18\\
QFET-2018-11
\end{flushright}

\vspace*{0.3truecm}

\begin{center}
{\Large \bf \boldmath 
Exploring $B\to\pi\pi, \pi K$ Decays at the 

\vspace*{0.2truecm}

High-Precision Frontier}
\end{center}

\vspace{0.9truecm}

\begin{center}
{\bf Robert Fleischer\,${}^{a,b}$  Ruben Jaarsma\,${}^{a}$, Eleftheria Malami\,${}^a$  and  K. Keri Vos\,${}^{c}$}

\vspace{0.5truecm}

${}^a${\sl Nikhef, Science Park 105, NL-1098 XG Amsterdam, Netherlands}

${}^b${\sl  Department of Physics and Astronomy, Vrije Universiteit Amsterdam,\\
NL-1081 HV Amsterdam, Netherlands}

${}^c${\sl Theoretische Physik 1, Naturwissenschaftlich-Technische Fakult\"at, \\
Universit\"at Siegen, D-57068 Siegen, Germany}

\end{center}

\vspace*{1.7cm}


\begin{abstract}
\noindent
The $B\to\pi\pi,\pi K$ system offers a powerful laboratory to probe strong and weak interactions. Using the isospin 
symmetry, we determine hadronic $B\to\pi\pi$ parameters from data where new measurements of direct CP violation in 
$B^0_d\to\pi^0\pi^0$ resolve a discrete ambiguity. With the help of the $SU(3)$ flavour symmetry, the $B\to\pi\pi$ parameters can be 
converted into their $B\to\pi K$ counterparts, thereby allowing us to make predictions of observables. A particularly interesting 
decay is $B^0_d\to\pi^0 K_{\rm S}$ as it exhibits mixing-induced CP violation. Using an isospin relation, complemented with a robust 
$SU(3)$ input, we calculate correlations between the direct and mixing-induced CP asymmetries of $B^0_d\to\pi^0 K_{\rm S}$, 
which are the theoretically cleanest $B\to\pi K$ probes. Interestingly, they show tensions with 
respect to the Standard Model. Should this $B\to\pi K$ puzzle originate from New Physics, electroweak penguins offer an attractive 
scenario for new particles to enter. We present a strategy to determine the parameters characterising these topologies and 
obtain the state-of-the-art picture from current data. In the future, this method will allow us to reveal the $B\to\pi K$ dynamics 
and to obtain insights into the electroweak penguin sector with unprecedented precision.
\end{abstract}


\vspace*{2.1truecm}

\vfill

\noindent
June 2018

\end{titlepage}





\newpage

\thispagestyle{empty}

\tableofcontents

\newpage




\setcounter{page}{1}


%
%
%
\section{Introduction}
For decades, the $B$-meson system has been an exciting playground for theorists and experimentalists to test the flavour- 
and CP-violating sector of the Standard Model (SM) \cite{BG}, which is encoded in the Cabibbo--Kobayashi--Maskawa (CKM)
matrix \cite{cab,KM}. After an era of pioneering measurements at the $B$ factories with the BaBar and Belle experiments 
as well as the Tevatron, 
the experimental stage is currently governed by the Large Hadron Collider (LHC) with its dedicated $B$-decay experiment LHCb. 
In the near future, Belle II at the KEK Super $B$ Factory will join these explorations, allowing for exciting new opportunities
\cite{Belle-II}, which will be complemented by the LHCb upgrade \cite{LHCb}. 

\begin{figure}[t]
	\centering
	\subfloat[Colour-suppressed tree]{\label{fig:feyn-cstree} \includegraphics[width=0.45\textwidth]{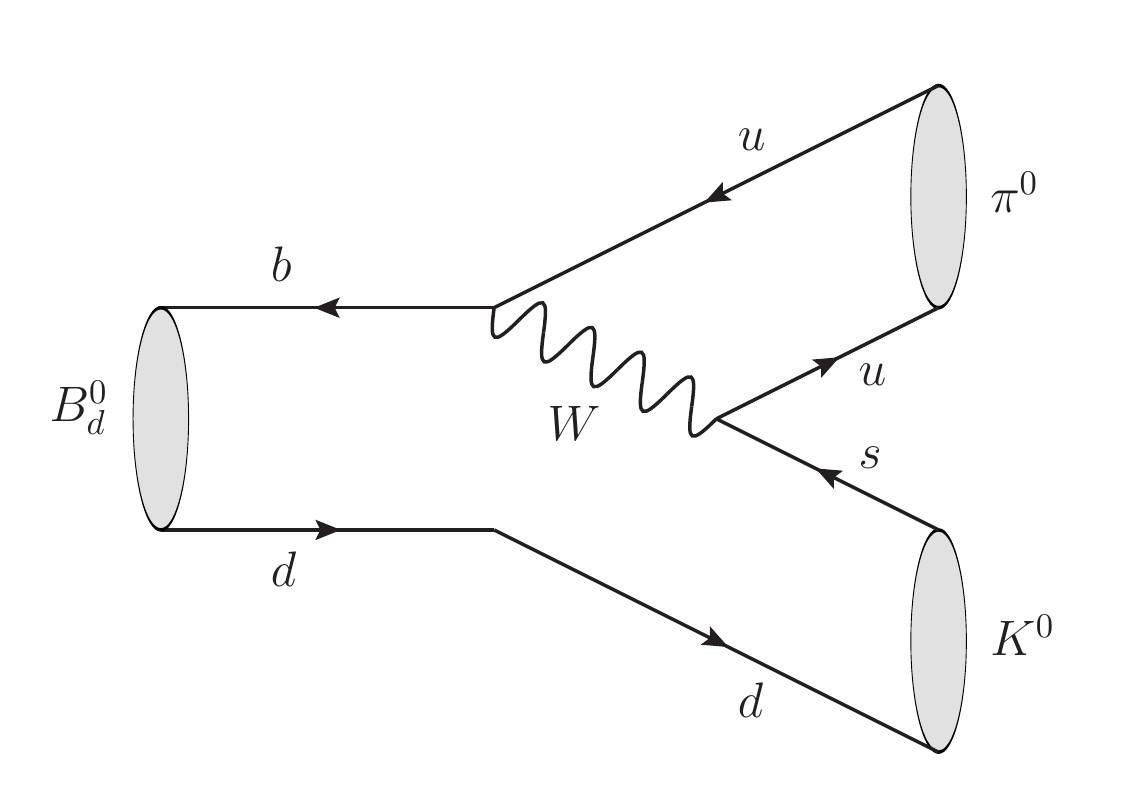}}
	\subfloat[QCD penguin]{\label{fig:feyn-qcdp} \includegraphics[width=0.45\textwidth]{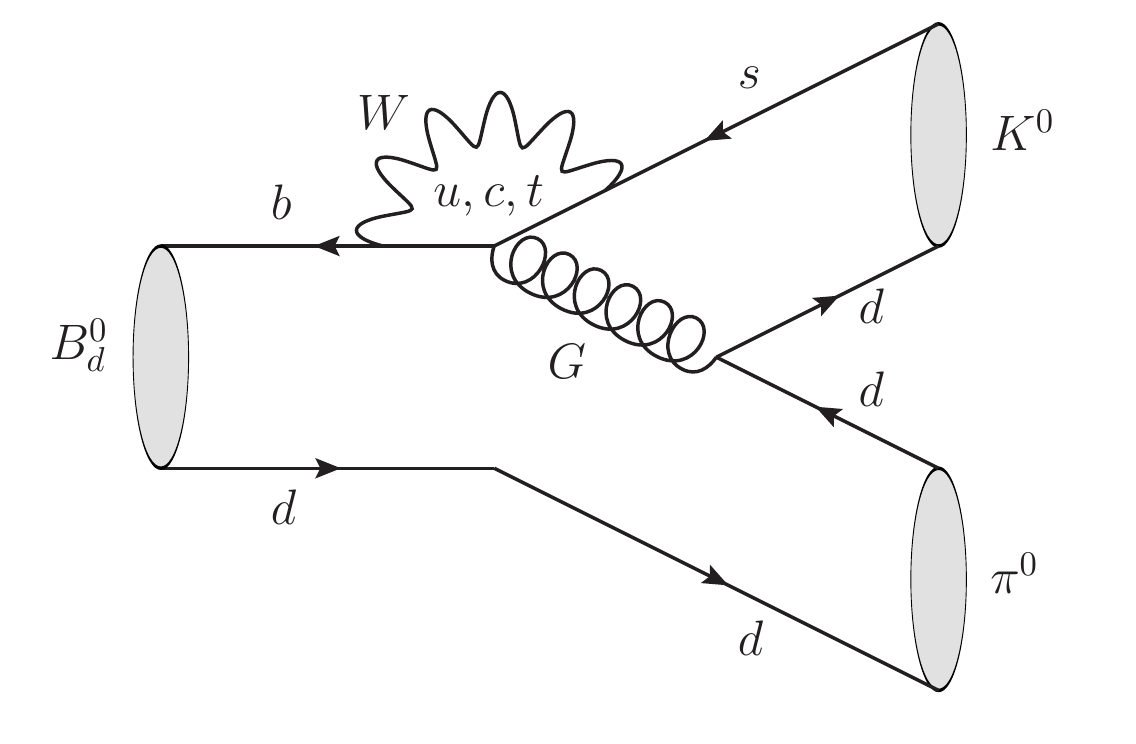}}\\
	\subfloat[Colour-allowed EW penguin]{\label{fig:feyn-caewp} \includegraphics[width=0.45\textwidth]{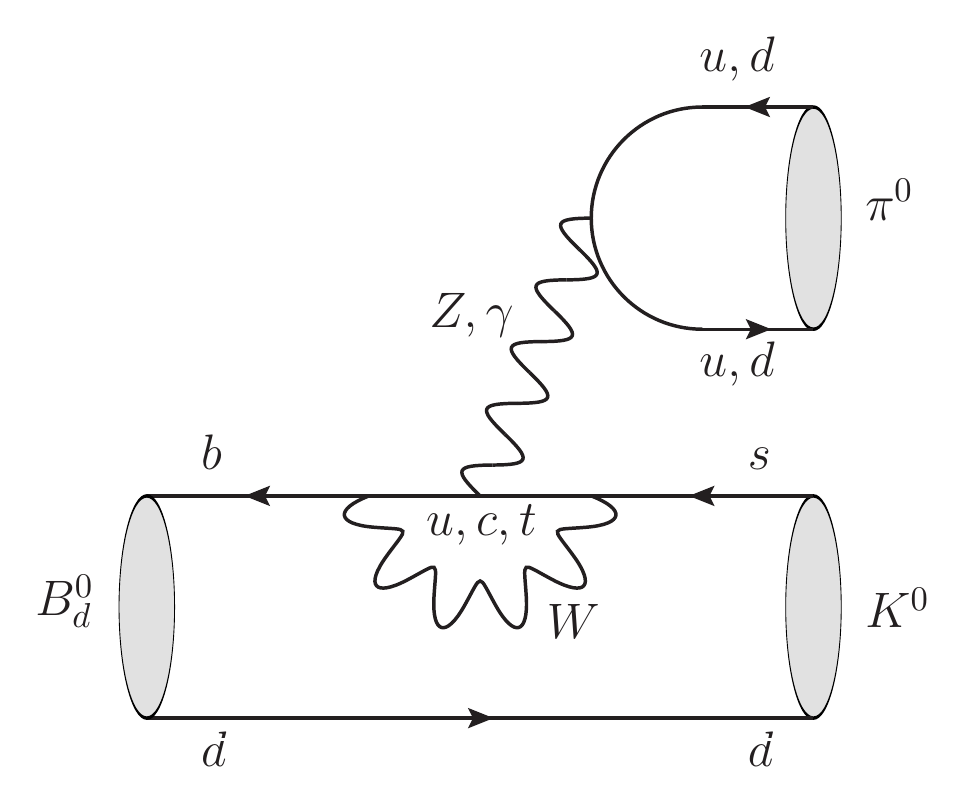}}
	\subfloat[Colour-suppressed EW penguin]{\label{fig:feyn-csewp} \includegraphics[width=0.45\textwidth]{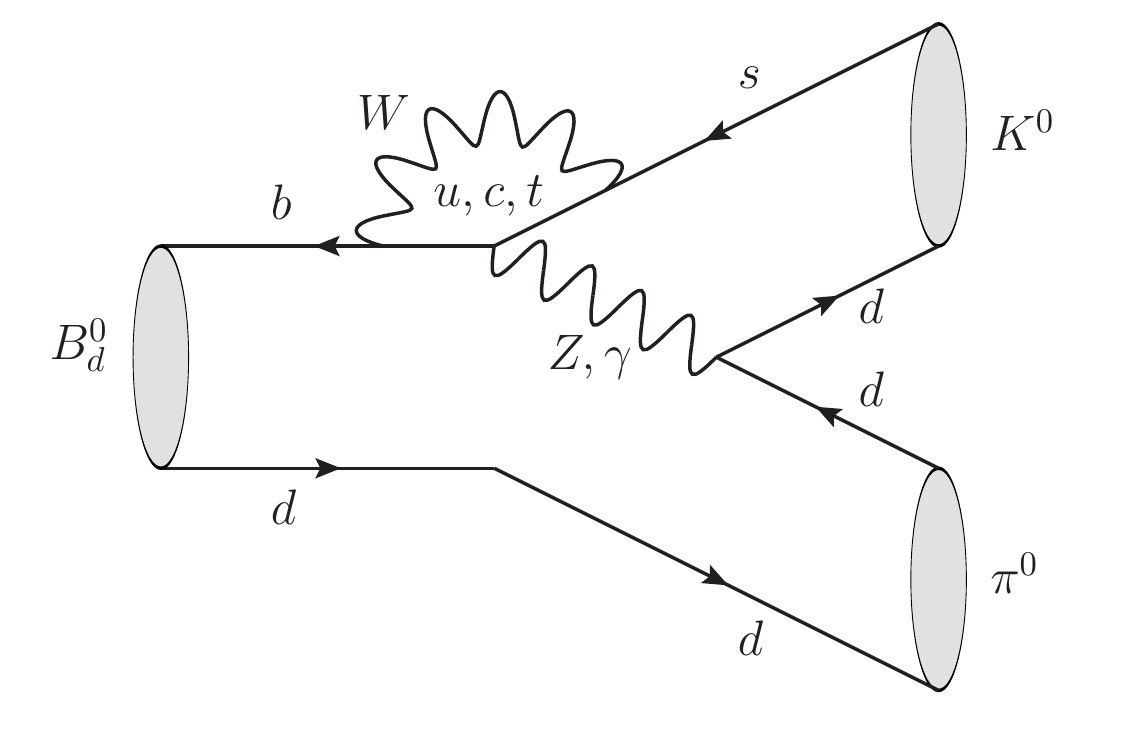}}
	\caption{Topologies contributing to the $B_d^0 \to \pi^0 K^0$ channel.}
	\label{fig:feynman-diagrams}
\end{figure}

In this endeavour, $B\to\pi K$ channels are a particularly interesting decay class (for a selection of original references, see Refs.~\cite{NQ, GHLR, GRL, RF-96, BF-98,Neu-98, BeNe, FRS, groro, BGV}). These modes are dominated by QCD penguin 
topologies as the tree contributions are strongly suppressed by the tiny CKM matrix element $|V_{ub}|$.  In the case of 
$B^+\to\pi^0K^+$ and $B^0_d\to \pi^0K^0$, colour-allowed electroweak (EW) penguin topologies enter at the same level as 
colour-allowed tree amplitudes, contributing ${\cal O}(10\%)$ to the decay amplitudes. As an illustration, we show the decay topologies that contribute to the $B_d^0 \to \pi^0 K^0$ channel in Fig.~\ref{fig:feynman-diagrams}. Since New Physics (NP) may well enter 
through EW penguins \cite{BFRS-1,BFRS-2,BEJLLW-1,BEJLLW-2,RF-FPCP,BDLRR}, these $B\to \pi K$ modes are especially promising. Examples of specific models are given by NP scenarios with extra $Z'$ bosons 
\cite{BEJLLW-1,BEJLLW-2,RF-FPCP,BDLRR}, which are receiving a lot of attention 
in view of anomalies in rare $B$-decay data (see Ref.~\cite{ANSS} and references therein).

In general, NP contributions are associated with new sources of CP violation that can be probed through CP-violating
observables. In this respect, $B^0_d\to\pi^0 K_{\rm S}$ is a particularly interesting decay as it is the only $B\to\pi K$ 
mode exhibiting mixing-induced CP violation \cite{RF-95,FJPZ}. 
This phenomenon emerges from interference between $B^0_d$--$\bar B^0_d$ 
mixing and decay processes of $B^0_d$ and $\bar B^0_d$ mesons into the $\pi^0K_{\rm S}$ final state.  As we will demonstrate
in this paper, the mixing-induced CP asymmetry of $B^0_d\to\pi^0 K_{\rm S}$ plays an outstanding role for testing the SM with the
$B\to\pi K$ system. This paper complements Refs.~\cite{FJV18,Moriond-18}, 
where we gave a compact presentation of the main results discussed in detail below.

Analyses of non-leptonic $B$ decays are in general very challenging due to hadronic matrix elements of four-quark operators entering 
the corresponding low-energy effective Hamiltonians. In the case of the $B\to\pi K$ decays, the flavour symmetries of 
strong interactions imply relations between the $B\to\pi K$ amplitudes and those of the $B\to \pi\pi, KK$ systems, which allow us to
eliminate the hadronic amplitudes or to determine them from experimental data for the latter decays. 

In our analysis, we aim at keeping the theoretical assumptions about strong interactions as minimal as possible, and shall use 
results from QCD factorization (QCDF) to include $SU(3)$-breaking corrections \cite{BeNe}.
 A central role is played by an isospin relation 
between amplitudes of neutral $B\to\pi K$ decays. Complementing it with an $SU(3)$ input to just fix a certain normalisation, 
this relation allows us to 
calculate a correlation between the direct and mixing-induced CP asymmetries of the $B^0_d\to\pi^0 K_{\rm S}$ mode \cite{FJPZ}. 
We find an intriguing tension with the SM, implying either that the current central values of the relevant 
observables will change in the future or signals of NP contributions which involve in particular new sources of CP violation. 

In order to clarify this situation and to reveal the dynamics underlying the EW penguin contributions of the $B\to\pi K$ decays, 
we develop a new strategy to determine the corresponding parameters. It utilises again the isospin relation between the neutral 
$B\to\pi K$ decays as well as its counterpart for the charged modes. As the experimental picture is shaper for the latter case, 
we perform a detailed analysis of these modes, resulting in the currently most stringent constraints on the EW penguin parameters. 
In the future, these quantities can be determined with the help of measurements of the mixing-induced CP asymmetry of 
$B^0_d\to\pi^0 K_{\rm S}$. We illustrate the promising potential of this method by discussing a variety of scenarios. 
The Belle II experiment offers exciting prospects for future measurements of the CP asymmetries in $B^0_d\to \pi^0 K_{\rm S}$ 
\cite{Belle-II}, which will allow us to enter a new territory in terms of precision.  Concerning $B\to\pi K, \pi\pi$ modes with charged 
pions and kaons in the final states, the LHCb upgrade will also have an important impact for the implementation of the new strategy.  
 
The outline of this paper is as follows: in Section~\ref{sec:Btopipi}, we discuss the hadronic parameters following from the
current $B\to\pi\pi$ data, where an important new ingredient is given by measurements of direct CP violation in $B^0_d\to\pi^0\pi^0$.
Having these parameters at hand, we apply the $SU(3)$ flavour symmetry to calculate their $B\to\pi K$ counterparts in 
Section~\ref{sec:BtopiK}, exploring also the impact of $SU(3)$-breaking corrections.
In Section~\ref{sec:1}, we utilize the isospin symmetry to calculate correlations between the CP asymmetries of 
$B^0_d\to \pi^0K_{\rm S}$ and discuss the intriguing picture following from the current measurements. In Section~\ref{sec:qphidet}, 
we present the details of the new method to determine the EW penguin parameters, apply it to the current data and demonstrate 
that we can match the expected experimental precision in the era of Belle II and the LHC upgrade(s) with the theoretical 
uncertainties. Finally, we summarize our conclusions in Section~\ref{sec:Conclusions}.


%
%
%
\section{The $\boldsymbol{B\to \pi\pi}$ system}\label{sec:Btopipi}
\subsection{Amplitude structure}
The $B\to\pi\pi$ system has been studied extensively in Ref.~\cite{BFRS-2}. Here we present an update of the determination 
of the hadronic parameters from the corresponding data, which we will use as input parameters in the 
$B\to\pi K $ analysis. The amplitudes of the charged and neutral $B\to\pi\pi$ decays satisfy the following 
isospin relation \cite{GL}:
\begin{equation}
\sqrt{2}A(B^+\to\pi^+\pi^0)=A(B^0_d\to\pi^+\pi^-)+\sqrt{2}A(B^0_d\to\pi^0\pi^0),
\end{equation} 
and have contributions from colour-allowed tree (${\cal T}$), colour-suppressed tree (${\cal C}$), penguin (${\cal P}$), exchange ($\mathcal{E}$), and penguin-annihilation (${\cal PA}$) topologies. The amplitudes can be parametrised
in the following way  \cite{BFRS-1,BFRS-2}:
\begin{align}
\sqrt{2} A(B^+\to \pi^+\pi^0) {}&=  - \tilde{T} e^{i\gamma}  (1+x e^{i\Delta})(1 + \tilde{q} e^{-i \beta}e^{-i\gamma}) \label{eq:ampli1} \\
\label{eq:ampli2}A(B^0_d\to \pi^-\pi^+) {}&= -\tilde{T} (e^{i\gamma}-de^{i\theta}) \\
\label{eq:ampli3}\sqrt{2}A(B^0_d\to \pi^0\pi^0) {}&= P \left[1+ \frac{x}{d} e^{i\gamma}e^{i(\Delta-\theta)} + \tilde{q} \left(\frac{1+x e^{i\Delta}}{d}e^{-i\theta}e^{-i\beta}\right)\right] \ ,
\end{align}
where 
\begin{align}
\tilde{T} {}&= \lambda^3 A R_b (\mathcal{T} - \mathcal{P}_{tu} + \mathcal{E} - \mathcal{PA}_{tu}) \ , \\
P{}&=\lambda^3 A (\mathcal{P}_{t}-\mathcal{P}_{c}) \ .
\end{align}
We use the notation ${\cal P}_{tq}$ and  ${\cal PA}_{tq}$ for the difference between penguin and penguin-annihilation topologies with internal $t$ and $q$ quarks, respectively, and introduce 
\begin{align}
\label{eq:ddef}d e^{i \theta} &\equiv - \frac{1}{R_b}\frac{\mathcal{P}_{tc} + \mathcal{PA}_{tc}}{\mathcal{T}- \mathcal{P}_{tu} +\mathcal{E} - \mathcal{PA}_{tu}} \ , \\
\label{eq:xdef} x e^{i \Delta} &\equiv \frac{\mathcal{C} + \mathcal{P}_{tu} - \mathcal{E} + \mathcal{PA}_{tu}}{\mathcal{T}- \mathcal{P}_{tu} +\mathcal{E}- \mathcal{PA}_{tu}} \ . 
\end{align}
For the considerations below, it is interesting to consider also the ratio
\begin{equation}\label{eq:rcpipi}
r_{\rm c}^{\pi\pi}e^{i \delta_{\rm c}^{\pi\pi}} \equiv \epsilon R_b \left[\frac{\mathcal{T}+\mathcal{C}}{\mathcal{P}_{tc}+ \mathcal{PA}_{tc}}\right] = -
\frac{\epsilon}{de^{i\theta}}(1+x e^{i\Delta}) \ ,
\end{equation}
where 
\begin{equation}
\epsilon \equiv \frac{\lambda^2}{1-\lambda^2} = 0.0535 \pm 0.0002
\end{equation}
involves the Wolfenstein parameter $\lambda \equiv |V_{us}|=0.22543 \pm 0.00042$, and 
\begin{equation}
R_b \equiv \left(1-\frac{\lambda^2}{2}\right)\frac{1}{\lambda}\left|\frac{V_{ub}}{V_{cb}}\right| = 0.390 \pm 0.030
\end{equation}
measures one side of the UT. Furthermore, $A\equiv |V_{cb}|/\lambda^2= 0.8227^{+ 0.0066}_{- 0.0136}$ is another CKM
factor \cite{Wol83,BLO} (for the numerical values, see Ref.~\cite{Cha15}). The UT angle $\gamma$ can be determined in a 
theoretically clean way from pure tree decays of the kind $B\rightarrow D^{(*)}K^{(*)}$ \cite{gw,ADS} 
(for an overview, see \cite{FR-gam}). In our numerical analyses, we use 
\begin{equation}\label{eq:gammaval}
\gamma = (70 \pm 7)^\circ \ ,
\end{equation}
which is an average of the experimental results compiled by the CKMfitter \cite{Cha15} and UTfit \cite{UTfit} collaborations 
and agrees with Ref.~\cite{BIG}. In the future, the uncertainty of the $\gamma$ determination from pure 
$B\rightarrow D^{(*)}K^{(*)}$ tree decays 
can be reduced to the $1^\circ$ level thanks to Belle II and the LHCb upgrade \cite{Belle-II,LHCb}.

\begin{table}[t]
	\centering
	\begin{tabular}{l | r r  r r }
		Mode & ${\mathcal Br}$$[10^{-6}]$& $A_\text{CP}^f$ & $S_\text{CP}^f$ & Ref. 	 \\
		\hline\hline 	{	\vspace{-0.3cm}}\\
				$B^0_d\to \pi^+ \pi^-$ & $5.12\pm 0.19 $  &$0.31 \pm 0.05$ & $-0.66 \pm 0.06$ & \cite{Amh14,PDG}\\
		$B^0_d\to \pi^0  \pi^0$ & $1.59\pm 0.18$
		& $0.33\pm 0.22$& $-$ & \cite{Lee13, Belle17} \\
				$B^+\to \pi^+ \pi^0$ &$ 5.5\pm 0.4\emph{•}$ & $0.03\pm 0.04 $ & $-$ &\cite{PDG} \\
	\end{tabular}
	\caption{Overview of the currently available $B \to \pi \pi$ measurements. Note that the branching ratios are actually CP-averaged quantities. }
	\label{tab:CPspipi}
\end{table}

In the $B\to\pi\pi$ system, the EW penguin topologies play a very minor role and are described by
\begin{equation}\label{eq:tildeq}
\tilde{q} \equiv \left|\frac{P_{EW} + { P^{{\rm C}}_{EW}}}{T+C}\right| \sim 1.3 \times 10^{-2} \left|\frac{V_{td}}{V_{ub}}\right| \sim 3 \times 10^{-2},
\end{equation} 
where 
\begin{equation}\label{T-C}
T = \lambda^3 A R_b \mathcal{T}, \quad C = \lambda^3 A R_b \mathcal{C}.
\end{equation}
For completeness, we have included them in Eqs.~(\ref{eq:ampli1}--\ref{eq:ampli3}) using the isospin symmetry of the strong interactions \cite{BF-98, Gro98}. Their effect on the determination of the hadronic parameters $d, \theta$ and $x, \Delta$ is negligible given the current uncertainties \cite{BFRS-2}. In the future, these contributions could be taken into account through a more
sophisticated analysis.

\subsection{Observables and hadronic parameters}\label{sec:Bpipi-Par}
In Table~\ref{tab:CPspipi}, we list the $B\to \pi\pi$ data used in our analysis. Here and in the following considerations,
the branching ratios are actually CP-averaged quantities. The branching ratio of $B^0_d \to \pi^0\pi^0$ quoted in Table~\ref{tab:CPspipi} is an average of the BaBar measurement \cite{Lee13}
\begin{equation}
\mathcal{B}r(B^0_d \to \pi^0\pi^0) = (1.83 \pm 0.21 \pm 0.13 ) \times 10^{-6}
\end{equation}
and the recent Belle result \cite{Belle17}
\begin{equation}
\mathcal{B}r(B^0_d \to \pi^0\pi^0) =( 1.31 \pm 0.19 \pm 0.18) \times 10^{-6},
\end{equation}
following from the procedure by the Particle Data Group (PDG) \cite{PDG}. Recently, a new LHCb measurement of the CP asymmetries in $B_d^0 \to \pi^- \pi^+$ came out\cite{Aaij:2018tfw}, where values close to the averages in Table~\ref{tab:CPspipi}
were reported.

For the charged $B$-meson decays, we introduce direct CP asymmetries as
\begin{equation}\label{eq:cpasy}
A_\text{CP}^f  \equiv \frac{\Gamma(B^- \rightarrow \bar{f})-\Gamma(B^+ \rightarrow f)}{\Gamma(B^- \rightarrow \bar{f})+\Gamma(B^+ \rightarrow f)}.
\end{equation}
In case of the decay of a neutral $B_d^0$ meson into a final state that is an eigenstate of the CP operator, we have
the following time-dependent decay rate asymmetry:
\begin{equation}\label{CP-Asy-t}
\mathcal{A}_\text{CP}(t) \equiv \frac{\Gamma(\bar{B}_d^0(t) \rightarrow f) - \Gamma(B_d^0(t) \rightarrow f)}{\Gamma(\bar{B}_d^0(t) \rightarrow f) + \Gamma(B_d^0(t) \rightarrow f)} = A_\text{CP}^f \cos(\Delta M_dt) + S_\text{CP}^f\sin(\Delta M_dt) \ , 
\end{equation}
where the time dependence comes from the oscillations between the $B_d^0$ and  $\bar{B}_d^0$ states, and \mbox{$\Delta M_d\equiv M^{(d)}_{\rm H}-M^{(d)}_{\rm L}$} denotes the mass differences between the 
``heavy" and ``light" $\uBd$ mass eigenstates, respectively \cite{Fle02}. The observable $A_\text{CP}^f$ describes direct CP violation
as in Eq.~(\ref{eq:cpasy}), while $S_\text{CP}^f$ measures mixing-induced CP violation. In Eq.~(\ref{CP-Asy-t}), we neglect 
the decay width difference $\Delta \Gamma_d = (\Gamma^{(d)}_{\rm H}-\Gamma^{(d)}_{\rm L})/\Gamma_d={\cal O}(10^{-3})$. 
First measurements of the direct CP asymmetry of the $B^0_d\to\pi^0\pi^0$ channel are available. In Table~\ref{tab:CPspipi}, 
we quote the average of the BaBar result \cite{Lee13}
\begin{equation}
A_{\rm CP}^{\pi^0\pi^0} = 0.43 \pm 0.26 \pm 0.05 
\end{equation}
and the recent Belle measurement \cite{Belle17}
\begin{equation}
A_{\rm CP}^{\pi^0\pi^0} = 0.14 \pm 0.36 \pm 0.12 \ .
\end{equation}

Using the parameters in Eqs.~(\ref{eq:ddef}) and (\ref{eq:xdef}), we obtain 
\begin{align}
\label{eq:cpasym}
A_{\rm CP}^{\pi^-\pi^+}&=\frac{2 d \sin\theta\sin\gamma}
{1-2 d\cos\theta\cos\gamma+d^2}\:, \\
\label{eq:cpasymmix}
S_{\rm CP}^{\pi^-\pi^+}&=-\left[
\frac{d^{ 2}\sin\phi_d -2  d \cos\theta\sin(\phi_d+\gamma)+\sin(\phi_d+2\gamma)}
{1-2 d\cos\theta\cos\gamma+d^2}\right] 
\end{align}
and
\begin{align}
\label{eq:cpasym00}
A_{\rm CP}^{\pi^0\pi^0}&=\frac{-2 dx \sin(\theta-\Delta)\sin\gamma}
{d^{ 2}+2 dx\cos(\theta-\Delta)\cos\gamma+x^2}\:, \\
\label{eq:cpasymmix00}
S_{\rm CP}^{\pi^0\pi^0}&=-\left[
\frac{d^{ 2}\sin\phi_d +2  d x\cos(\theta-\Delta)\sin(\phi_d+\gamma)+x^2\sin(\phi_d+2\gamma)}
{d^{ 2}+2 dx\cos(\theta-\Delta)\cos\gamma+x^2}\right]\ ,
\end{align}
where 
\begin{equation}\label{phid-expr}
\phi_d=(43.2\pm1.8)^\circ
\end{equation}
denotes the CP-violating $B^0_d$--$\bar B^0_d$ mixing phase. The numerical value follows from an analysis of 
CP violation in $B^0_d\to J/\psi K_{\rm S}$ \cite{PDG}, including corrections from doubly Cabibbo-suppressed penguin 
effects \cite{DeBrFl}. 

Using the value of $\gamma$ in Eq.~\eqref{eq:gammaval} and the current experimental values of the CP asymmetries 
in $\Bdtopipi$, we find the following hadronic parameters \cite{BIG, FJV-S}: 
\begin{equation} \label{eq:dthetaBIG}
d= 0.58 \pm 0.16, \; \; \theta = (151.4 \pm 7.6 )^\circ .
\end{equation}

In order to get a handle on $x$ and $\Delta$,  it is useful to introduce the ratios
\begin{equation}
R_{+-}^{\pi\pi} \equiv 2 \frac{M_{B^+}}{M_{B_d}} \frac{\Phi(m_\pi/M_{B_d}, m_\pi/M_{B_d})}{\Phi(m_{\pi^0}/M_{B^+}, m_\pi/M_{B^+})}\left[\frac{\mathcal Br(B^+ \to \pi^+ \pi^0)}{{\mathcal Br}(B_d^0 \to \pi^+ \pi^-)}\right]\frac{\tau_{B_d^0}}{\tau_B^+} \myexp{exp} 2.00 \pm 0.16 
\end{equation}
and
\begin{equation}
R_{00}^{\pi\pi} \equiv 2 \frac{\Phi(m_\pi/M_{B_d}, m_\pi/M_{B_d})}{\Phi(m_{\pi^0}/M_{B_d}, m_{\pi^0}/M_{B_d})} \left[\frac{{\mathcal Br}(B_d^0 \to \pi^0 \pi^0)}{{\mathcal Br}(B_d^0 \to \pi^+ \pi^-)}\right] \myexp{exp} 0.621 \pm 0.074 \ ,
\end{equation}
where 
\begin{equation}
\Phi(X, Y) = \sqrt{[1-(X+Y)^2][1-(X-Y)^2]}
\end{equation}
is the usual phase-space function. We have used the CP-averaged branching ratios in Table~\ref{tab:CPspipi} and $\tau_{B^+}/\tau_{B_d^0}=1.076 \pm 0.004$ \cite{PDG}. Needless to note, the branching ratios still have large uncertainties, which influence $R_{00}^{\pi\pi}$ accordingly. 
\begin{figure}[t]
	\centering
	\includegraphics[width = 0.5\linewidth]{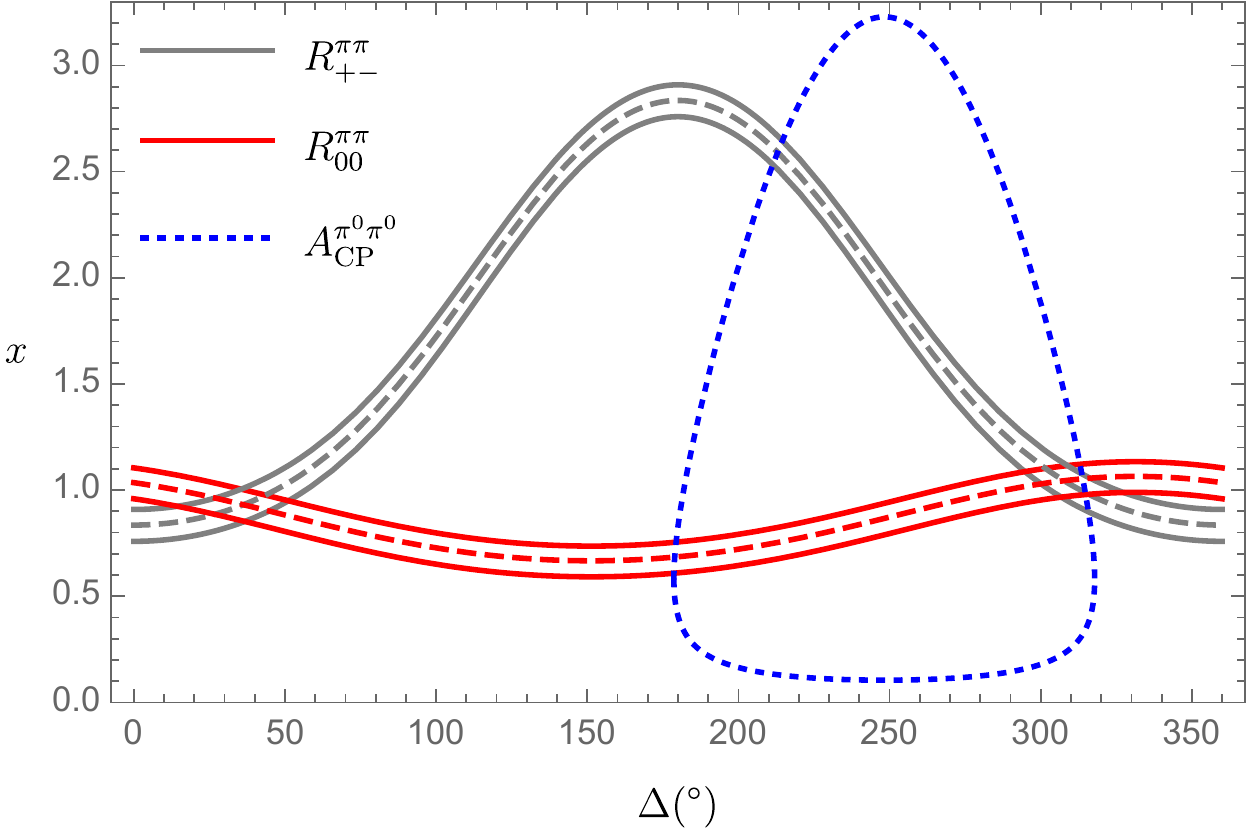}
	\caption{Determination of $x$ and $\Delta$ from the current data for the ratios $R_{+-}^{\pi\pi}$ and $R_{00}^{\pi\pi}$. The two-fold ambiguity can be resolved through the direct CP asymmetry $A_{\text{CP}}^{\pi^0\pi^0}$, where the current experimental central 
	value results in the dotted blue line. }\label{fig:xdel1}
\end{figure}
In terms of the hadronic parameters introduced above, we obtain
\begin{equation}
R_{+-}^{\pi\pi} = \frac{1+ 2x \cos\Delta + x^2}{1-2 d\cos\theta\cos\gamma+d^2}
\end{equation}
and
\begin{equation}
R_{00}^{\pi\pi} =  \frac{d^2+ 2dx \cos(\Delta-\theta)\cos\gamma + x^2}{1-2 d\cos\theta\cos\gamma+d^2}\ ,
\end{equation}
which give
\begin{equation}
x= - \cos\Delta \pm \sqrt{r_\pi R_{+-}^{\pi\pi} -\sin^2\Delta}
\end{equation}
and
\begin{equation}
x = - d \cos\gamma \cos(\Delta - \theta) \pm \sqrt{r_\pi R_{00}^{\pi\pi} - (1-\cos^2(\Delta-\theta))d^2} \ ,
\end{equation}
respectively, with
\begin{equation}
r_\pi = 1-2 d \cos\theta \cos\gamma +d^2.
\end{equation}

Using the current measurements, we illustrate the corresponding contours in Fig.~\ref{fig:xdel1}. Interestingly, the emerging 
twofold ambiguity for $x$ and $\Delta$ can be resolved through the direct CP asymmetry of the $B^0_d \to \pi^0\pi^0$ 
channel, resulting in 
\begin{equation}\label{eq:xDel}
x= 1.04 \pm 0.09 \ , \quad \Delta = -(52.3 \pm 19.3 )^\circ \ .
\end{equation} 
In a previous analysis \cite{BFRS-2}, the ambiguity was resolved with the help of the $SU(3)$ flavour symmetry and the $B^\pm\to\pi^0K^\pm$ channel, where the unphysical solution would result in a large direct CP asymmetry that is excluded
by experimental data. The clean new constraint following from the $A_{\rm CP}^{\pi^0\pi^0}$ is consistent with these considerations. 

Using the parameters determined above, we find the following SM predictions for the CP asymmetries of the $B^0_d\to\pi^0\pi^0$
channel: 
\begin{equation}
A_{\rm CP}^{\pi^0\pi^0}|_{\rm SM} = 0.44 \pm 0.21 \ , \;\; S_{\rm CP}^{\pi^0\pi^0}|_{\rm SM} = 0.81 \pm 0.32 \ ,
\end{equation}
which depend strongly on the value of $\mathcal{B}r(B_d^0\to \pi^0\pi^0)$. In comparison with Ref.~\cite{BFRS-2}, the prediction for the mixing-induced CP asymmetry  moved up by almost $1\sigma$. The Belle II collaboration expects to reach an uncertainty for the $B_d^0\to \pi^0\pi^0$ branching ratio of $\pm 0.03 \rm{(stat.)} \pm 0.05 \rm{(syst.)}$ \cite{Belle-II}, whereas the expected uncertainties of the direct and mixing-induced CP asymmetries are $\pm 0.04$ and $\pm 0.33$, respectively \cite{Belle-II}.

Finally, we  determine the ratio in Eq.~\eqref{eq:rcpipi} as
\begin{equation} \label{eq:rcdeltacpipi}
r_{\rm c}^{\pi\pi} = 0.17 \pm 0.04 \ , \; \;  \delta_{\rm c}^{\pi\pi} = (1.9 \pm 7.5)^\circ \ ,
\end{equation}
where we have used Eqs.~(\ref{eq:dthetaBIG})~and~(\ref{eq:xDel}). We note that $\delta_{\rm c}^{\pi\pi}$ takes a remarkably small
value, which is driven by the determination of the strong phase $\Delta$. The counterpart of this quantity in the $B\to\pi K$ system 
will play an important role in the later discussion. It is interesting to note that the pattern of the values in Eq.~(\ref{eq:rcdeltacpipi}) 
is in accordance with the corresponding QCDF predictions \cite{BBNS,BeNe}.
The determination of the hadronic $B\to \pi\pi$ parameters discussed above is actually clean from the theoretical  point of view as it depends only on isospin relations and the experimental values of $\gamma$ and $\phi_d$.

\begin{table}[t]
	\centering
	\begin{tabular}{l | r r  r }
		Mode & ${\mathcal Br}$$[10^{-6}]$& $A_\text{CP}$ & $S_\text{CP}$ 	 \\
		\hline\hline 	{	\vspace{-0.3cm}}\\
		$\bar{B}^0_d\to \pi^+ K^-$ & $19.6\pm 0.5$  & $ -0.082 \pm 0.006$ & $-$\\
		$\bar{B}^0_d\to \pi^0 \bar{K}^0$ & $9.9\pm 0.5$  &$ 0.00 \pm 0.13$ & $0.58 \pm 0.17$\\
		$B^+\to \pi^+ K_S$ &$ 23.7\pm 0.8$ & $-0.017\pm 0.016 $ & $-$ \\
		$B^+\to \pi^0 K^+ $ &$ 12.9\pm 0.5$ & $0.037\pm 0.021 $ & $-$ \\
			\end{tabular}
	\caption{Overview of the current measurements in the $B\to \pi K$ system \cite{PDG}.}
	\label{tab:CPs}
\end{table}

\section{ The $\boldsymbol{B \to \pi K}$ system}\label{sec:BtopiK}
\subsection{Amplitude structure}
We now focus on the $B\to\pi K$ system. The $B^+\to \pi^+K^0$ and $B^0_d\to \pi^-K^+$ amplitudes have only 
colour-suppressed EW penguins, while $B^+\to \pi^0K^+$ and $B^0_d\to \pi^0K^0$ have in addition contributions 
from colour-allowed EW penguin topologies. The EW penguin contributions are described by the following parameter:
\begin{equation}
q e^{i\phi} e^{i\omega} \equiv - \left(\frac{{\hat{P}}'_{EW} + {\hat{ P}^{'{\rm C}}_{EW}}}{\hat{T}' +\hat{C}'} \right),
\end{equation}
where $\phi$ and $\omega$ are CP-violating and CP-conserving phases, 
and ${\hat{P}'}_{EW}$ ($\hat{T}'$) and $\hat{P}^{'{\rm C}}_{EW}$ ($\hat{C}'$) denote colour-allowed and colour-suppressed 
EW penguin (tree) amplitudes, respectively. 

In the SM, the contribution of the EW penguins can be calculated by using the general expressions for the 
corresponding four-quark operators. The Wilson coefficients of the EW penguin operators $Q_7$ and $Q_8$ are tiny and their contributions can be neglected. The remaining $Q_9$ and $Q_{10}$ operators are Fierz equivalent to the current--current operators $Q_1$ and $Q_2$. Applying then the $SU(3)$ flavour symmetry to the hadronic matrix elements, 
we obtain the following result\cite{BF-98, NR, RF-95}:
\begin{equation}\label{eq:qdef}
	q e^{i\phi} e^{i\omega} \equiv   \frac{-3}{2\lambda^2 R_b}\left[\frac{C_9(\mu) + C_{10}(\mu)}{C_1(\mu) + C_2(\mu)}
	\right]R_q = (0.68\pm0.05) R_q \ , 
\end{equation}
where the $C_i(\mu)$ are perturbative Wilson coefficients \cite{BBNS}.
We observe that the strong phase $\omega$ vanishes in the $SU(3)$ limit. The smallness of this phase is actually a 
model-independent feature, as noted in Ref.~\cite{NR}. In the remainder of this paper, we use $\omega =0^\circ$. 
Making numerical 
studies, we find that values of $\omega$ up to $10^\circ$ would not have an impact on our analysis.  The parameter $R_q$ 
describes  $SU(3)$-breaking effects. Following Ref.~\cite{FJPZ}, we allow for corrections of $30\%$ by taking 
$R_q = 1.0\pm 0.3$.  As a theory benchmark scenario, we assume
\begin{equation}\label{eq:theoryupRq}
R_q = 1.00 \pm 0.05 \ ,
\end{equation}
which is based on expected future progress for lattice calculations of the relevant quantities as discussed in more 
detail in Ref.~\cite{FJPZ}. Since the CP-violating phase $\phi$ vanishes in the SM,  a sizeable  value would be a 
``smoking gun" signal for the presence of NP.

Following Ref.~\cite{BFRS-2}, we parametrize the amplitudes as 
\begin{align}\label{eq:ampli}
	A(B^+\to \pi^+K^0) {}&= - P' \left[1+\rho_c e^{i\theta_c}e^{i\gamma} -\frac{1}{3} \hat{a}_C e^{i \Delta_{\hat 
C}} q e^{i\omega} e^{i\phi}\; r_c e^{i\delta_c} \right] \nonumber \\
	\sqrt{2} A(B^+\to \pi^0K^+) {}&=  P' \left[1+\rho_c e^{i\theta_c}e^{i\gamma} - \left\{ e^{i\gamma} - \left(1-\frac{1}{3} \hat{a}_C e^{i \Delta_{\hat C}}\right)  qe^{i\phi}e^{i\omega}\right\}r_c e^{i \delta_c} \right] \nonumber\\
	A(B^0_d\to \pi^-K^+) {}&= P' \left[1 +\frac{2}{3} a_C e^{i\Delta_C} q e^{i\omega} e^{i\phi}\; r_c e^{i\delta_c}- r e^{i\delta}e^{i\gamma}\right]\nonumber \\
	\sqrt{2}A(B^0_d\to \pi^0K^0) {}&= - P' \left[1-r e^{i\delta}e^{i\gamma}+\left\{e^{i\gamma} -\left(1-\frac{2}{3} a_C e^{i\Delta_C}\right)qe^{i\phi}e^{i\omega}\right\} r_c e^{i\delta_c}\right]  \ .
\end{align}
Here we include the colour-suppressed EW penguin topologies through the parameters 
$a_C$ and $\hat{a}_C$ as well as the CP-conserving phases $\Delta_C$ and $\Delta_{\hat C}$ for the $B_d^0$ and 
$B^+$ decays, respectively. These quantities, which enter with the EW penguin parameters $q$ and $\phi$, are related 
by the isospin symmetry as
\begin{equation}
a_C = \hat{a}_C,  \quad  \Delta_C = \Delta_{\hat{C}} \ ,
\end{equation}
where
\begin{equation}
a_C e^{i \Delta_C} \equiv \frac{\hat{ P}^{'{\rm C}}_{EW}}{\hat{ P}^{'}_{EW} + \hat{ P}^{'{\rm C}}_{EW}} \ .
\end{equation}
The overall normalization of the decay amplitudes in Eq.~(\ref{eq:ampli}) is given by
\begin{equation} 
	P'
	\equiv \frac{\lambda^3 A}{\sqrt{\epsilon}} (\mathcal{P}^\prime_t- \mathcal{P}^\prime_c) \ ,
\end{equation}
where the primes indicate that we are dealing with $\bar{b} \to \bar{s}$ transitions. The $B^+\to \pi^+K^0$ amplitude differs from 
$|P'|$ only through the colour-suppressed EW penguin contributions and 
the doubly Cabibbo-suppressed hadronic parameter
\begin{equation}\label{eq:rhocdef}
	\rho_c e^{i\theta_c} \equiv \left(\frac{\lambda^2R_b}{1-\lambda^2}\right) 
	\left[\frac{\mathcal{P}^\prime_t- \mathcal{\tilde P}^\prime_u - \mathcal{A}'}{\mathcal{P}^\prime_t- \mathcal{P}^\prime_c }\right] \ ,
\end{equation}
where $\mathcal{\tilde P}^\prime_u$ is a QCD penguin  and $ \mathcal{A}'$ an annihilation amplitude. This parameter
can be determined through the $U$-spin symmetry of strong interactions from data for the $B^+\to K^+ \bar{K}^0$ decay
\cite{Fle02, BFRS-2}. The most recent analysis gives the following result \cite{BIG}:
\begin{equation}\label{eq:rhocval}
\rho_{\rm c} = 0.03 \pm 0.01\ , \;\;  \theta_{\rm c} = (2.6 \pm 4.6)^\circ \ , 
\end{equation}
which agrees with the expected order of magnitude of the doubly Cabibbo-suppressed $\rho_{\rm c}$. In particular, no 
anomalously large final-state interaction effects are indicated by the data. It is interesting to note that also the small direct 
CP asymmetry of $B^+\to \pi^+ K^0$ is in agreement with this pattern. 
The remaining hadronic parameters are given by
\begin{equation}\label{eq:rc}
	r_{\rm c} e^{i\delta_{\rm c}}  \equiv \left(\frac{\lambda^2R_b}{1-\lambda^2}\right) \left[\frac{\mathcal{T}^\prime +\mathcal{C}^\prime}{\mathcal{P}^\prime_t- \mathcal{P}^\prime_c }\right]
	 \equiv \frac{\hat{T}' + \hat{C}'}{P'}
\end{equation}
\begin{equation}
	r e^{i\delta}   \equiv \left(\frac{\lambda^2R_b}{1-\lambda^2}\right) \left[\frac{\mathcal{T}^\prime -(\mathcal{P}^\prime_t- \mathcal{P}^\prime_u) }{\mathcal{P}^\prime_t- \mathcal{P}^\prime_c }\right] 
	\equiv \frac{\hat{T}' - \hat{P}_{tu}'}{P'}\ ,
\end{equation}
where the normalized amplitudes 
\begin{equation}
\hat{T}' = |V_{ub}V_{us}^*|\mathcal{T}' \quad\mbox{and}\quad
\hat{C}' = |V_{ub}V_{us}^*| \mathcal{C}',
\end{equation}
describe, in analogy to Eq.~(\ref{T-C}), the colour-allowed and colour-suppressed tree-diagram contributions, respectively.

\subsection{Determination of the hadronic parameters}\label{sec:dethad}
The $B \to \pi K$ system is related to the $B \to \pi \pi$ modes through the $SU(3)$ flavour symmetry of strong interactions, which 
allows us to convert the $B\to\pi\pi$ parameters determined in Subsection~\ref{sec:Bpipi-Par} into their $B\to\pi K$ counterparts 
\cite{BFRS-1, BFRS-2}. As EW penguins play a negligible role in the $B\to\pi\pi$ system, the resulting hadronic $B\to\pi K$ 
parameters are essentially not affected by possible NP contributions to the EW penguin sector. 

A complication arises from exchange ($E$) and penguin-annihilation ($PA$) topologies, which are present in the $B \to \pi\pi$ 
system but do not contribute to the $B \to \pi K$ modes. These contributions are dynamically suppressed and expected to play 
a minor role. Using data for $B_s^0 \to \pi^-\pi^+$ and $B_d^0\to K^- K^+$ decays \cite{Aaij:2016elb}, which exclusively emerge from such topologies, 
and the $SU(3)$ flavour symmetry, the $E$ and $PA$ contributions can be constrained. As discussed in detail in Ref.~\cite{BIG}, this 
results in effects at the few percent level of the overall $B\to\pi K$ amplitudes. In the future, measurements of CP asymmetries
in the $B_s^0 \to \pi^-\pi^+$ and $B_d^0\to K^- K^+$ decays will result in more precise determinations of these effects \cite{BIG},
allowing us to take these corrections into account.

Let us now determine the hadronic $B\to \pi K$ parameters from $B\to\pi\pi$ decays. First, we discuss $r_{\rm c}$ and 
$\delta_{\rm c}$ with their counterparts in the $B\to \pi\pi$ system as given in Eq.~\eqref{eq:rcpipi}. In the limit of the $SU(3)$ 
flavour symmetry, we have
\begin{equation} \label{eq:rcpipiuspin}
r_{\rm c}e^{i \delta_{\rm c}} = r_{\rm c}^{\pi\pi} e^{i \delta_{\rm c}^{\pi\pi}},
\end{equation}
which is not affected by factorizable $SU(3)$-breaking corrections if contributions from the colour-suppressed tree topology are neglected. The $B^+\to \pi^+ \pi^0$ decay allows us to determine the $|\mathcal{T}+\mathcal{C}|$ amplitude, 
which can be converted into its $B\to\pi K$ counterpart using
\begin{equation}
|\mathcal{T}' + \mathcal{C}'| = R_{T+C} |\mathcal{T} + \mathcal{C}|,
\end{equation}
where $R_{T+C}$ parameterizes $SU(3)$-breaking effects. We can write this quantity  as
\begin{equation}
R_{T+C} = \left|\frac{\mathcal{T}'}{\mathcal{T}}\right|\left|\frac{1+\kappa'}{1+\kappa}\right| 
\end{equation}
with
\begin{equation}
\kappa^{(\prime)} \equiv \frac{\mathcal{C}^{(\prime)}}{\mathcal{T}^{(\prime)}} \ ,
\end{equation}
which is expected to take a value at the 0.3 level. Within the factorization framework, we obtain
\begin{equation} \label{eq:rtcfact}
R_{T+C}|_\text{fact}\equiv \left|\frac{\mathcal{T}'}{\mathcal{T}}\right|_\text{fact} = \frac{f_K}{f_\pi} = 1.1928 \pm 0.0026 \ ,
\end{equation}
where we have used the numerical value of $f_K/f_\pi$ given in Ref.~\cite{Rosner:2015wva}. 

Finally, we may determine $|P'|$ from the $B^+\to \pi^+ K^0$ branching ratio, yielding the following relation \cite{BFRS-2}:
\begin{equation}\label{r_c}
r_{\rm c} = \sqrt{2} \left|\frac{V_{us}}{V_{ud}}\right| R_{T+C}
\sqrt{r_\rho \left[\frac{\Phi(m_\pi/m_{B^+},m_{K^0}/m_{B^+})}{\Phi(m_\pi/m_{B^+},m_{\pi^0}/m_{B^+})}\right] 
\left[ \frac{\mathcal{B}r(B^+\to \pi^+ \pi^0)}{\mathcal{B}r(B^+\to \pi^+ K^0)} \right] } \ ,
\end{equation}
where we take also the small correction from $\rho_{\rm c}$ through 
\begin{equation}
r_\rho = 1 + 2\rho_{\rm c}\cos\theta_{\rm c}\cos\gamma + \rho_{\rm c}^2
\end{equation}
with the values in Eq.~(\ref{eq:rhocval}) into account. Using $R_{T+C}$ from Eq.~\eqref{eq:rtcfact} and the most 
recent values of the CKM matrix elements in Ref.~\cite{Cha15} yields
\begin{equation} \label{eq:rcFrompippi0}
r_{\rm c} = 0.19 \pm 0.01 \ .
\end{equation}

\begin{figure}[t]
	\centering
	\subfloat[]{\label{fig:combi-full} \includegraphics[height=0.4\textwidth]{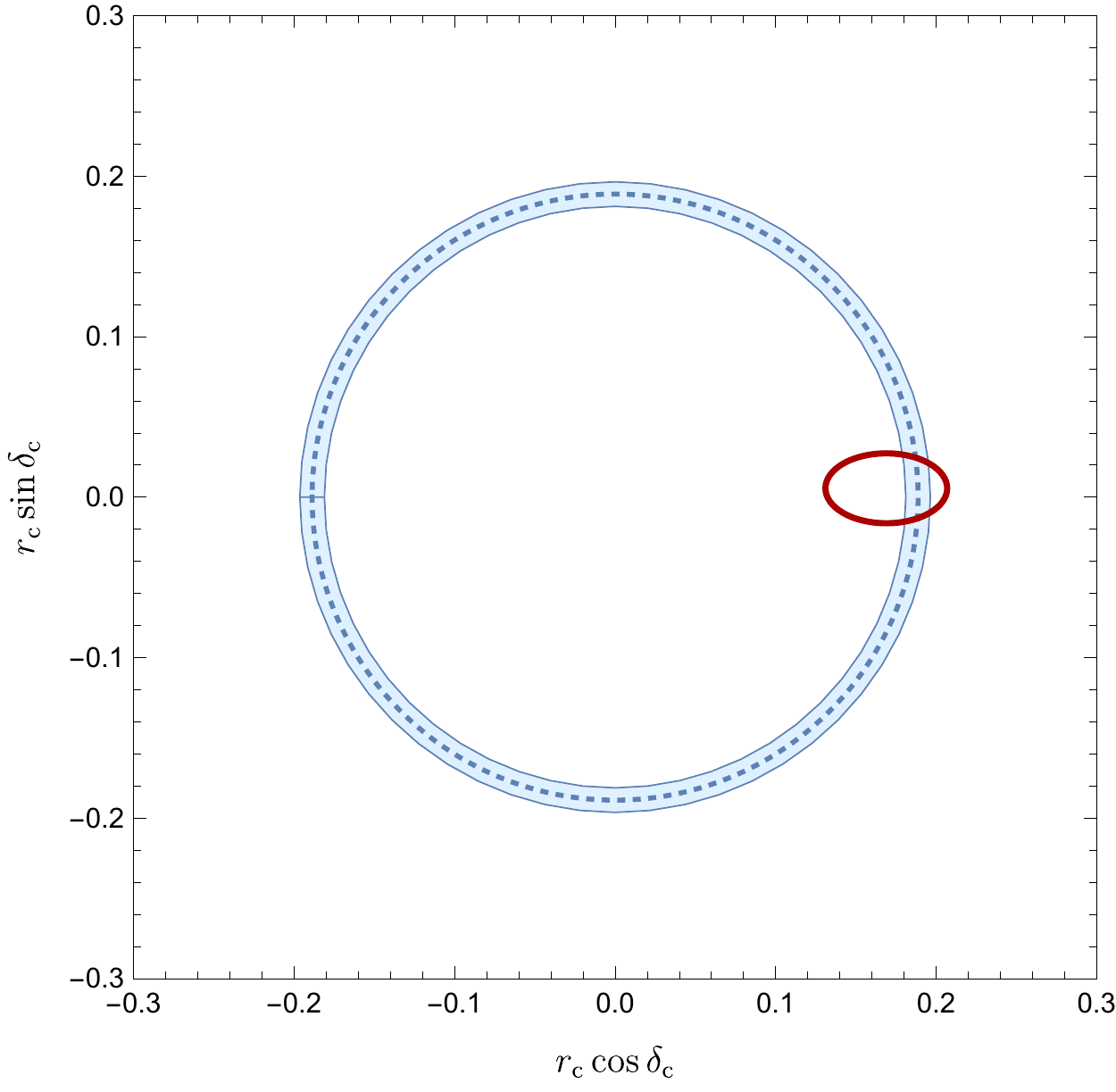}}
	\subfloat[]{\label{fig:combi-zoom} \includegraphics[height=0.4\textwidth]{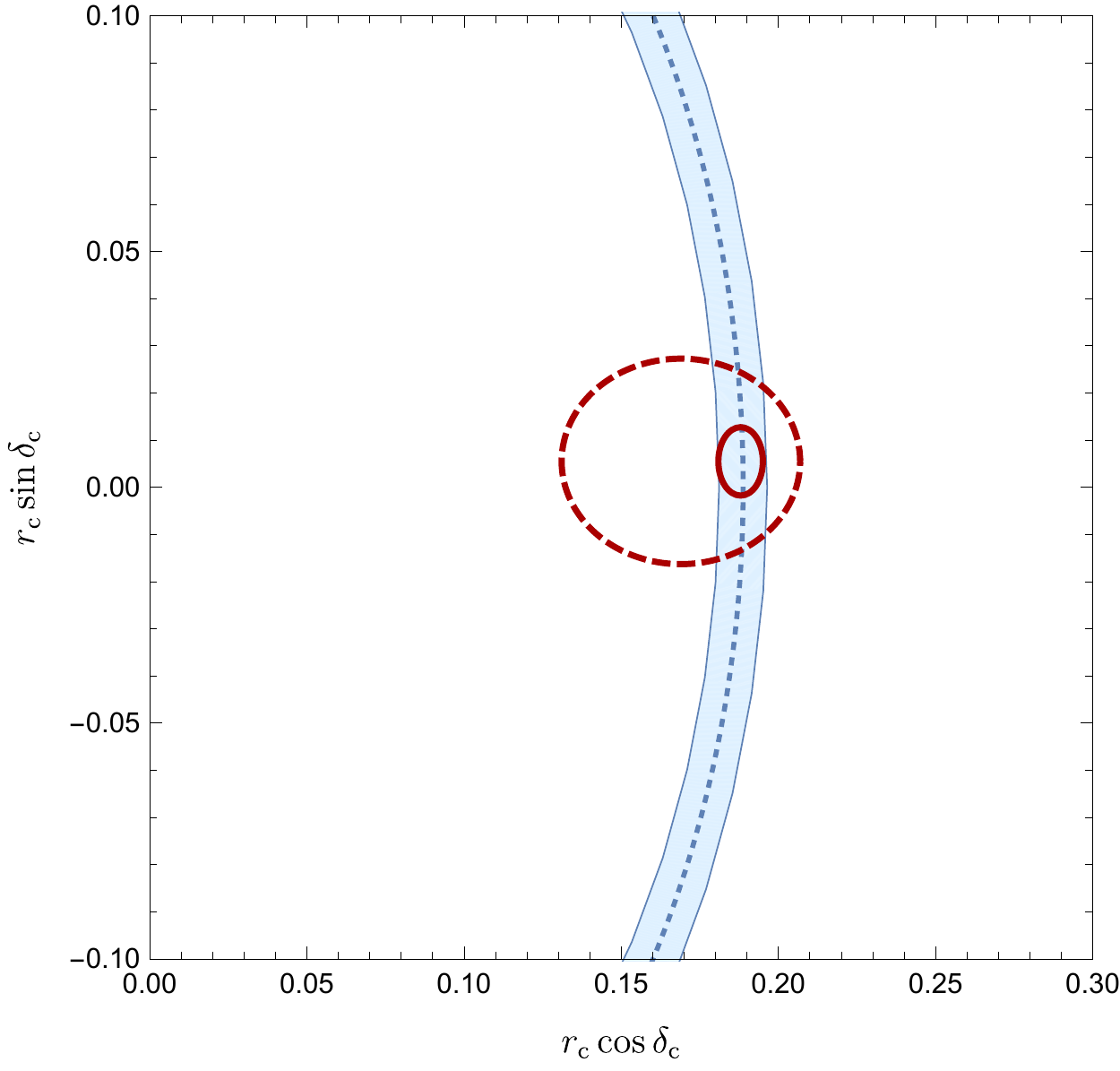}}
	\caption{Constraints on $r_{\rm c}$ and $\delta_{\rm c}$: (a) the blue circle depicts the $1\sigma$ constraints from Eq.~(\ref{r_c}) with $R_{T+C}|_{\rm fact}$ while the red ellipse follows from the $B\to \pi\pi$ data yielding the results in Eq.~\eqref{eq:rcdeltacpipi}; (b) scenario for the expected future precision as discussed in the text.}
	\label{fig:combi}
\end{figure}

In Fig.~\ref{fig:combi-full}, we compare this determination with $r_{\rm c}^{\pi\pi}$ and $\delta_{\rm c}^{\pi\pi}$ in Eq.~\eqref{eq:rcdeltacpipi}. Here the latter parameters give the red ellipse, whereas the blue circle follows from Eq.~(\ref{eq:rcFrompippi0}). In Fig.~\ref{fig:combi-zoom} we zoom in on the red ellipse and show the precision that can be obtained for $r_{\rm c}^{\pi\pi}$ and $\delta_{\rm c}^{\pi\pi}$ in the era of Belle II and the LHCb upgrade, using the expected uncertainty for the $B\to\pi\pi$ observables \cite{Belle-II}, as well as $\gamma = (70 \pm 1)^\circ$ \cite{LHCb, Belle-II} and $\phi_d = (43.2 \pm 0.6)^\circ$ \cite{DeBrFl}. We have shifted the ellipse to get agreement with the blue contour, and observe that both constraints have
actually similar precision. In order to guide the eye, we have also added the dashed red ellipse which corresponds to the current data.

The impressive agreement between the two determinations in Fig.~\ref{fig:combi-full} does not indicate non-factorizable $SU(3)$-breaking corrections within the current experimental precision. In order to quantify this feature, we reverse Eq.~(\ref{r_c}) and use it to determine $R_{T+C}$ from the value of $r_{\rm c}$ in Eq.~(\ref{eq:rcdeltacpipi}), yielding
\begin{equation}\label{RTC-det}
R_{T+C} = 1.07 \pm 0.23 .
\end{equation}
This value agrees with 
\begin{equation} \label{eq:rtcwithnf}
R_{T+C} = 1.2 \pm 0.2 \ 
\end{equation}
given in Ref.~\cite{FJPZ}, where non-factorizable $SU(3)$-breaking corrections as large as
$100\%$ of the factorizable effects in Eq.~(\ref{eq:rtcfact}) were considered. 

Utilizing non-perturbative QCD rum rule techniques and allowing for non-factorizable effects, the parameter $R_{T+C}$
was calculated in Ref.~\cite{Khodjamirian:2003xk}:
\begin{equation} \label{eq:rtcqcdsr}
	R_{T+C} = \left|(1.21_{-0.014}^{+0.015}) + (0.008_{-0.015}^{+0.013})i\right| = 1.21 \pm 0.015.
\end{equation}
Interestingly, a small CP-conserving strong phase arises in this calculation, resembling a picture in analogy to 
Eq.~(\ref{eq:rcdeltacpipi}). Moreover, the agreement between Eqs.~(\ref{eq:rtcfact}) and (\ref{eq:rtcqcdsr}) indicates that non-factorizable effects have actually a small impact on this parameter. 

In the following discussion, $R_{T+C}$ is a key quantity. For the numerical analyses of the current data, we shall use the value
in Eq.~(\ref{eq:rtcwithnf}). In view of the discussion in the previous paragraph, the corresponding uncertainty is conservative. 
As a future benchmark scenario, we follow Ref.~\cite{FJPZ}, and assume
\begin{equation}\label{eq:theoryuprtc} 
R_{T+C}= 1.22 \pm 0.02  
\end{equation}
as a result from expected progress in lattice QCD calculations.

We include non-factorizable corrections to the relation in Eq.~(\ref{eq:rcpipiuspin}) via
\begin{equation}
r_{\rm c}e^{i \delta_{\rm c}} = \xi_{SU(3)}^{r_{\rm c}}r_{\rm c}^{\pi\pi} e^{i(\Delta_{SU(3)}^{r_{\rm c}} + \delta_{\rm c}^{\pi\pi})} \ ,
\end{equation}
where $\xi_{SU(3)}^{r_{\rm c}}$ and $\Delta_{SU(3)}^{r_{\rm c}}$ parametrize the $SU(3)$-breaking effects. Considering non-factorizable corrections of up to $20 \%$ through
\begin{equation}\label{eq:uspincor}
\xi_{SU(3)}^{r_{\rm c}} = 1.0 \pm 0.2, \quad \Delta_{SU(3)}^{r_{\rm c}} = (0 \pm 20)^\circ 
\end{equation}
 yields
\begin{align} \label{eq:rcval}
r_{\rm c} = 0.17 \pm 0.04|_\text{input} \pm 0.03|_{SU(3)} = 0.17 \pm 0.05 \ , \nonumber \\
\delta_{\rm c} = (1.9 \pm 7.5|_\text{input} \pm 20.0|_{SU(3)})^\circ = (1.9 \pm 21.4)^\circ \ ,
\end{align}
where we give the errors of the individual input parameters and add them in quadrature.

Let us now determine the parameters $r$ and $\delta$ which enter the amplitude of the $B_d^0\to \pi^-K^+$ channel. They
are related to their \Bdtopipi counterparts through the $SU(3)$ relation
\begin{equation} \label{eq:rdeltafromBtopipmpipm}
re^{i\delta} = - \frac{e^{-i\Delta_{SU(3)}^d}}{\xi_{SU(3)}^d} 
\left[\frac{\epsilon}{d} e^{-i\theta} \right]\ ,
\end{equation}
where $\xi_{SU(3)}^d$ and $\Delta_{SU(3)}^d$ describe non-factorizable $SU(3)$-breaking corrections. 
Allowing again for such effects of $20\%$ through
\begin{equation}
\xi_{SU(3)}^d = 1.0 \pm 0.2, \quad \Delta_{SU(3)}^d = (0 \pm 20)^\circ,
\end{equation}
we find
\begin{align} \label{eq:rdelfromdtheta}
r = 0.09 \pm 0.03|_\text{input} \pm 0.02|_{SU(3)} = 0.09 \pm 0.03  \ , \nonumber \\
\delta = ( 28.6\pm 7.6|_\text{input} \pm 20.0|_{SU(3)})^\circ = (28.6 \pm 21.4)^\circ \ .
\end{align}
These quantities have also been obtained with the help of data for the $B^0_s \to \pi^+ K^-$ channel, 
which is the $U$-spin partner of the $B^0_d\to\pi^-K^+$ decay \cite{BIG}:
\begin{align} \label{eq:rdeltaBsPiK}
r = 0.10 \pm 0.01|_\text{input} \pm 0.02|_{SU(3)} = 0.01 \pm 0.02 \ , \nonumber \\
\delta = (24.6 \pm 3.3|_\text{input} \pm 20.0|_{SU(3)})^\circ = (24.6 \pm 20.3)^\circ \ .
\end{align}
The impressive agreement between Eqs.~(\ref{eq:rdelfromdtheta})~and~(\ref{eq:rdeltaBsPiK}) does not indicate any anomalously large $SU(3)$-breaking effects or contributions from exchange and penguin-annihilation topologies.

\boldmath
\subsection{Observables and dynamics} \unboldmath \label{sec:btopiK-obs}
\subsubsection{Branching ratios}

\begin{table}
  \centering
    \begin{tabular}{l | r }
    Parameter & Value  \\
    \hline \hline
   $\gamma$ & $(70 \pm 7)^\circ$  \\
      $\phi_d$ & $(43.2 \pm 1.8)^\circ$  \\
    \hline
        $r$ & $0.09 \pm 0.03 $ \\
    $\delta $ & $(28.6\pm 21.4)^\circ$ \\
    $r_{\rm c}$ & $0.17 \pm 0.05$ \\
    $\delta_{\rm c} $ & $(1.9 \pm 21.4)^\circ$ \\
        $\rho_{\rm c}$ & $0.03 \pm 0.01 $ \\
    $\theta_{\rm c} $ & $(2.6\pm 4.6)^\circ$ 
    \end{tabular}
    \caption{Input and hadronic $B\to\pi K$ parameters obtained from the current $B\to \pi\pi$ data, including 
    uncertainties from $SU(3)$-breaking effects as discussed in Sec.~\ref{sec:dethad}.}
    \label{tab:sumpara}
\end{table}

It is useful to introduce the following ratios of the branching ratios of the four $B\to \pi K$ modes \cite{BF-98, NR, FM}:
\begin{align}\label{eq:R}
R {}&\equiv  \left[\frac{{\mathcal Br}(B_d^0 \to \pi^- K^+) }{{\mathcal Br}(B^+ \to \pi^+ K^0) }	
\right]\frac{\tau_{B^+}}{\tau_{B^0_d}} \myexp{exp} 0.89\pm 0.04 \ , \\
\label{eq:Rc}
R_{\rm c}{}&\equiv 2 \left[\frac{{\mathcal Br}(B^+\to\pi^0 K^+)}{{\mathcal Br}(B^+\to\pi^+K^0)}\right] \myexp{exp} 1.09\pm 0.06 \ , \\
\label{eq:Rn}
R_{\rm n}{}&\equiv \frac{1}{2} \left[\frac{{\mathcal Br}(B_d^0\to\pi^- K^+}{{\mathcal Br}(B^0_d\to\pi^0K^0)}\right]  
\myexp{exp} 0.99\pm 0.06 \ ,
\end{align}
where the values are obtained from the current data summarized in Table~\ref{tab:CPs}. The ratios $R_{\rm c}$ and 
$R_{\rm n}$ depend on the EW penguin parameters $q$ and $\phi$, while $R$ only involves colour-suppressed 
EW penguins. Using the expressions in Eq.~\eqref{eq:ampli}, we can express these ratios in terms of the hadronic 
parameters introduced above. 

It is instructive to use the fact that $r$ and $r_{\rm c}$ are 
small parameters of $\mathcal{O}(0.1)$, and make expansions in terms of $r_{\rm (c)}$, which yields
\begin{eqnarray}
R_{\rm c}& = & 1- 2 \, r_{\rm c}\cos\delta_{\rm c}(\cos\gamma-q\cos\phi)+{\cal O}(r_{\rm c}^2) \ ,\label{Rc-expr} \\
R_{\rm n} & = & 1-2r_{\rm c}\cos\delta_{\rm c}(\cos\gamma-q\cos\phi) + {\cal O}(r_{({\rm c})}^2) \ .
\end{eqnarray}
We note an interesting relation:
\begin{equation}\label{eq:Rminus}
R_{\rm c}-R_{\rm n}=0+{\cal O}(r_{({\rm c})}^2) = 0.10\pm0.08,
\end{equation}
where the numerical value follows from the experimental results in Eqs.~\eqref{eq:Rc} and \eqref{eq:Rn}. Consequently, the relation 
is actually satisfied by the data at the $1\sigma$  level. 

\subsubsection{Colour-suppressed electroweak penguins}\label{ssec:CSEWP}
In the case of the observable $R$, we obtain
\begin{equation}\label{R-expr}
R=1-2\,r\cos\delta\cos\gamma+2 \, r_{\rm c} \, \tilde a_{\rm C} \, q \cos\phi -2 \rho_{\rm c} \cos\theta_{\rm c} \cos\gamma+ {\cal O}(r_{\rm (c)}^2, \rho_c^2) \ ,
\end{equation}
where 
\begin{equation} \label{eq:actil}
 \tilde{a}_C  \equiv a_C \cos(\delta_{\rm c} + \Delta_C) 
\end{equation}
describes the colour-suppressed EW penguin topologies. The direct CP asymmetry of the $B_d^0\to \pi^- K^+$ channel 
takes the form 
\begin{equation}\label{ACPpi-K+}
A_{\text{CP}}^{\pi^-K^+}\equiv\mathcal{A}^{\rm dir}_{\rm CP} (B_d^0\to \pi^- K^+)= \frac{4}{3} r_c\; \tilde{a}_S q \sin\phi - 
2 r \sin\delta \sin\gamma  +{\cal O}(r_{\rm (c)}^2) 
\end{equation}
with
\begin{equation}\label{eq:astil}
\tilde{a}_S  \equiv  a_C \sin(\delta_c + \Delta_C) \ .
\end{equation}
The parameter $\tilde{a}_S$ enters also the direct CP asymmetries of the other $B\to\pi K$ decays. For small phases $\delta_c$ 
(see Eq.~(\ref{eq:rcval})) and $\Delta_C$, the sine term results in a strong suppression of $\tilde{a}_S$.
Having the hadronic parameters in Subsection~\ref{sec:dethad} at hand, $R$ and $A_{\text{CP}}^{\pi^-K^+}$ 
allow the determination of the colour-suppressed EW penguin contributions $\tilde{a}_C$ and $\tilde{a}_S$. 
Neglecting  
sub-leading terms, we find
\begin{align}
\tilde{a}_S \;q \sin\phi   & = \frac{3 (A_{\rm CP}^{\pi^-K^+}  + 2r\sin\delta \sin\gamma)}{4 r_c }   \ , \\ 
 \tilde{a}_C \;q \cos\phi  {}& = \frac{R - 1 + 2r \cos\delta \cos\gamma+ 2\rho_c \cos\theta_c \cos\gamma}{2 r_c }  \label{eq:actildet}\ . 
\end{align}
Using the parameters in Table~\ref{tab:sumpara} gives
\begin{equation}\label{eq:tildeacsm}
\tilde a_{\rm C} \, q \cos\phi = -0.10 \pm 0.15,\quad \tilde a_{\rm S} \, q \sin\phi = -0.005\pm 0.274 \ .
\end{equation}
Assuming the SM value of $q$ in Eq.\ \eqref{eq:qdef}, we obtain
\begin{equation}
\tilde a_{\rm C}|_{\rm SM}=-0.15 \pm 0.23,
\end{equation}
which supports the expectation that colour-suppressed EW penguins play a minor role.

\subsubsection{Direct CP asymmetries and sum rules}\label{sec:SR}
Performing again expansions in the small $r_{\rm c}$ as well as the tiny $\rho_{\rm c}$ yields
\begin{align}\label{eq:cp1}
A_{\text{CP}}^{\pi^+K^0}\equiv A_{\rm CP}^{\rm dir}(B^+ \to \pi^+ K^0){}& = 
2\rho_{\rm c}\sin\theta_{\rm c}\sin\gamma-\frac{2}{3}\tilde{a}_S qr_{\rm c}\sin\phi + {\cal O}(r_{\rm c}^2,\rho_{\rm c}^2) \ , \nonumber \\
A_{\text{CP}}^{\pi^0K^
+} \equiv A^{\rm dir}_{\rm CP}(B^+\to \pi^0 K^+) {}&= 2\rho_{\rm c}\sin\theta_{\rm c}\sin\gamma -
2r_{\rm c}\sin\delta_{\rm c}[\sin\gamma-q\sin\phi] \nonumber \\
& - \frac{2}{3}\tilde{a}_{S}qr_{\rm c} \sin\phi +{\cal O}(r_{\rm c}^2,\rho_{\rm c}^2) \ , \nonumber \\
A^{\pi^0 K^0}_{\text{CP}}\equiv 
A^{\rm dir}_{\rm CP}(B^0_d\to \pi^0 K^0)  &= 
2r_{\rm c}\sin\delta_{\rm c}[\sin\gamma-q\sin\phi]+\frac{4}{3}\tilde{a}_{S}qr_{\rm c} \sin\phi \nonumber \\
&- 2r\sin\delta\sin\gamma + {\cal O}(r_{({\rm c})}^2) \ ,
\end{align}
which complement the expression in Eq.~(\ref{ACPpi-K+}). Interestingly, the contribution from $\tilde{a}_S$ vanishes in the 
case of $\phi=0^\circ$, which includes also the SM. 

Using the information encoded in the CP-averaged branching ratios, we obtain
\begin{eqnarray}
\label{eq:sum-rule-I}
\Delta_{\rm SR}^{({\rm I})} &=& A_\text{CP}^{\pi^\pm K^\mp} +  A_\text{CP}^{\pi^\pm K^0} 
\frac{\mathcal{B} r(B^{+}\to\pi^{+}  K^0)}{\mathcal{B} r (B^0_d\to\pi^-  K^+)} \frac{\tau_{B^0}}{\tau_{B^+}} 
- A_\text{CP}^{\pi^0K^\pm} \frac{2 {\mathcal{B} r(B^{+}\to\pi^0  K^+)}}{\mathcal{B}r(B^0_d\to\pi^-  K^+)} 
\frac{\tau_{B^0}}{\tau_{B^+}} \nonumber \\ 
&-& A_\text{CP}^{\pi^0 K^0} \frac{2  \mathcal{B}r(B^0_d\to\pi^0  K^0)}{\mathcal{B}r(B^0_d\to\pi^-  K^+)} 
= 0 + {\cal O}(r_{({\rm c})}^2,\rho_{\rm c}^2) \ ,
\end{eqnarray}
which offers an interesting test of the SM. This sum rule was actually pointed out in Refs.~\cite{GR,gro}. 
 Evaluating the sub-leading terms gives
\begin{equation} \label{eq:SR-theo-1}
\Delta_{\rm SR}^{({\rm I})}
 = 2qr_{\rm c}\left[\frac{r\sin(\delta_{\rm c}-\delta)+\rho_{\rm c}\sin(\delta_{\rm c}-\theta_{\rm c})}{1-2r\cos\delta\cos\gamma+r^2}\right]\sin(\gamma-\phi). 
 \end{equation}
 If we use the hadronic parameters in Subsection~\ref{sec:dethad} and the SM values of $(q,\phi)$, we obtain the
 SM prediction
 \begin{equation}\label{eq:SR-1-SM}
\Delta_{\rm SR}^{({\rm I})}|_{\rm SM} = -0.009\pm 0.013.
\end{equation}
On the other hand, the current data in Table~\ref{tab:CPs} give
\begin{equation}\label{SR-I-pred}
	\Delta_{\rm SR}^{({\rm I})}|_{\rm exp} = -0.15 \pm 0.14,
\end{equation}
which is consistent with zero and the SM prediction within the uncertainties.

In addition, we also study another sum rule \cite{GR,gro}:
\begin{equation}\label{eq:sum-rule-II}
\Delta_{\rm SR}^{({\rm II})} \equiv A_{\rm CP}^{\pi^-K^+} + A_{\rm CP}^{\pi^+ K^0} - A_{\rm CP}^{\pi^0 K^+} - A_{\rm CP}^{\pi^0 K^0} , 
\end{equation}
which can be written as
\begin{align} \label{eq:sum-rule-II-theo}
\Delta_{\rm SR}^{({\rm II})} &= 2r_{\rm c}\Big\{\sin(2\delta_{\rm c})\Big[\sin(2\gamma)-2q\sin(\gamma+\phi)+q^2\sin(2\phi)\Big]r_{\rm c} \\
&-\sin(\delta_{\rm c}+\delta)\left[\sin(2\gamma)-q\sin(\gamma+\phi)\right]r \nonumber \\
&-\sin(\delta_{\rm c}+\theta_{\rm c})\left[\sin(2\gamma)-q\sin(\gamma+\phi)\right]\rho_{\rm c}\Big\} + {\cal O}(r_{({\rm c})}^3,\rho_{\rm c}^3). \nonumber
\end{align}
In contrast to Eq.~(\ref{eq:SR-theo-1}), this expression has a $q^2$ term, thereby showing different sensitivity to a modified 
EW penguin sector.  We find the SM prediction
\begin{equation}\label{eq:SR-2-SM}
\Delta_{\rm SR}^{({\rm II})}|_{\rm SM} = -0.003\pm 0.028,
\end{equation}
while the current data give
\begin{equation}\label{SR-2-exp}
\Delta_{\rm SR}^{({\rm II})}|_{\rm exp} = -0.14\pm 0.13.
\end{equation}
Comparing Eq.~\eqref{eq:SR-2-SM} with Eq.~\eqref{eq:SR-1-SM}, we observe that the uncertainty of the second sum rule is larger.
This feature is caused by a more pronounced dependence of Eq.~(\ref{eq:sum-rule-II-theo}) on the relevant decay
parameters.
 
The current experimental value of the direct CP asymmetry of the $B^0_d\to\pi^0K^0$ channel suffers from a 
large uncertainty which actually governs the errors of Eqs.~(\ref{SR-I-pred}) and (\ref{SR-2-exp}). In fact, the PDG value 
in Table~\ref{tab:CPs} is an average of BaBar \cite{Aub08} and Belle \cite{Fuj08} measurements which show different signs.
On the other hand, we may use the sum rules to predict the direct CP violation in $B^0_d\to\pi^0K^0$. Using the cleaner sum rule in Eq.~(\ref{eq:sum-rule-I}) and taking into account higher-order effects from Eq.~(\ref{eq:SR-theo-1}), we find 
\begin{equation}\label{eq:Adirsumrule}
A_\text{CP}^{\pi^0 K^0} = -0.14 \pm 0.03.
\end{equation}
A similar result, with a slightly larger error, follows from Eqs.~\eqref{eq:sum-rule-II}~and~\eqref{eq:sum-rule-II-theo}. The prediction in Eq.~\eqref{eq:Adirsumrule} lies within the $1\sigma$ range of the experimental value in Table~\ref{tab:CPs} but has a much smaller uncertainty. 
We shall use Eq.~(\ref{eq:Adirsumrule}) as the reference value of $A_\text{CP}^{\pi^0 K^0}$.

\boldmath
\subsection{Vanishing CP violation in the electroweak penguin sector} \label{sec:phi0}
\unboldmath
An interesting case is given by $\phi=0^\circ$, which includes the SM but allows also for NP contributions through 
non-SM values of $q$. In view of the discussion in Subsection~\ref{ssec:CSEWP}, we neglect contributions from colour-suppressed
EW penguin topologies. The observables take then the following forms \cite{BFRS-2}:
\begin{align} 
R {}& =  \frac{1  - 2 r \cos\delta\cos\gamma + r^2}{ 1 + \rho_c^2 + 2 \rho_c \cos\gamma \cos \theta_c} \ ,  \\
R_{\rm n} {}&= \frac{1}{b} (1- 2 r\cos\delta\cos\gamma + r^2)\ ,  \\
R_{\rm c}{}& = 1 + \frac{r_c^2 r_q - 2 \rho_c r_c \cos(\delta_c - \theta_c)(1 - q \cos\gamma) -2 (-q +  \cos\gamma)r_c \cos\delta_c}{1 + 
\rho_c^2 + 2\rho_c \cos\gamma\cos\theta_c}\ , 
\end{align}%
where $r_q = 1 - 2 q \cos\gamma + q^2$ and \cite{BFRS-2}
\begin{align}\label{eq:bburas}
b \equiv {}& 1-2 r \cos\delta\cos\gamma +r^2 + 2 r_c \cos\delta_c (-q + \cos\gamma) +
 2 r \cos(\delta-\delta_c) r_c ( -1 + q\cos\gamma)\nonumber \\ 
{}&  + r_c^2 (1+ q^2 -2 q \cos\gamma)\ .
\end{align}
For the CP asymmetries, we find
\begin{align}\label{eq:cpmix}
A_{\text{CP}}^{\pi^-K^+} {}& = \frac{-2 r \sin\delta \sin\gamma}{1 - 2 r\cos\delta \cos\gamma +r^2 } \ , \\ 
A_{\text{CP}}^{\pi^+K^0} {}& = \frac{2\rho_c \sin\theta_c \sin\gamma}{ 1+ 2 \rho_c\cos\theta_c \cos\gamma + \rho_c^2 } \ , \\
A_{\text{CP}}^{\pi^0K^
+}  {}&= - \frac{2 r_c \sin\delta_c\sin\gamma  -2 \rho_c \sin\theta_c \sin\gamma + 
2 q \rho_c r_c \sin(\delta_c - \theta_c) }{R_c (1 + \rho_c^2 + 2 \rho_c \cos\gamma \cos\theta_c)} \ . 
\end{align}
In contrast to Ref.~\cite{BFRS-2}, we include the tiny parameter $\rho_c$.

In the case of $B_d^0 \to \pi^0 K_{\rm S}$, the final state is a CP-odd eigenstate.\footnote{As usual, we neglect tiny CP violation in the neutral kaon system.} Interference between $B_d^0$--$\bar{B}_d^0$ mixing and decays of $B_d^0$ or $\bar{B}_d^0$ mesons
into the $\pi^0 K_{\rm S}$ final state gives rise to a mixing-induced CP asymmetry, which satisfies the following general relation
\cite{FFM,FJPZ}:
\begin{equation}\label{eq:SpiKs}
S_{\rm CP}^{\pi^0K_{\rm S}}  = \sqrt{1- (A^{\pi^0K_{\rm S}}_{\rm CP})^2} \sin(\phi_d -\phi_{00}) ,
\end{equation}
where 
\begin{equation}\label{ACPdir0K0}
A^{\pi^0K_{\rm S}}_{\rm CP} = A^{\pi^0K^0}_{\rm CP} = \frac{2\sin\gamma}{b}
\Big[-r \sin \delta + r_c  (q r \sin(\delta-\delta_c) + \sin\delta_c )  \Big]
\end{equation}
 is the direct CP asymmetry, $\phi_d$ denotes the $B^0_d$--$\bar B^0_d$ mixing phase given in Eq.~(\ref{phid-expr}), and
\begin{equation}\label{eq:phi00}
\phi_{00}\equiv \rm{arg}(\bar{A}_{00} A^*_{00}) 
\end{equation}
measures the angle between the decay amplitude $A_{00}=A(B_d^0\to \pi^0 K^0)$ and its CP conjugate 
$\bar{A}_{00} = A(\bar{B}_d^0\to \pi^0 \bar{K}^0)$. It is useful to introduce
\begin{equation}\label{s2b}
(\sin\phi_d)_{\pi^0 K_{\rm S}} \equiv  \sin(\phi_d -\phi_{00}) = \frac{S_{\rm CP}^{\pi^0K_{\rm S}}  }{\sqrt{1- (A^{\pi^0K_{\rm S}}_{\rm CP})^2}} = \left[1+ \frac{1}{2} (A^{\pi^0K_{\rm S}}_{\rm CP})^2 + \ldots \right] S_{\rm CP}^{\pi^0K_{\rm S}}.
\end{equation}
Thanks to the functional dependence of this expression, for the sum rule prediction in Eq.~(\ref{eq:Adirsumrule}), 
the direct CP asymmetry has only a tiny numerical impact at the 1\% level. Consequently, $(\sin\phi_d)_{\pi^0 K_{\rm S}}$ 
is fully governed by the mixing-induced CP asymmetry.

The amplitude parametrizations given above yield
\begin{equation}\label{eq:phi00easy}
\left.\tan \phi_{00}\right|_{\phi=0^\circ} = 2 \left(\frac{Q }{W} \right) \sin \gamma \ ,
\end{equation}
where
\begin{align}
Q {}& =  r \cos\delta- r_c \cos\delta_c +q r_c^2 - q r r_c \cos(\delta - \delta_c) \nonumber\\ 
{}&- 
 (r_c^2 -2 r r_c \cos(\delta-\delta_c) +r^2)\cos\gamma \ , \nonumber  \\
W {}& = 1- 2(q r_c^2 - r_c \cos\delta_c + r \cos\delta - q r r_c \cos(\delta-\delta_c)) \cos\gamma - 2 q r_c \cos(\delta_c)  \nonumber \\
{}&+ (r_c^2 -2 r r_c \cos(\delta-\delta_c) +r^2)\cos(2\gamma) + q^2 r_c^2 \ .
\end{align}
As the direct CP asymmetry $A^{\pi^0K_{\rm S}}_{\rm CP}$ can be measured, the prediction of the mixing-induced CP asymmetry
$S_{\rm CP}^{\pi^0K_{\rm S}}$ requires knowledge of the angle $\phi_{00}$, which will play a central role in the remainder of this 
paper.  It is interesting to note that Eq.~(\ref{eq:phi00easy}) involves cosine functions of CP-conserving strong phases which are very robust for smallish phases. On the other hand, the direct CP asymmetry in Eq.~(\ref{ACPdir0K0}) depends on sine functions which are much more sensitive to the values of the strong phases. 

\begin{figure}[t]
	\centering
	\includegraphics[width=0.4\textwidth]{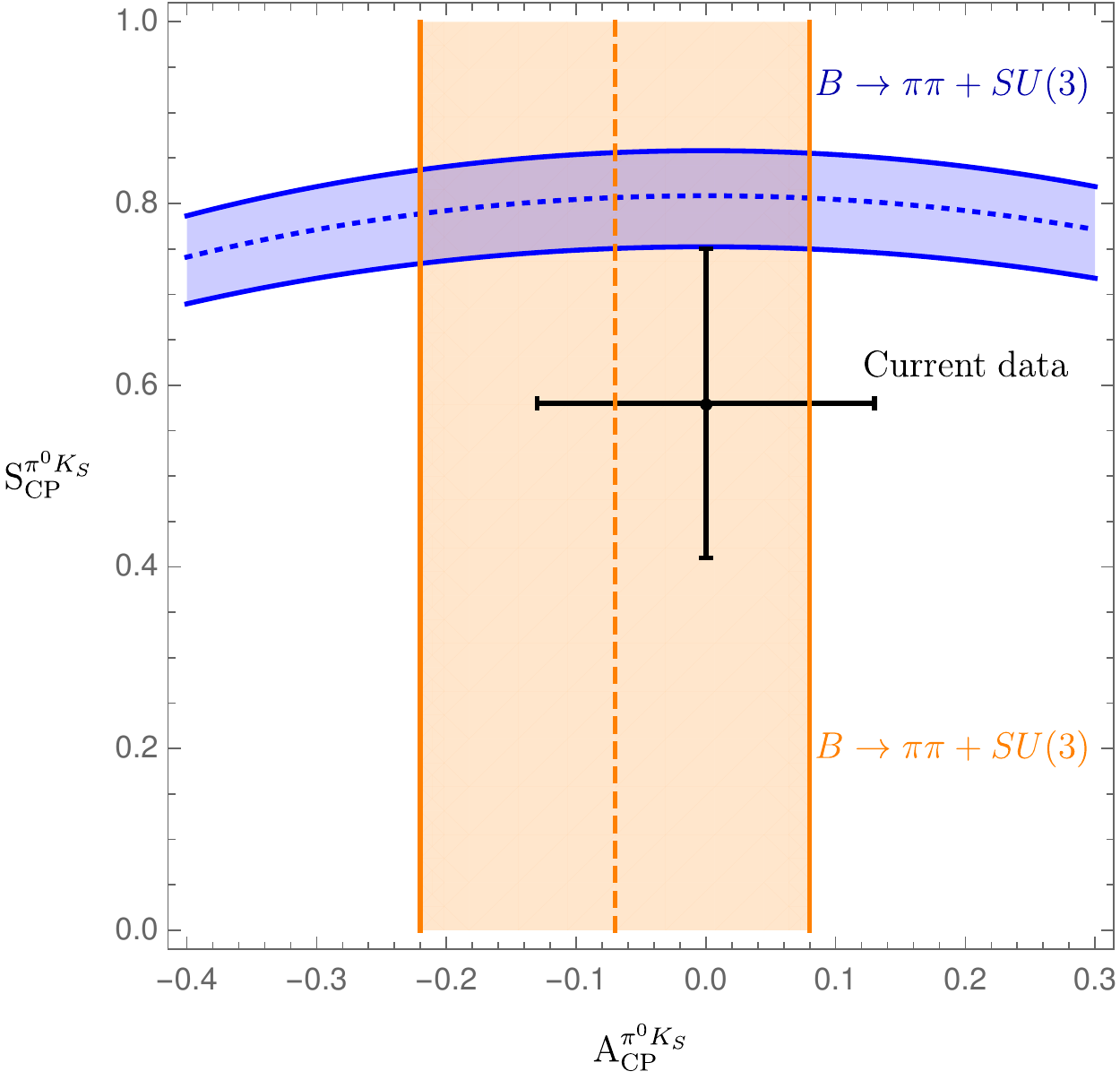}
	\caption{Predictions of the CP-violating $B_d^0 \to \pi^0 K_{\rm S}$ observables for the SM values of $q$ and $\phi$ 
	and the hadronic parameters fixed through the current $B \to \pi\pi$ data and the $SU(3)$ flavour symmetry. 
	}\label{fig:BtopiKpuzzle-had}
\end{figure}

Using the SM value of $q$ in Eq.~(\ref{eq:qdef}) and the hadronic parameters in Table~\ref{tab:sumpara} yields
\begin{equation} \label{eq:tanphi00-SM}
 \phi_{00}|_{\rm SM} =  (-10.8 \pm 5.2 )^\circ ,
\end{equation} 
where the error is dominated by the uncertainty of $R_q$ discussed in Section~\ref{sec:BtopiK}. For $\phi=0^\circ$, colour-suppressed EW penguins do not contribute to $\phi_{00}$ at leading order. We have checked that including them has indeed a very minor impact. 
Moreover, we obtain
\begin{equation}
(\sin\phi_d)_{\pi^0 K_{\rm S}}|_{\text{SM}} = 0.81 \pm 0.06.
\end{equation}
The expression in Eq.~(\ref{ACPdir0K0}) yields
\begin{equation}\label{eq:ASMpred}
A_{\rm CP}^{\pi^0 K_{\rm S}}|_{\text{SM}} = -0.07 \pm 0.15,
\end{equation}
which allows us to convert $(\sin\phi_d)_{\pi^0 K_{\rm S}}$ into the mixing-induced CP asymmetry
\begin{equation}\label{eq:SSMpred}
S_{\rm CP}^{\pi^0 K_{\rm S}}|_{\text{SM}} = 0.81 \pm 0.07.
\end{equation}
In Fig.~\ref{fig:BtopiKpuzzle-had}, we summarise the current situation in the 
$A_{\rm CP}^{\pi^0 K_{\rm S}}$--$S_{\rm CP}^{\pi^0 K_{\rm S}}$ plane. Here the blue contour corresponds to 
Eq.~(\ref{eq:SpiKs}) with $\phi_{00}$ given in
Eq.~(\ref{eq:tanphi00-SM}), whereas the vertical band represents Eq.~(\ref{eq:ASMpred}).  The black cross 
shows the current data. We observe that the mixing-induced CP asymmetry exhibits a deviation from the measured value
at the $1\,\sigma$ level. The value of $A_{\rm CP}^{\pi^0 K_{\rm S}}$ is in full agreement with the data and the sum rule 
prediction in Eq.~(\ref{eq:Adirsumrule}) having a much smaller uncertainty.

In Table~\ref{tab:BpiKpred}, we summarize the SM predictions and experimental data for  the various $B \to \pi K$ observables.
The errors are dominated by the currently large uncertainty of the $SU(3)$-breaking parameter $R_q$. For the 
branching ratios, we 
use the measured branching ratio of $B^+ \to \pi^+ K^0$ to fix the normalization $|P'|$. We observe that all predictions are well within 
the current experimental measurements. The excellent agreement of $R$ and $A_{\text{CP}}^{\pi^-K^+}$ with the measurements 
reflects the smallness of the colour-suppressed EW penguin contributions found in 
Subsection~\ref{ssec:CSEWP}, where these observables were used to determine the colour-suppressed EW penguin parameters. The largest deviation arises in the ratio $R_{\rm n}$, where there is a tension of a bit more 
than $1\sigma$ significance.

\begin{table}
  \centering
    \begin{tabular}{l | r | r}
    Observable & SM Prediction & Experiment \\
    \hline \hline
        $R$ & $0.93 \pm 0.03$ & $0.89 \pm 0.04$ \\
    $R_{\rm n}$ & $1.13 \pm 0.10$ & $0.99 \pm 0.06$ \\
        $R_{\rm c}$ & $1.11 \pm 0.08$ & $1.09 \pm 0.06$ \\
    \hline
      $A_\text{CP}^{\pi^\pm K^\mp}$ & $-0.085 \pm 0.064$ & $-0.082 \pm 0.006$ \\ 
      $A_\text{CP}^{\pi^\pm K^0}$ & $0.003 \pm 0.005$ & $-0.017 \pm 0.016$ \\
    $A_\text{CP}^{\pi^0K^\pm}$ & $-0.007 \pm 0.11$ & $0.037 \pm 0.021$ \\   
    $A_\text{CP}^{\pi^0 K_S}$ & $-0.07 \pm 0.15$ & $0.00 \pm 0.13$ \\
    $S_\text{CP}^{\pi^0 K_S}$ & $0.81 \pm 0.07$ & $0.58 \pm 0.17$ \\
  
    \hline
      $\mathcal{B}r(B_d^0\to\pi^- K^+)\times10^6$ & $20.6 \pm 0.7$ & $19.6 \pm 0.5$ \\
    ${\mathcal{B}r(B^+ \to \pi^+ K^0)\times10^6}$ & \rm{Normalization} & $23.7 \pm 0.8$ \\
    $\mathcal{B}r(B^+\to\pi^0 K^+)\times10^6$ & $13.1 \pm 1.0$ & $12.9 \pm 0.5$ \\
    $\mathcal{B}r(B^0_d\to\pi^0 K^0)\times10^6$ & $9.1 \pm 0.9$ & $ 9.9 \pm 0.5$    
    \end{tabular}
    \caption{SM predictions of the $B \to \pi K$ observables and comparison with the current experimental data.     	 }
    \label{tab:BpiKpred}
\end{table}

\section{Correlations between CP asymmetries of $\boldsymbol{B^0_d\to\pi^0 K_{\rm S}}$} \label{sec:1}
\subsection{Preliminaries}
The mixing-induced CP asymmetry of the $B_d^0\to\pi^0K_{\rm S}$ channel is a particularly interesting probe for testing
the SM. In the previous section, we have used hadronic parameters which were determined from $B\to\pi\pi$ data by means
of the $SU(3)$ flavour symmetry, resulting in the picture shown in Fig.~\ref{fig:BtopiKpuzzle-had}. Interestingly, we
can obtain a much more precise correlation in the $A_{\rm CP}^{\pi^0 K_{\rm S}}$--$S_{\rm CP}^{\pi^0 K_{\rm S}}$ plane
\cite{FJPZ}, as we will discuss in this section.

The starting point is given by the following isospin relations \cite{NQ, GHLR}:
\begin{equation}\label{eq:isospinrel}
\sqrt{2} A(B_d^0\to \pi^0 K^0) +
A(B^0\to \pi^- K^+) = - (\hat{T}' +\hat{C}')e^{i\gamma} + \left(\hat P'_{EW}+{\hat{ P}^{'{\rm C}}_{EW}}\right) \equiv 3A_{3/2} ,
\end{equation} 
\begin{equation}\label{eq:isospinrel-CP}
\sqrt{2} A(\bar B_d^0\to \pi^0 \bar K^0) +
A(\bar B^0\to \pi^+ K^-) = 3\bar{A}_{3/2}   ,
\end{equation} 
where the isospin $I=3/2$ amplitude $A_{3/2}$ and its CP-conjugate $\bar{A}_{3/2}$ are given by
\begin{align}\label{eq:A32}
3 A_{3/2} {}& \equiv 3 |A_{3/2}|e^{i\phi_{3/2}}=- \left[\hat{T}'+\hat{C}'\right](e^{i\gamma}-qe^{+i\phi}) \ , \\
3 \bar{A}_{3/2} {}& \equiv 3 |\bar{A}_{3/2}|e^{i\bar{\phi}_{3/2}}=- \left[\hat{T}'+\hat{C}'\right](e^{-i\gamma}-qe^{-i\phi}) \ .
\label{eq:A32-CP}
\end{align}
Here we have used $\omega=0$ and observe the relation 
\begin{equation}\label{phi-diff}
\bar{\phi}_{3/2}=-\phi_{3/2}.
\end{equation}
The absolute value of the amplitude $\hat{T}'+\hat{C}'$ can be fixed through the measured branching ratio of the 
$B^\pm\to\pi^0\pi^\pm$ decay with the help of the $SU(3)$ flavour symmetry \cite{GRL}:
\begin{equation}\label{eq:tcabs}
|\hat{T}'+\hat{C}'|= R_{T+C}\left|\frac{V_{us}}{V_{ud}}\right|\sqrt{2} |A(B^+\to\pi^+ \pi^0)| \ .
\end{equation}

\begin{figure}[t]
\centering
\includegraphics[width=0.5\linewidth]{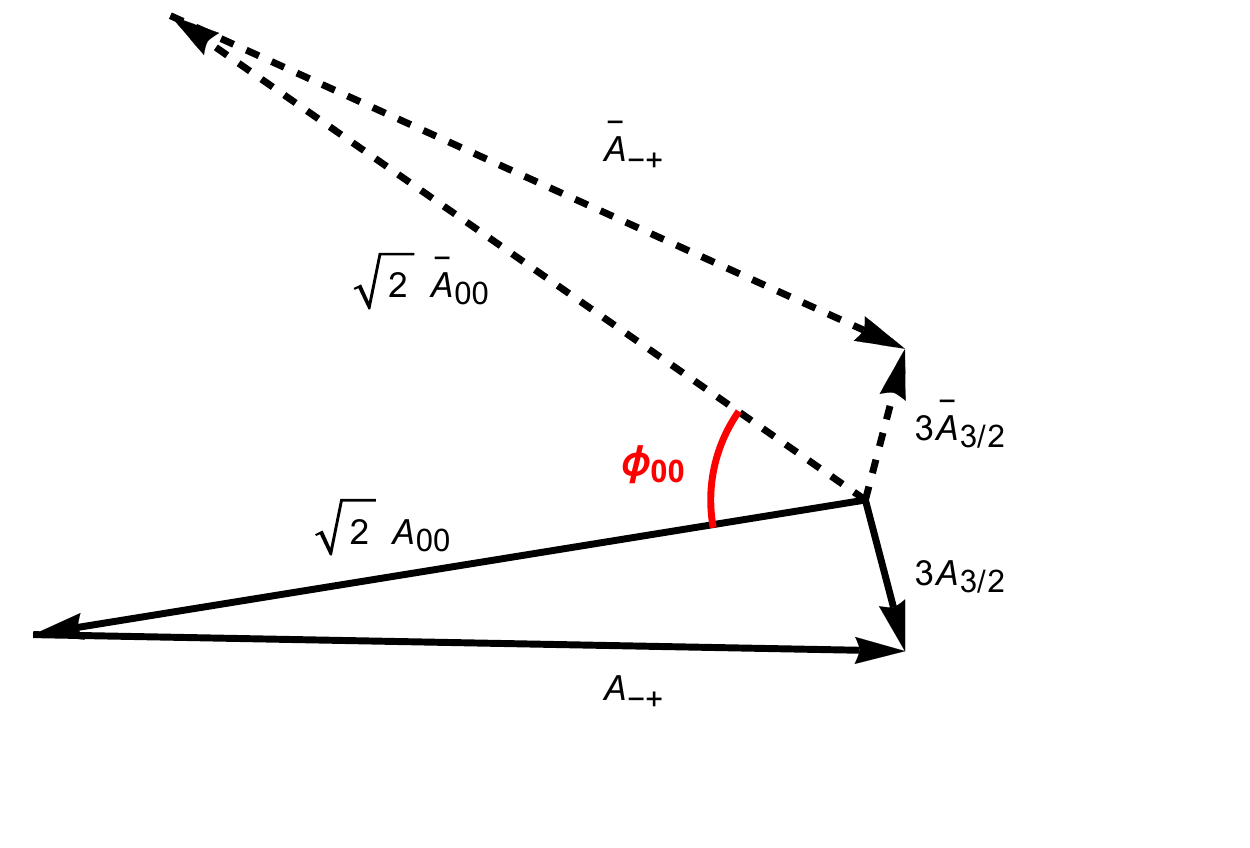}
\caption{Illustration of the amplitude triangles following from the isospin relations in Eqs.~\eqref{eq:isospinrel} and
\eqref{eq:isospinrel-CP}. The solid triangle corresponds to the $B^0_d \to \pi^0 K^0$, $B^0_d\to\pi^-K^+$ decays while the dashed 
one represents the  CP-conjugate processes.}\label{fig:isospin}
\end{figure}

As was pointed out in Ref.~\cite{FJPZ}, using measured CP-averaged branching ratios and direct CP asymmetries,
the amplitude relations in Eqs.~(\ref{eq:isospinrel}--\ref{eq:tcabs}) allow us to determine the angle $\phi_{00}$ for given 
values of the EW penguin parameters $q$ and $\phi$, in particular also for the SM case as described by Eq.~(\ref{eq:qdef}). 
Having $\phi_{00}$ at hand, the expression in Eq.~(\ref{eq:SpiKs}) allows us to calculate a contour in the 
$A_{\rm CP}^{\pi^0 K_{\rm S}}$--$S_{\rm CP}^{\pi^0 K_{\rm S}}$ plane. The corresponding correlation relies only on 
the clean isospin relations in Eqs.~(\ref{eq:isospinrel}) and (\ref{eq:isospinrel-CP}) and the $SU(3)$ input given by 
$R_{T+C}$ in Eq.~(\ref{eq:tcabs}), which is a very  robust parameter as discussed in Section~\ref{sec:dethad}.

It is instructive to have a closer look at the corresponding analysis. The isospin relation in Eq.~\eqref{eq:isospinrel} can be represented by an amplitude triangle in the complex plane as depicted in Fig.~\ref{fig:isospin}. For given EW penguin parameters
$q$ and $\phi$, such as in the SM which we consider in the following discussion, the amplitudes $A_{3/2}$ and $\bar{A}_{3/2}$ 
are fixed. Using the direct asymmetries $ A_{\text{CP}}^{\pi^0 K_{\text{S}}} $ and $A_{\text{CP}}^{\pi^- K^+}$ taking the forms
\begin{equation}
A_{\text{CP}}^{\pi^0 K_{\text{S}}}= \frac{|\bar{A}_{00}|^2 - |A_{00}|^2}{|\bar{A}_{00}|^2 + |A_{00}|^2},\ \ \ \ \  A_{\text{CP}}^{\pi^- K^+} = \frac{|\bar{A}_{-+}|^2 - |A_{-+}|^2}{|\bar{A}_{-+}|^2 + |A_{-+}|^2}
\end{equation}
with $|A_{-+}| \equiv |A(B_d^0 \rightarrow \pi^- K^+)|$ and ${|\bar{A}_{-+}|} \equiv |A(\bar{B}^0_d \rightarrow \pi^+ K^- )|$, and the CP-averaged branching ratios allows the determination of the absolute values of the individual amplitudes.  
Finally, we determine $\phi_{00}$ and subsequently $ S_{\text{CP}}^{\pi^0 K_{\text{S}}}$ using Eq.~\eqref{eq:SpiKs}. Since the triangles can be flipped around the $A_{3/2}$ and $\bar{A}_{3/2}$ axes, we obtain a four-fold 
ambiguity for $\phi_{00}$ and correspondingly for $ S_{\text{CP}}^{\pi^0 K_{\text{S}}}$.

\begin{figure}[t]
\centering
\subfloat[]{
\includegraphics[width=0.5\textwidth]{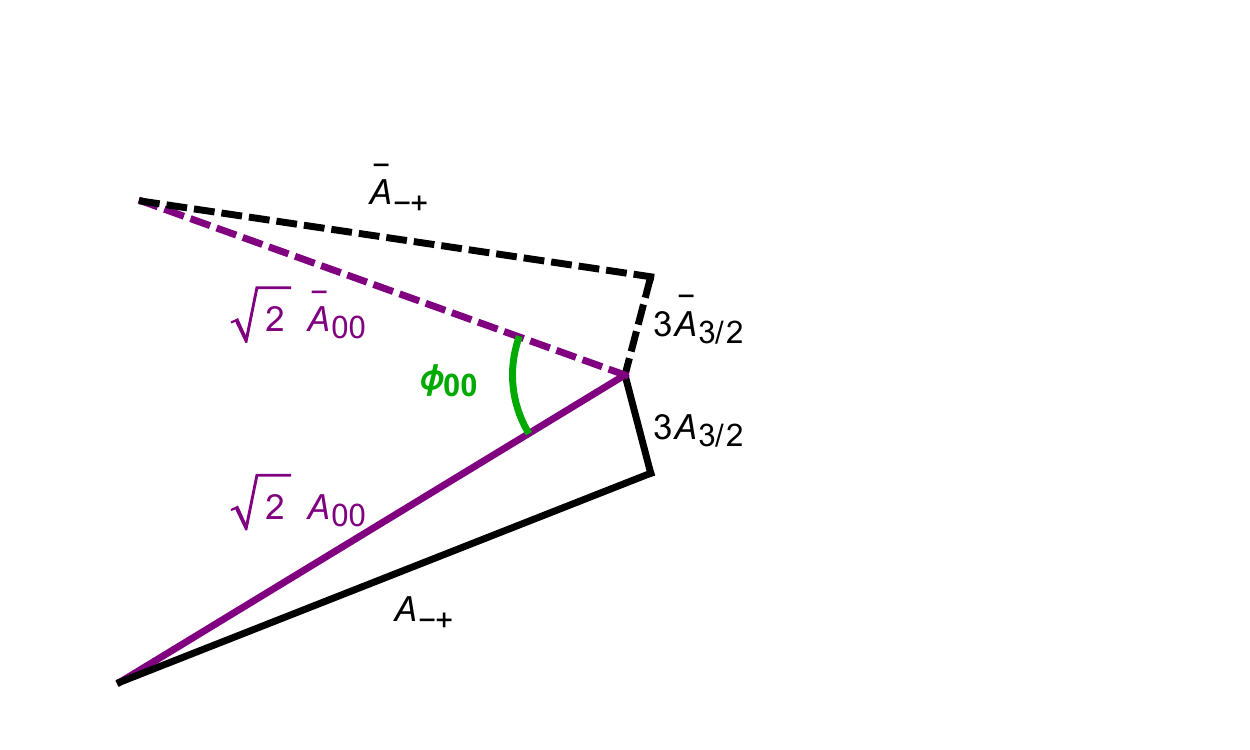}}
\centering
\subfloat[]{
\includegraphics[width=0.5\textwidth]{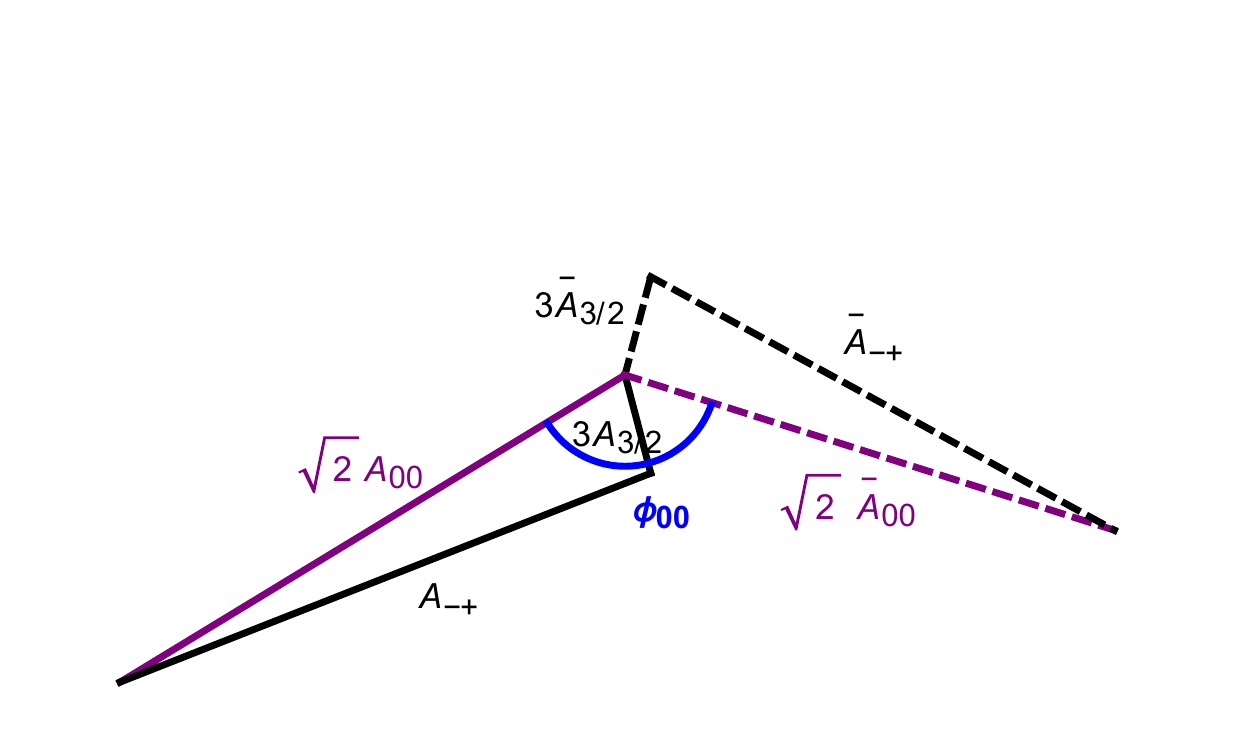}}\\
\centering
\subfloat[]{
\includegraphics[width=0.5\textwidth]{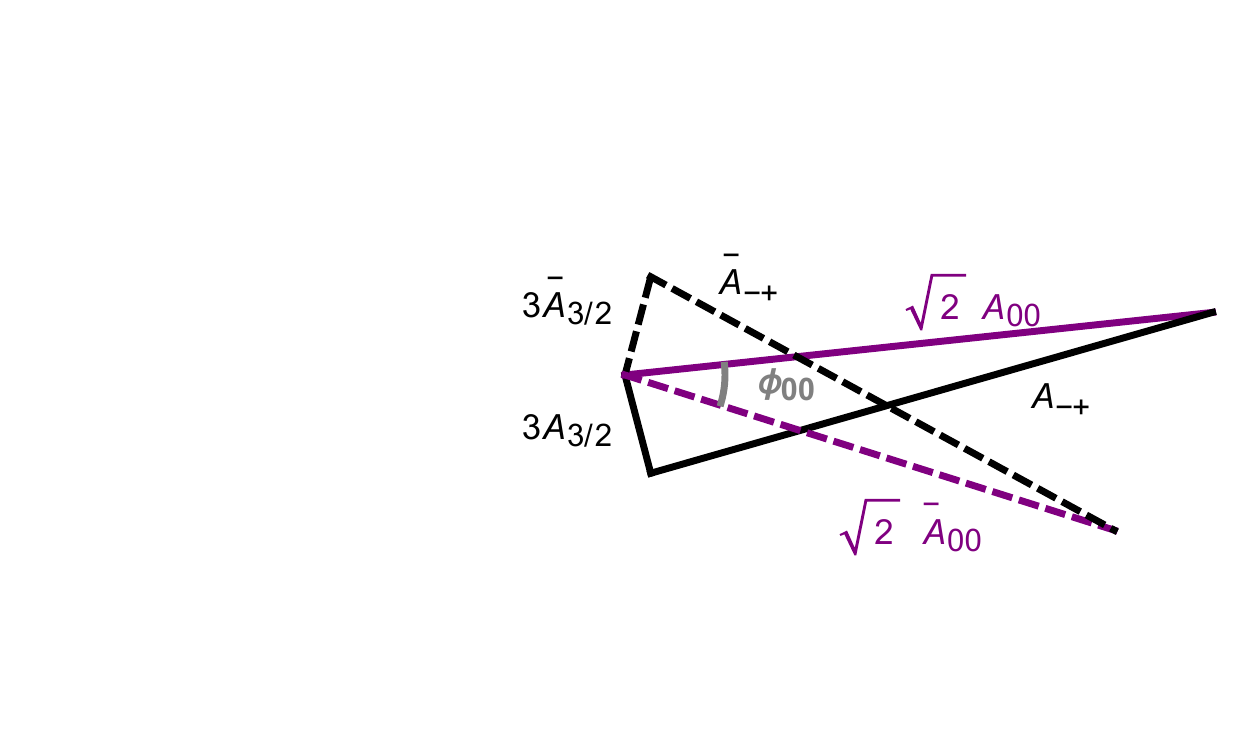}}
\subfloat[]{
\includegraphics[width=0.5\textwidth]{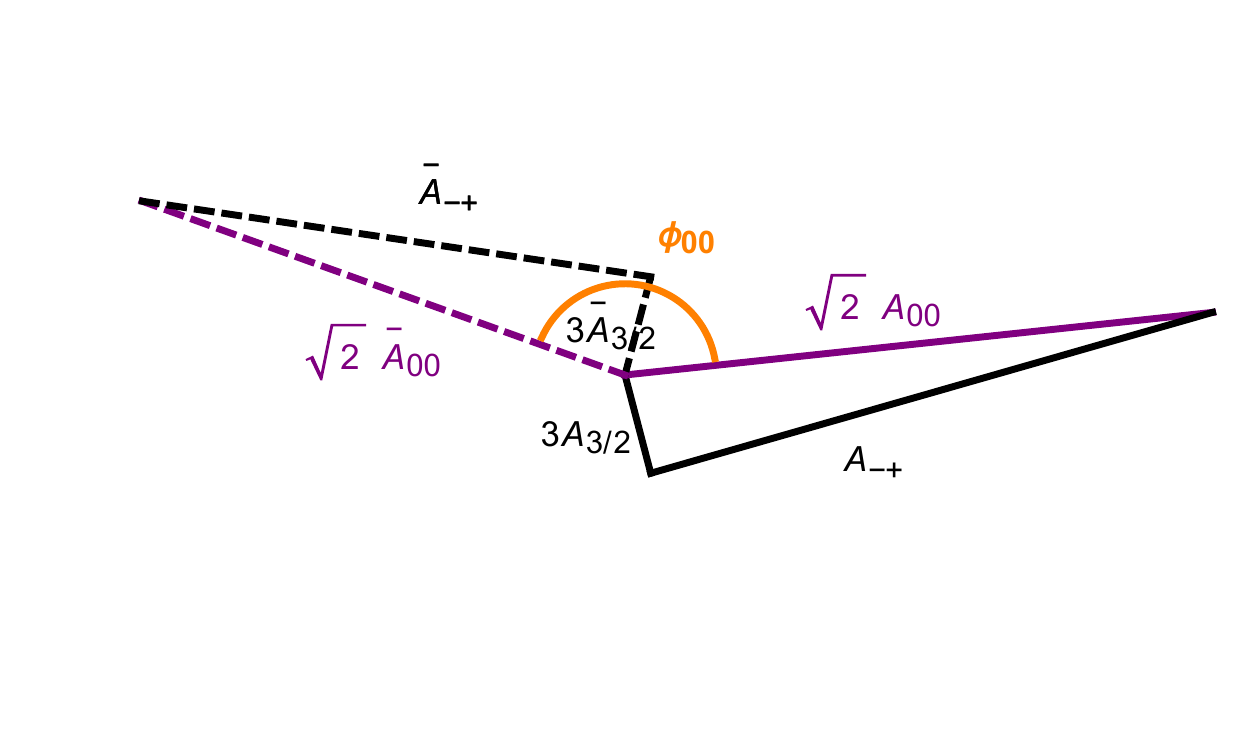}}
\caption{The four orientations of the amplitude triangles for current data and $A_{\text{CP}}^{\pi^0 K_{\text{S}}}$ in Eq.~\eqref{eq:Adirsumrule}. Varying $A_{\text{CP}}^{\pi^0 K_{\text{S}}}$, the 
triangle configurations correspond to the contours in the $A_{\text{CP}}^{\pi^0 K_{\text{S}}}$--$S_{\text{CP}}^{\pi^0 K_{\text{S}}}$ plane shown in Fig.~\ref{fig:conts} with the same colour.} 
\label{fig:phiorient}
\end{figure}

Let us illustrate this method by taking $A_{\text{CP}}^{\pi^0 K_{\text{S}}}$ from the sum rule in Eq.~\eqref{eq:Adirsumrule} 
and central values
of the measured observables. The four orientations of the resulting triangles are shown in Fig.~\ref{fig:phiorient}, and correspond
to the angles $\phi_{00}$ and mixing-induced CP asymmetries $S_{\text{CP}}^{\pi^0 K_{\text{S}}}$ given in Table~\ref{tab:SAphi}. The triangles are drawn in arbitrary units, since only the shape of the triangles is important for the determination of $\phi_{00}$.

\begin{table}[b]
\centering
 \begin{tabular}{c | c || c | c}
  \ \ \ $\phi_{00}$ \ \ \ & \ \ \ $S_{\text{CP}}^{\pi^0 K_{\text{S}}}$ \ \ \ &\ \ \  $\phi_{00}$\ \ \ & \ \ \ $S_{\text{CP}}^{\pi^0 K_{\text{S}}}$ \ \ \  \\
   \hline \hline
       $-49.8^\circ$ & $0.989$ & $-22.9^\circ$ & $0.903$ \\
    \hline
      $128.9^\circ$ & $-0.988$ & $145.5^\circ$  & $-0.967$\\
   \end{tabular}
    \caption{The angles $\phi_{00}$ and the corresponding mixing-induced CP asymmetries $S_{\text{CP}}^{\pi^0 K_{\text{S}}}$ following 
    from the triangle construction for current data using $A_{\text{CP}}^{\pi^0 K_{\text{S}}}$ in Eq.~(\ref{eq:Adirsumrule}).}
    \label{tab:SAphi}
\end{table}

\begin{figure}[t]
	\centering
	\subfloat[]{\label{fig:cont1} \includegraphics[width=0.49\textwidth]{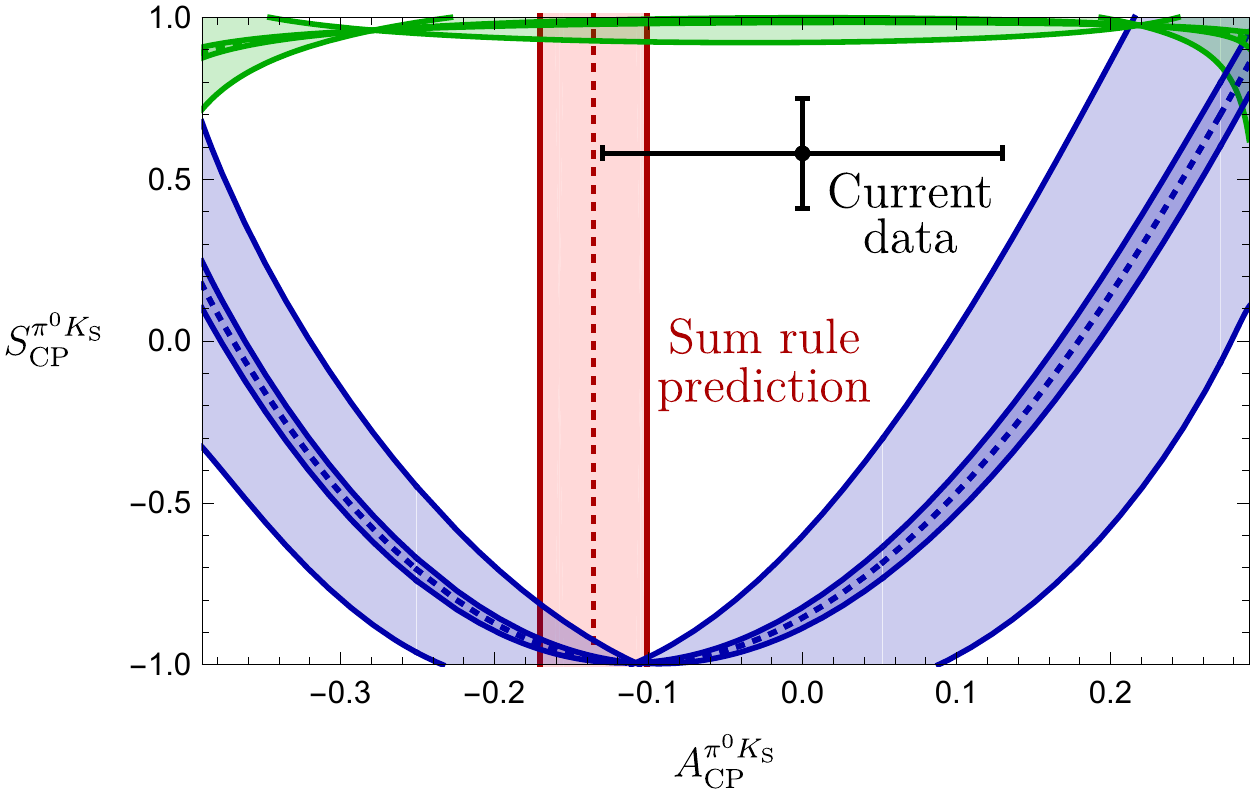}}
	\subfloat[]{\label{fig:cont2} \includegraphics[width=0.49\textwidth]{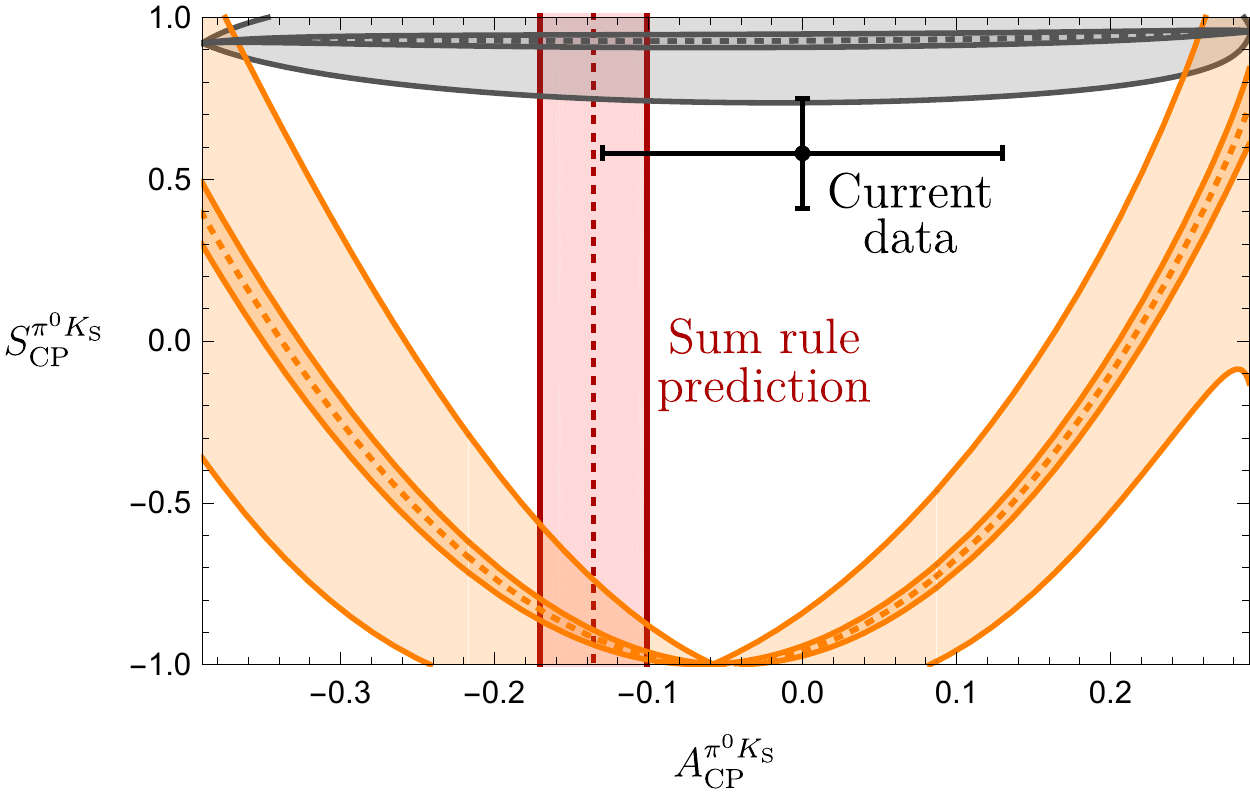}}\\
		\subfloat[]{\label{fig:an1} \includegraphics[width=0.49\textwidth]{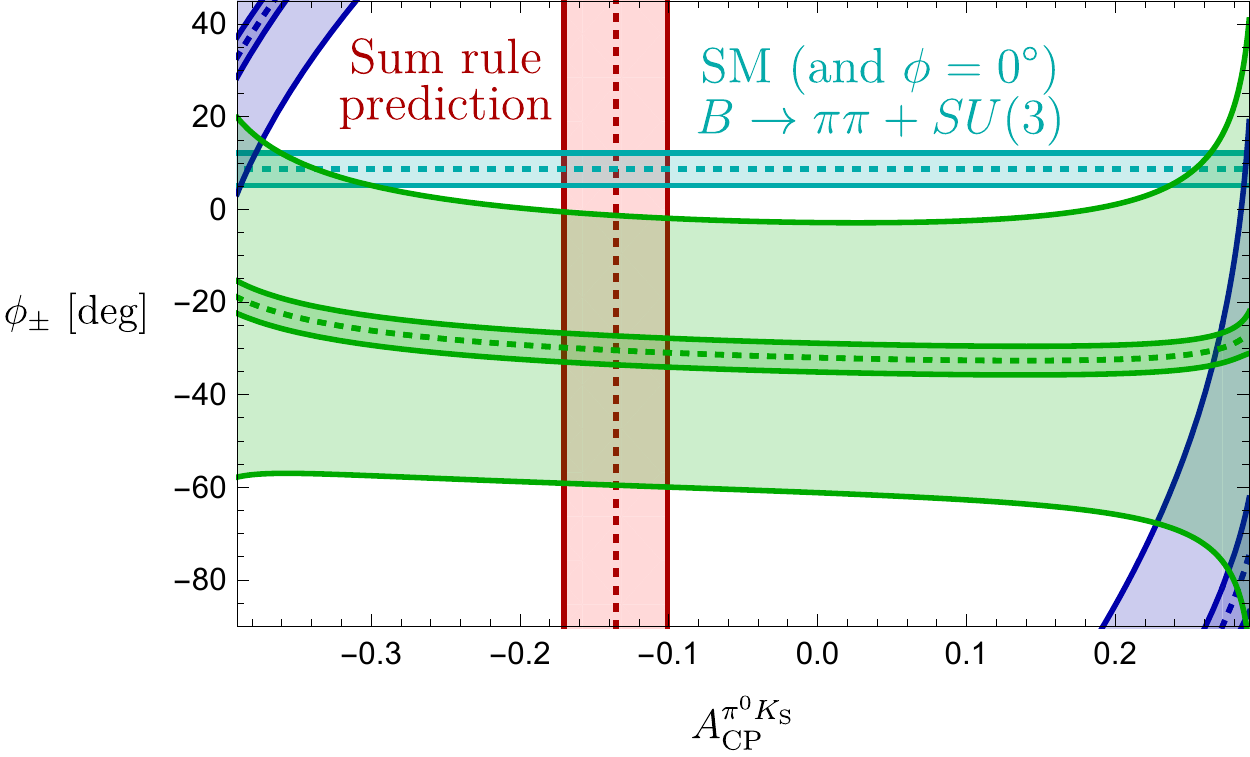}}
	\subfloat[]{\label{fig:an2} \includegraphics[width=0.49\textwidth]{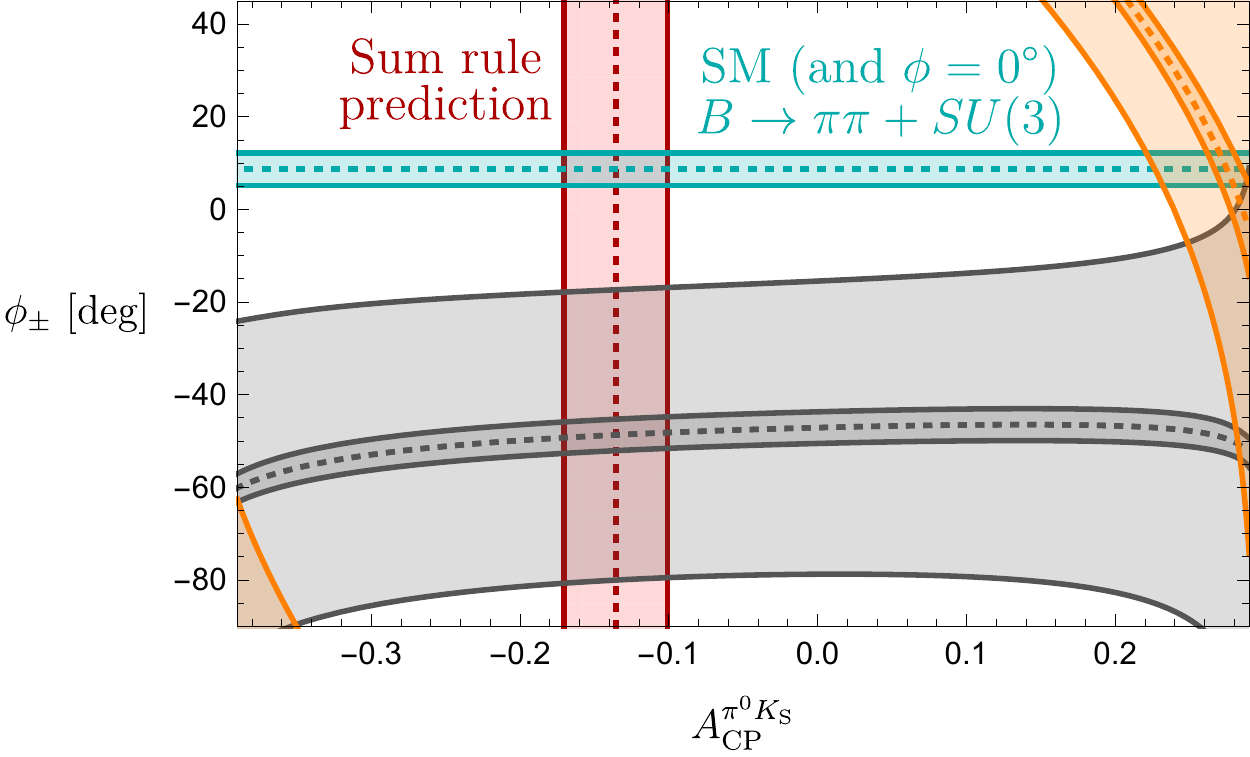}}
	\caption{(a, b) Correlations between the CP asymmetries of $B_d^0 \to \pi^0 K_{\rm S}$ following from the isospin triangles
	 illustrated in Fig.~\ref{fig:phiorient}. The vertical band (red) gives the sum rule prediction in Eq.~\eqref{eq:Adirsumrule}. (c, d) For each of the triangle configurations we show the associated angle $\phi_{\pm}$. The horizontal band (blue) gives the current SM constraint in Eq.~\eqref{eq:phipmsm}. The narrow bands correspond to
	a future scenario discussed in the text. }
	\label{fig:conts}
\end{figure}

 If we now vary the direct CP asymmetry of $B_d^0 \to \pi^0 K_S$, we obtain a correlation between 
 $S_{\text{CP}}^{\pi^0 K_{\text{S}}}$ and $A_{\text{CP}}^{\pi^0 K_{\text{S}}}$ \cite{BFRS-2, FJPZ}. This results in the 
 four contours in the 
$A_{\text{CP}}^{\pi^0 K_{\text{S}}}$--$S_{\text{CP}}^{\pi^0 K_{\text{S}}}$ plane shown in Figs.~\ref{fig:cont1} and \ref{fig:cont2}, where we have also taken the experimental errors and the uncertainties of $R_{T+C}$ and $R_q$ into account. The four contours correspond to the configurations in Fig.~\ref{fig:isospin} where $\phi_{00}$ is labelled with the same colour. We have also 
included the current experimental data point for the CP asymmetries from Table~\ref{tab:CPs}, and the vertical band refers to the 
sum rule value of $A_{\text{CP}}^{\pi^0 K_{\text{S}}}$  in Eq.~(\ref{eq:Adirsumrule}). In addition, the narrow bands illustrate a future scenario including only the expected theory uncertainties for $R_q$ and $R_{T+C}$ in Eqs.~\eqref{eq:theoryupRq} and 
Eq.~\eqref{eq:theoryuprtc}, respectively. 

\subsection{Discrete ambiguities} \label{sec:resolve-ambiguities}
The four-fold ambiguity arising from the different orientations of the amplitude triangles can be distinguished through the strong phase $\delta_{\rm c}$ \cite{FJPZ}. The values of $\delta_{\rm c}$ corresponding to different points on the four contours in Figs.~\ref{fig:cont1} and \ref{fig:cont2} can be found by parametrizing $A_{\rm CP}^{\pi^0K_{\rm S}}$ and $S_{\rm CP}^{\pi^0K_{\rm S}}$ in terms of the hadronic parameters using Eqs.~(\ref{eq:SpiKs})~and~(\ref{ACPdir0K0}). Employing now only the neutral $B \to \pi K$ observables, the hadronic parameters $r$, $\delta$ and $r_{\rm c}$ can be expressed in terms of the strong phase $\delta_{\rm c}$ using the ratio $R_{\rm n}$ and $A_{\rm CP}^{\pi^-K^+}$. In addition, we use $\mathcal{B}r(B_d^0\to\pi^-K^+)$ to fix the normalization $|P'|$ which then enters $r_{\rm c}$ via
\begin{equation}
r_{\rm c} 
= \sqrt{2}\left|\frac{V_{us}}{V_{ud}}\right| R_{T+C} \sqrt{\frac{\mathcal{B}r(B^+\to \pi^0 \pi^+)}{\mathcal{B}r(B_d^0 \to \pi^- K^+) }\frac{\tau_{B_d^0} }{\tau_{B^+} }} \sqrt{ 1+ r^2 -2 r \cos\delta \cos \gamma} \ ,
\end{equation}
where we used again the $SU(3)$ flavour symmetry.

Finally, we find that the contours in Fig.~\ref{fig:cont1} correspond to $|\delta_{\rm c}| < 90^\circ$, while those in Fig.~\ref{fig:cont2} 
give $|\delta_{\rm c}|> 90 ^\circ$. Using the range of $\delta_{\rm c}$ in Eq.~\eqref{eq:rcval}, only the contours in Fig.~\ref{fig:cont1}  are allowed. Going one step further, we can also consider the associated value of $r_{\rm c}$ for the contours in Fig.~\ref{fig:cont1}. We find that the lower contour implies very large values of $r_{\rm c}$ that are excluded from Eq.~\eqref{eq:rcval}, thereby leaving only the upper contour in Fig.~\ref{fig:cont1}. In comparison with Ref.~\cite{FJPZ}, we already obtain a much sharper picture due to the improved value of $\gamma$ in Eq.~\eqref{eq:gammaval}, which pushes $S_{\rm CP}^{\pi^0K_{\rm S}}$ close to its maximal value of $1$. We observe a  discrepancy between the data and the triangle constraint at the $2.5\;\sigma$ level.

As a new element, we employ the angle 
\begin{equation}\label{eq:pmcon}
\phi_\pm = \text{Arg}\left[\bar{A}_{-+} A_{-+}^* \right] 
\end{equation}
between the decay amplitudes $A_{-+} \equiv A(B^0_d\to \pi^- K^+)$ and $\bar{A}_{-+} \equiv A(\bar{B}_d^0\to \pi^+ K^-)$ 
\cite{FJV18}.
 In Figs.~\ref{fig:an1} and ~\ref{fig:an2}, we give $\phi_{\pm}$ for each of the four triangle configurations. The narrow band depicts the future theory scenario as discussed above. 

In the SM, we may actually calculate $\phi_\pm$ from the hadronic parameters in Eq.~\eqref{eq:rdelfromdtheta}. 
For $\phi=0^\circ$, we obtain
\begin{equation}
\left.\tan \phi_\pm\right|_{\phi=0^\circ} 
= \frac{-r^2 \sin 2\gamma + r\sin(\gamma-\delta) + r \sin(\gamma+\delta) + C_{\pm}}{1+r^2 \cos 2\gamma - 
r \cos(\gamma-\delta)-r \cos(\gamma+\delta) + B_{\pm} }\ ,
\end{equation}
where
\begin{align}
B_\pm {}& = \frac{4}{3} q r_{\rm c} [\tilde{a}_C - r\cos\gamma (\tilde{a}_C \cos\delta + \tilde{a}_S \sin\delta) + \frac{1}{3}qr_{\rm c}(\tilde{a}_C^2+\tilde{a}_S^2)] \nonumber \\
C_{\pm} {}& = \frac{4}{3} q r r_{\rm c} \sin\gamma (\tilde{a}_C \cos\delta + \tilde{a}_S \sin\delta) \ .
\end{align}
The $B_\pm$ and $C_{\pm}$ actually give tiny numerical contributions, and we find
\begin{equation}\label{eq:phipmsm}
\phi_{\pm} = 2 r \cos\delta\sin\gamma + \mathcal{O}(r_{(c)}^2) = 
(8.7 \pm 3.5)^\circ \ , 
\end{equation}
where we have used the values of $r, \delta$ and $\gamma$ in Table~\ref{tab:sumpara} to obtain the numerical result.  We have added this SM constraint to Fig.~\ref{fig:conts} and note that two of the contours in Figs.~\ref{fig:cont1} and \ref{fig:cont2} are excluded by the new constraint $\phi_\pm$. This can also be seen in the illustration of the triangles in Fig.~\ref{fig:phiorient}. In addition, the grey contour in Fig.~\ref{fig:an2} is in tension with the constraint on $\phi_\pm$. However, this specific configuration was already excluded because it implies $|\delta_c| > 90^\circ$. We therefore focus on the (green) upper contour in Fig.~\ref{fig:cont1}. However, 
for this configuration, we observe a tension with the SM prediction for $\phi_\pm$ which is currently at the $1\sigma$ level but 
may become much more pronounced as illustrated by the narrow band referring to a future scenario. 

\begin{figure}[t]
	\centering
	\includegraphics[width=0.4\textwidth]{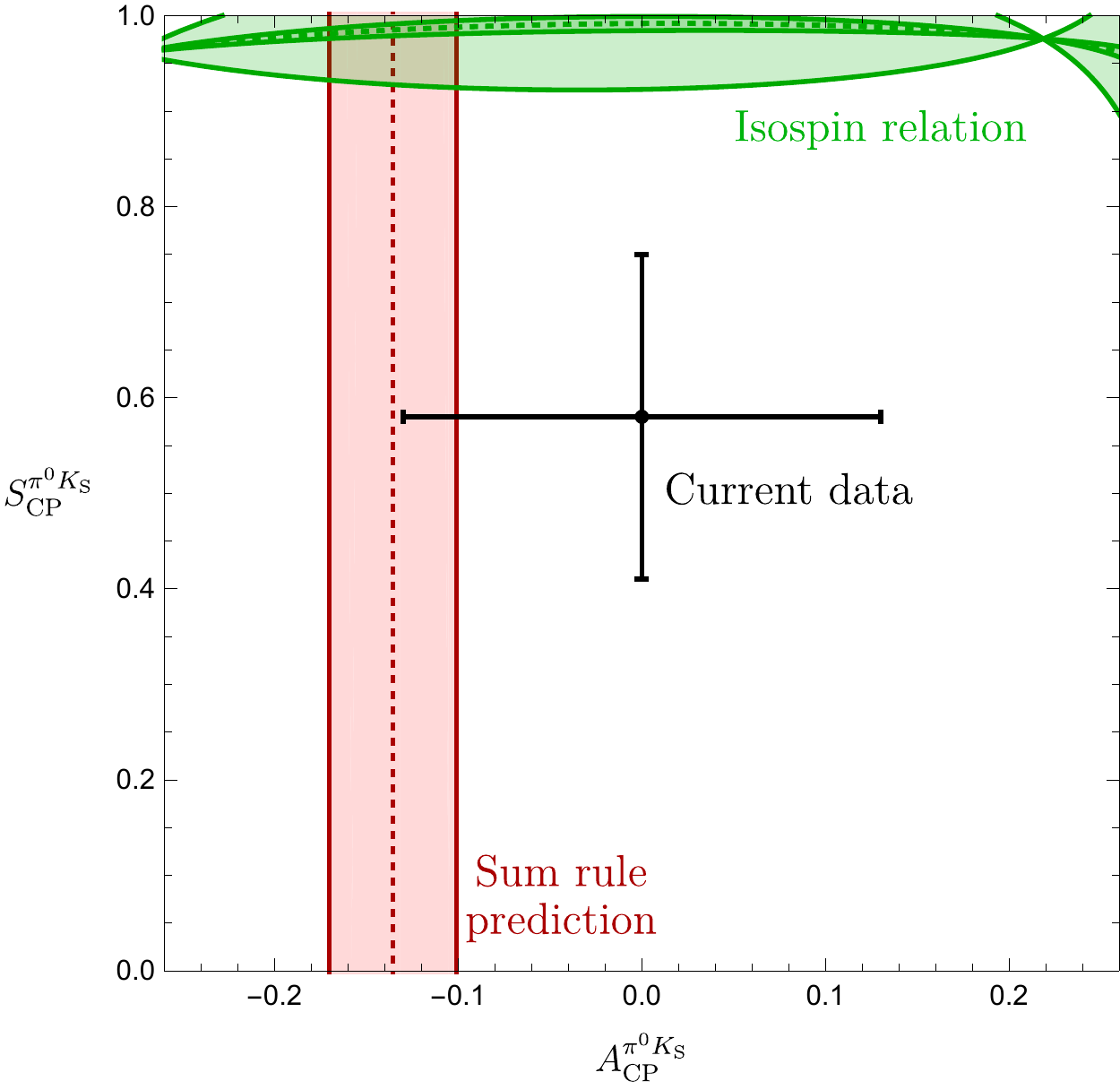}
	\caption{Illustration of the $B \to \pi K$ puzzle: the upper green band follows from the isospin analysis while 
	the vertical band shows the sum rule prediction in  Eq.~\eqref{eq:Adirsumrule}. 
	}\label{fig:BtopiKpuzzle}
\end{figure}

\begin{figure}[t]
\centering
\includegraphics[width = 0.405\linewidth]{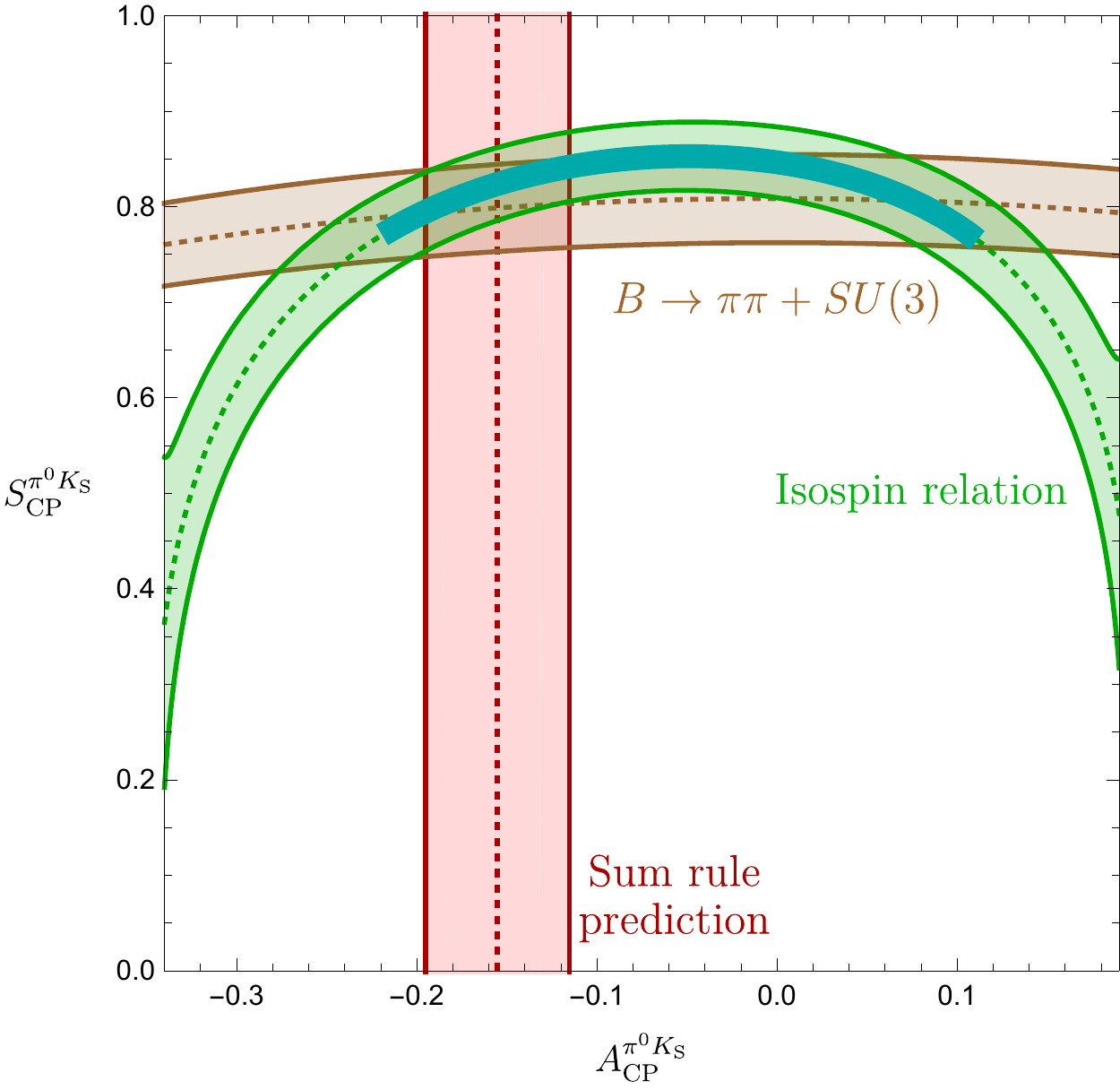}
	\caption{Reducing $\mathcal{B}r(B_d^0\to \pi^0 K^0)$ by $2.5\,\sigma$ gives a picture consistent with the SM, where the dark part agrees perfectly with the $\phi_\pm$ constraint in Eq.~\eqref{eq:phipmsm}.}
	\label{fig:solpuz}
\end{figure}

\boldmath
\subsection{How to resolve the ${B \to \pi K}$ puzzle?}\label{sec:sumruleqphi}
\unboldmath
In Fig.~\ref{fig:BtopiKpuzzle}, we summarize the intriguing picture following from the isospin triangles, showing only the contour 
remaining once the constraints from Section~\ref{sec:resolve-ambiguities} have been applied. In comparison with
 Fig.~\ref{fig:BtopiKpuzzle-had}, we obtain a much cleaner picture, requiring only $SU(3)$ input from $R_{T+C}$ and $R_q$, which are very robust as discussed in Section~\ref{sec:BtopiK}. On the other hand, Fig.~\ref{fig:BtopiKpuzzle-had} relies on the $SU(3)$ flavour 
 symmetry for the determination of the hadronic $B\to\pi K$ parameters from their $B\to\pi\pi$ counterparts.
 
Obviously, the observed tension with the SM in Fig.~\ref{fig:BtopiKpuzzle} could be resolved by a change of the data. However, it is far from trivial to fulfil all constraints simultaneously and an interesting question to explore how the data would have to change in
order to get agreement with the SM. In view of the large experimental uncertainty of the $B_d^0\to \pi^0 K^0$ branching ratio, this
quantity is a prime candidate. In fact, we find that lowering the central value of $\mathcal{B}r(B_d^0\to \pi^0 K^0)$ by about
$2.5\,\sigma$ gives a picture which is fully consistent with the SM, provided also the central value of the mixing-induced CP
asymmetry of $B_d^0\to \pi^0 K^0$ would move up by about $1 \,\sigma$. We illustrate the corresponding situation in 
Fig.~\ref{fig:solpuz}, where the part of the contour that is in agreement with the constraint on $\phi_\pm$ in Eq.~\eqref{eq:pmcon} is highlighted in cyan. In addition, the triangle determination agrees also with the brown SM band following from Eq.~(\ref{eq:SpiKs}) with $\phi_{00}$ in Eq.~(\ref{eq:phi00easy}) and the sum rule prediction in Eq.~\eqref{eq:Adirsumrule}.

On the other hand, the puzzling situation may also be a signal of NP effects in the EW penguin sector, thereby affecting the values
of $q$ and $\phi$. A particularly exciting aspect is the sensitivity to new sources of CP violation.

\section{Extracting the electroweak penguin parameters}\label{sec:qphidet}
\subsection{Preliminaries}
In the previous section, we have used the isospin relations in Eqs.~(\ref{eq:isospinrel}) and (\ref{eq:isospinrel-CP})
to calculate a correlation between the direct and mixing-induced CP asymmetries of the $B^0_d\to\pi^0K_{\rm S}$ channel, 
resulting in an intriguing picture for the current experimental data that may be an indication of a modified EW penguin sector. 
In view of this result and to test the corresponding SM sector, it would be very interesting to determine the EW penguin 
parameters $q$ and $\phi$ from experimental data and to compare the corresponding results with the SM prediction 
(see Eq.~\eqref{eq:qdef}). The parameter $R_q$ is then only needed for the SM prediction of $q$ while the CP-violating phase
$\phi$, which vanishes in the SM, may give a ``smoking-gun" signal of new sources of CP violation.

In order to achieve this goal, we apply again the isospin relations in Eqs.~(\ref{eq:isospinrel}) and (\ref{eq:isospinrel-CP})
for the neutral $B\to\pi K$ decays. These relations have also counterparts in the system of the charged $B\to\pi K$ decays, 
where the $B^+\to \pi^0 K^+$ mode receives significant contributions from colour-allowed EW penguin topologies. We have
\begin{equation} \label{eq:isospin2}
  \sqrt{2} A(B^+\to \pi^0 K^+) + A(B^+\to \pi^+ K^0) 
  = 3A_{3/2} \equiv 3|A_{3/2}| e^{i\phi_{3/2}}
 \end{equation}
\begin{equation} \label{eq:isospin2-CP}
  \sqrt{2} A(B^-\to \pi^0 K^-) + A(B^-\to \pi^- \bar K^0) 
  = 3\bar{A}_{3/2}  \equiv 3|\bar{A}_{3/2}| e^{i\bar{\phi}_{3/2}},
 \end{equation} 
 where the isospin amplitude $A_{3/2}$ and its CP-conjugate $\bar{A}_{3/2}$ are given in Eqs.~(\ref{eq:A32}) and 
 (\ref{eq:A32-CP}), respectively. 
 
 In view of the large experimental uncertainties of the CP-violating observables of the $B_d^0 \to \pi^0 K_{\rm S}$ channel, 
 let us focus on the charged $B\to\pi K$ decays. Using the corresponding CP-averaged branching ratios
 and direct CP asymmetries, the isospin relations in Eqs.~(\ref{eq:isospin2}) and (\ref{eq:isospin2-CP}) can be represented 
 as amplitude triangles in the complex plane for a given value of $|A_{3/2}|=|\bar{A}_{3/2}|$. The relative orientation of the
 triangles is fixed through the tiny angle 
 \begin{equation}\label{eq:phicval}
\phi_{\rm c} \equiv \text{Arg}\left[\bar{A}_{+0} A_{+0}^*\right] = {\cal O}(1^\circ)
\end{equation}
 between $A_{+0} \equiv A(B^+ \rightarrow \pi^+ K^0)$ and $\bar{A}_{+0} \equiv A(B^- \rightarrow \pi^- \bar{K}^0)$. 
 Employing Eq.~(\ref{eq:ampli}) for the $B^+\to \pi^+K^0$ amplitude and neglecting the colour-suppressed 
 EW penguin contributions (see Subsection~\ref{ssec:CSEWP}), we obtain 
 \begin{equation}
\tan\phi_{\rm c} = 
\frac{-\rho_{\rm c}^2 \sin 2\gamma - \rho_{\rm c} \sin(\gamma-\theta_{\rm c})-\rho_{\rm c} \sin(\gamma+\theta_{\rm c})}{1+\rho_{\rm c}^2 
\cos 2\gamma + \rho_{\rm c} \cos(\gamma-\theta_{\rm c})+
\rho_{\rm c} \cos(\gamma+\theta_{\rm c})} \ .
\end{equation}
Using then the values of the corresponding parameters  in Table~\ref{tab:sumpara} yields
\begin{equation}
\phi_{\rm c} = (-3.2 \pm 1.1)^\circ.
\end{equation}

\begin{figure}[t]
\centering
\includegraphics[width=0.7\linewidth]{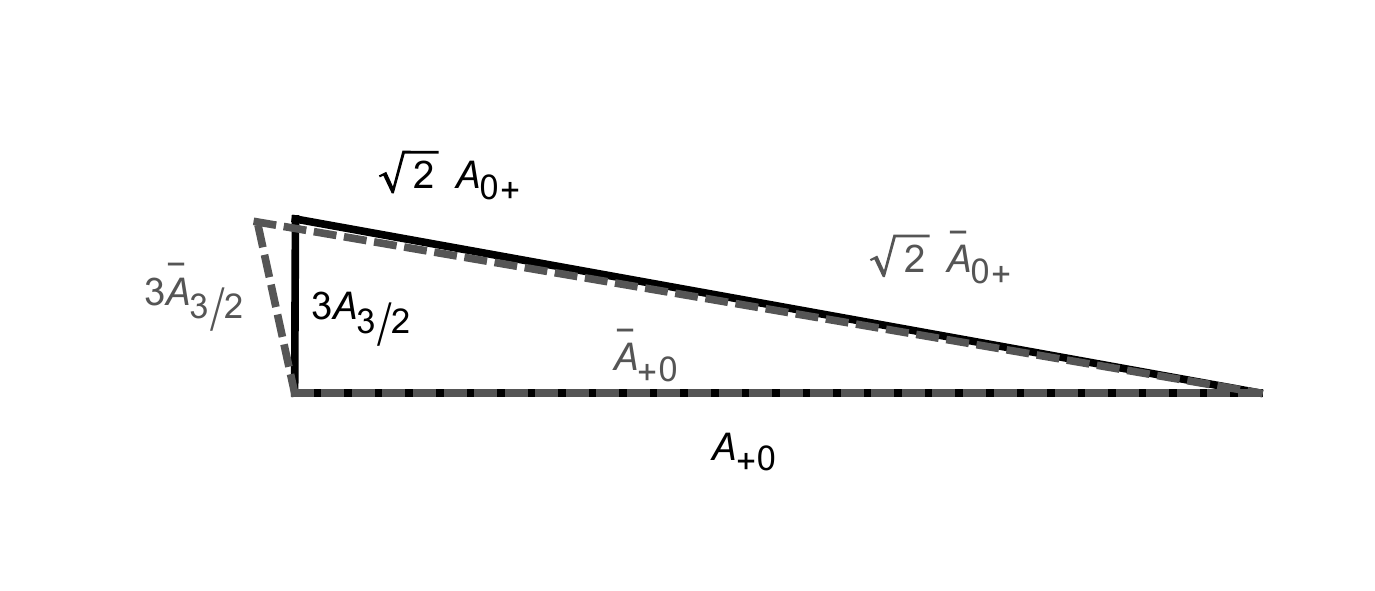}
\caption{Illustration of the isospin triangles for the charged $B \rightarrow \pi K$ decays with $|A_{3/2}|=|\bar A_{3/2}|$.}
\label{chargedtriangles}
\end{figure}

\begin{figure}[h!]
	\centering
	\subfloat[]{\label{fig:charged}\includegraphics[width = 0.5\linewidth]{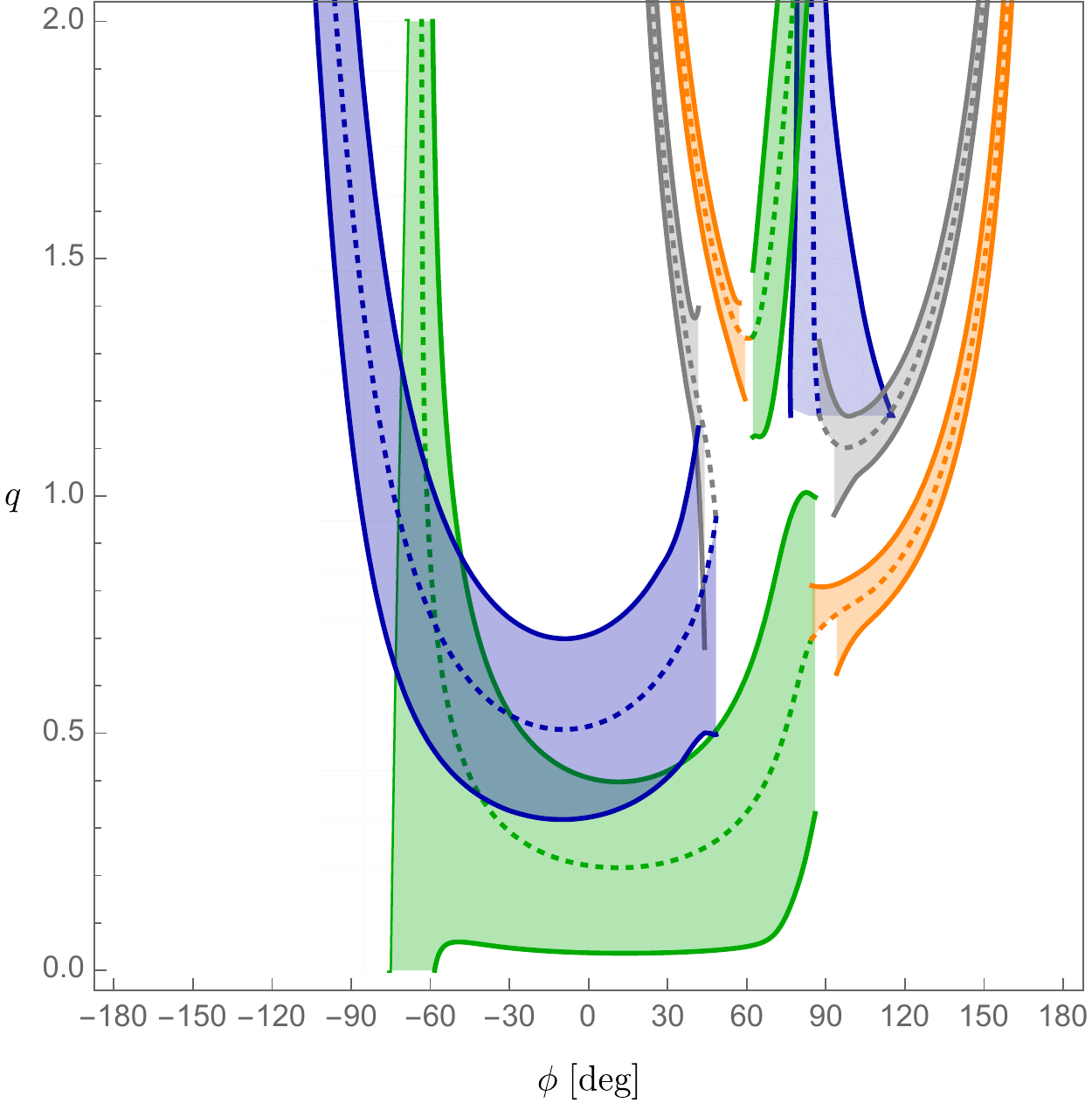}}
		\subfloat[]{\label{fig:chargedreduced}\includegraphics[width = 0.5\linewidth]{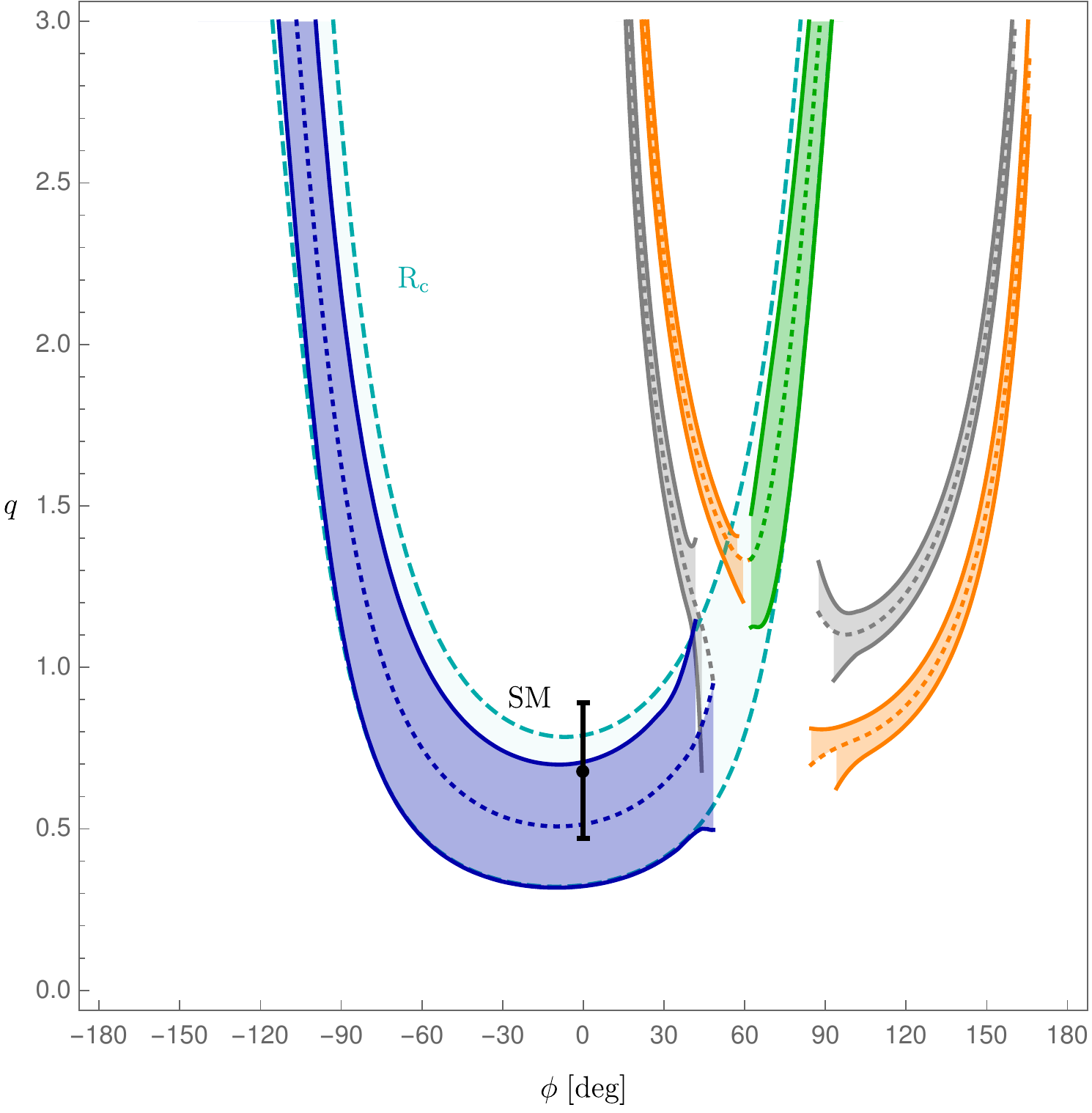}}\\
	\subfloat[]{\label{fig:charan1} \includegraphics[width=0.49\textwidth]{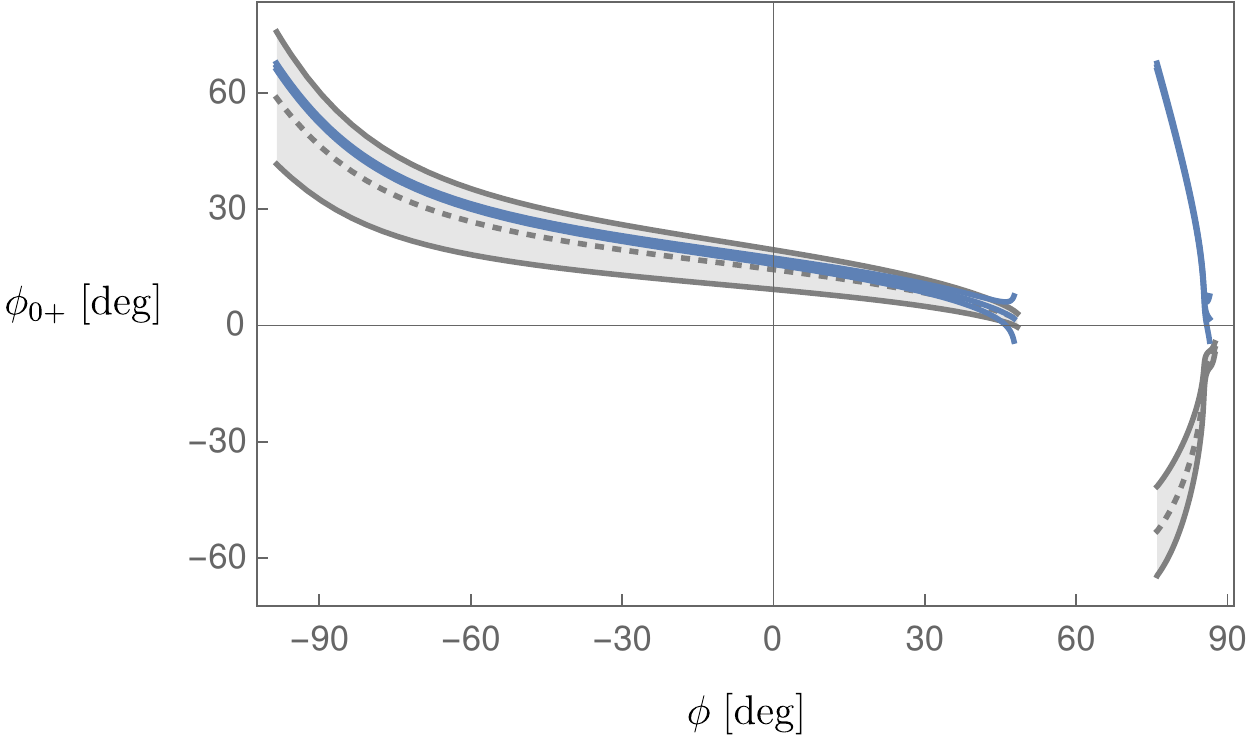}}
	\subfloat[]{\label{fig:charan2} \includegraphics[width=0.49\textwidth]{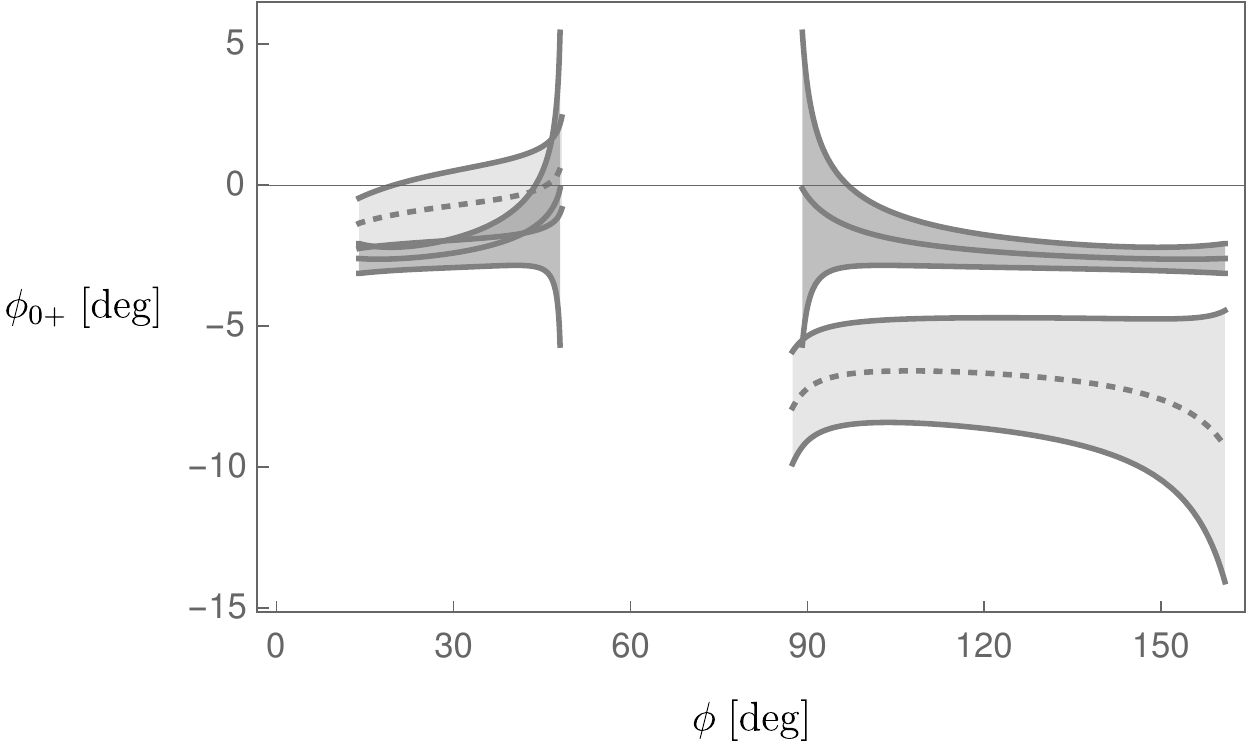}}\\
		\subfloat[]{\label{fig:charan3} \includegraphics[width=0.49\textwidth]{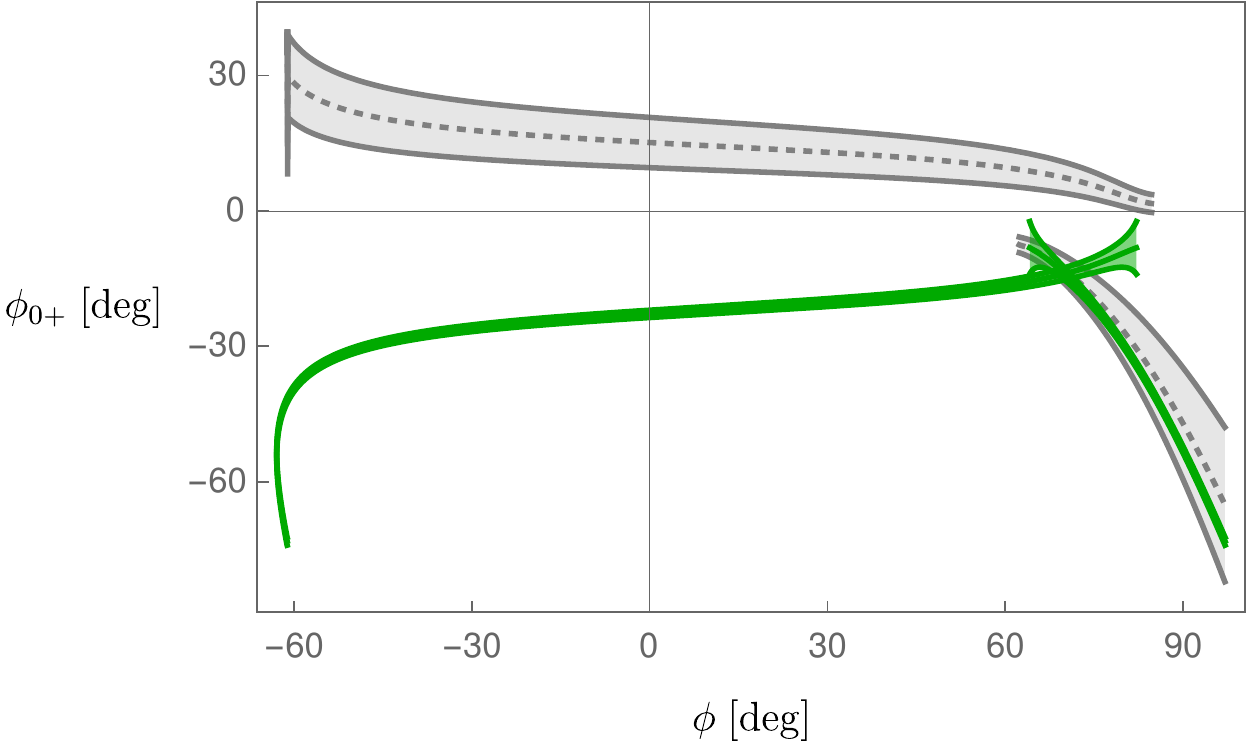}}
	\subfloat[]{\label{fig:charan4} \includegraphics[width=0.49\textwidth]{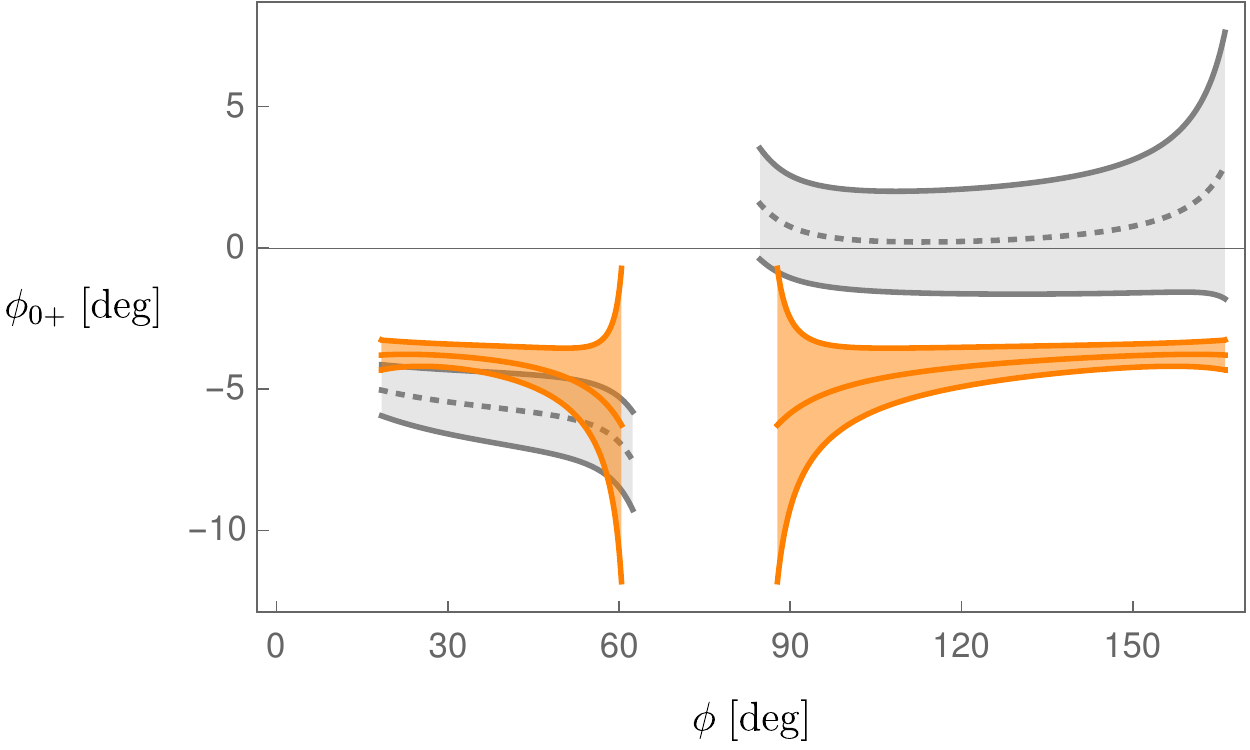}}
	\caption{(a) Contours in the $\phi$--$q$ plane for the current data for the charged $B\to \pi K $ decays. (b) Contours 
	remaining after imposing the constraints discussed in the text. (c--f) Theory constraints (grey) combined with the $\phi_{0+}$
	constraint following from the triangle constructions in the same colours. }
	\label{fig:angle2}
\end{figure}

In Fig.~\ref{fig:charged}, we illustrate the charged $B\to\pi K$ isospin triangles for the central values of the current data, 
assuming the SM values of the EW penguin parameters. The triangle construction allows us to determine the difference 
\begin{equation}
\Delta\phi_{3/2}\equiv \phi_{3/2}-\bar\phi_{3/2}
\end{equation}  
between the phases $\phi_{3/2}$ and $\bar{\phi}_{3/2}$ of the amplitudes  $A_{3/2}$ and $\bar A_{3/2}$, respectively, which
is given by $\Delta\phi_{3/2}=2\phi_{3/2}$ as can be seen in Eq.~(\ref{phi-diff}). Introducing  \begin{equation}\label{N-def}
N\equiv 3 |A_{3/2}|/|\hat{T}' +\hat{C}'| \ ,
\end{equation}
 we obtain
  \begin{equation}\label{q-det}
q=\sqrt{N^2-2c\cos\gamma-2s\sin\gamma+1}
\end{equation}
and
\begin{equation}
\tan\phi=\frac{\sin\gamma-s}{\cos\gamma-c}, \quad q\,\sin\phi=\sin\gamma-s
\end{equation}
with
\begin{equation}
c\equiv \pm N\cos(\Delta\phi_{3/2}/2), \;\;\; s\equiv \pm N\sin(\Delta\phi_{3/2}/2),
\end{equation}
 allowing us to calculate contours in the $\phi$--$q$ plane. In order to convert the given value of 
 $|A_{3/2}|=|\bar A_{3/2}|$ into the parameter $N$, we use again the $SU(3)$ relation in Eq.~(\ref{eq:tcabs}).
 
For the current charged $B\to \pi K$ decay data, we arrive at the contours shown in Fig.~\ref{fig:charged}. As was the case in Section~\ref{sec:1}, we have a four-fold ambiguity for $\Delta\phi_{3/2}$ since the triangles can be flipped around the $A_{3/2}$ and $\bar{A}_{3/2}$ axes. This is represented by the four different colours for the contours in Fig.~\ref{chargedtriangles}. Moreover, for every value of $\Delta\phi_{3/2}$, there are two contours in the $\phi$--$q$ plane due to solving a quadratic equation, giving two contours of every colour and eight contours in total. We find discontinuities of the contours around $q\sim1$, $\phi\sim70^\circ$, because $|A_{3/2}|$ cannot become arbitrarily small as then the amplitudes in Eq.~(\ref{eq:isospin2}) cannot form triangles anymore.

As in the neutral case, we can eliminate some contours by considering the angle
\begin{equation}
\phi_{0+} = \text{Arg}\left[ \bar{A}_{0+} A_{0+}^*\right] \ ,
\end{equation}
where $A_{0+} \equiv A(B^+ \to \pi^0 K^+)$ and  $\bar{A}_{0+} \equiv A(B^- \to \pi^0 K^-)$. We may now compare $\phi_{0+}$ as obtained from the triangle construction with its theoretical prediction:
\begin{equation}
\tan\phi_{0+} = 2 r_{\rm c} \left[\cos\delta_{\rm c} \sin\gamma - \left(\cos\delta_{\rm c} - \frac{1}{3} \tilde{a}_C \right) q \sin\phi\right] + \mathcal{O}(r_{({\rm c})}^2, \rho_c) \ ,
\end{equation}
where the colour-suppressed EW penguin parameter $\tilde{a}_C$ was defined in Eq.~\eqref{eq:actil}. These effects can be included using the ratio $R$ via Eq.~\eqref{eq:actildet}. Contrary to the SM case discussed above, now the theoretically allowed $\phi_{0+}$ depends on $q$ and $\phi$. At the same time, the $\phi_{0+}$ obtained from the triangle construction also depends on $\phi$. In Fig.~\ref{fig:angle2}, we show this angle for each of the eight branches of the triangle determinations in the same colour. In addition, in grey we show the theoretically allowed values of $\phi_{0+}$ as a function of $\phi$, using the exact expression but neglecting colour-suppressed EW penguin contributions. For this theoretical prediction, we use the $q$ as a function of $\phi$ from the associated triangle contour. This implies that each of the eight triangle contours has a different theoretical prediction for $\phi_{0+}$ as function of $\phi$. We observe that one of the contours in Fig.~\ref{fig:charan1} and one in Fig.~\ref{fig:charan3} is clearly excluded by the theoretical constraint on $\phi_{0+}$. We have removed those curves
in Fig.~\ref{fig:chargedreduced}. 

It is interesting to have a closer look at the ratio $R_{\rm c}$ of the CP-averaged branching ratios of the charged $B\to \pi K$ 
decays introduced in  Eq.\ \eqref{Rc-expr}. It allows us to derive the following exact expression:
\begin{equation}\label{eq:consRc}
q = \frac{-B_{R_{\rm c}} \pm \sqrt{B_{R_{\rm c}}^2-4A_{R_{\rm c}}C_{R_{\rm c}}}}{2A_{R_{\rm c}}} \ ,
\end{equation}
where 
\begin{align}
A_{R_{\rm c}} &\equiv r_{\rm c}^2 \ , \\
B_{R_{\rm c}} &\equiv  2r_{\rm c} \left[ \cos\delta_{\rm c}\cos\phi-\left(r_{\rm c}-\rho_{\rm c}\cos(\theta_{\rm c}-\delta_{\rm c})\right)\cos(\gamma-\phi) \right]\ , \\
C_{R_{\rm c}} &\equiv \left[1+2\rho_{\rm c}\cos\theta_{\rm c}\cos\gamma+\rho_{\rm c}^2\right]\left[1-R_{\rm c}\right] - 2\rho_{\rm c}r_{\rm c}\cos(\theta_{\rm c}-\delta_{\rm c}) \nonumber  \\
&-2r_{\rm c}\cos\delta_{\rm c}\cos\gamma + r_{\rm c}^2 .
\end{align}
Using the information for $r_{\rm c}$ and $\delta_{\rm c}$ in Eq.~(\ref{eq:rcval}) and including also the tiny $\rho_{\rm c}$ parameter 
as given in Eq.~(\ref{eq:rhocval}), the measured value of $R_{\rm c}$ can be converted into yet another contour in the
$\phi$--$q$ plane. In contrast to the analysis using the isospin relations, we require now also the strong phase $\delta_{\rm c}$.
In  Fig.~\ref{fig:chargedreduced}, we have added the resulting contour, which is in excellent agreement with two branches of the
isospin triangle construction. This curve is actually also consistent with the SM value of $q$ and $\phi$.

We note that the allowed parameter space for $q$ and $\phi$ following from the current data of the charged $B\to\pi K$ system is 
significantly reduced in comparison with the situation discussed in Ref.~\cite{FJPZ}. Moreover, we have presented a transparent way to
calculate the contours in the $\phi$--$q$ plane and do not have to make a fit to the data. The constraints on $q$ and $\phi$ have
actually a highly non-trivial structure that follows from the  isospin relation and can be understood in an analytic way. 
The only additional $SU(3)$ input is the quantity $R_{T+C}$ discussed in Section~\ref{sec:dethad}, which is required for the conversion of $|A_{3/2}|$ into the parameter $N$.

\begin{figure}[t]
	\centering
	\subfloat[]{\label{fig:errortri}\includegraphics[height = 0.33\linewidth]{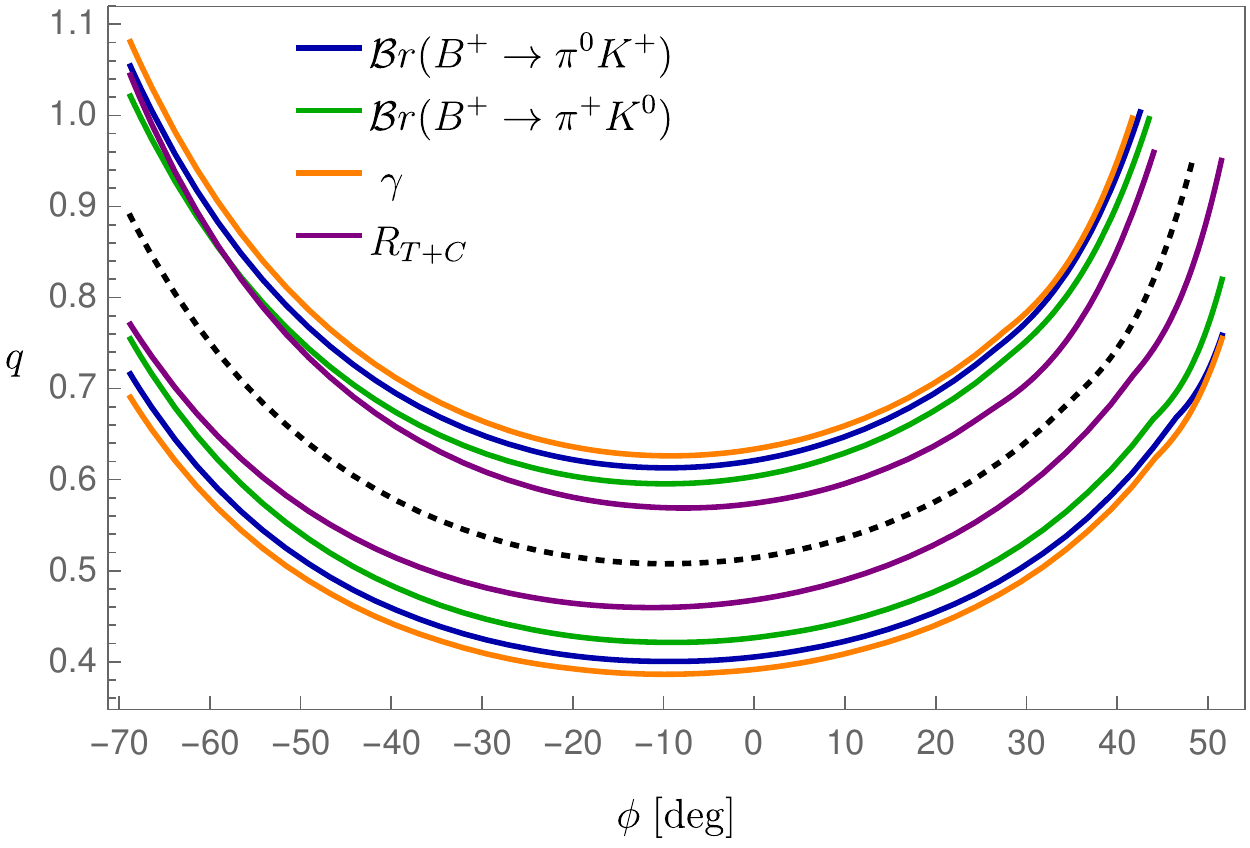}} \qquad
	\subfloat[]{\label{fig:piechart}\includegraphics[height = 0.33\linewidth]{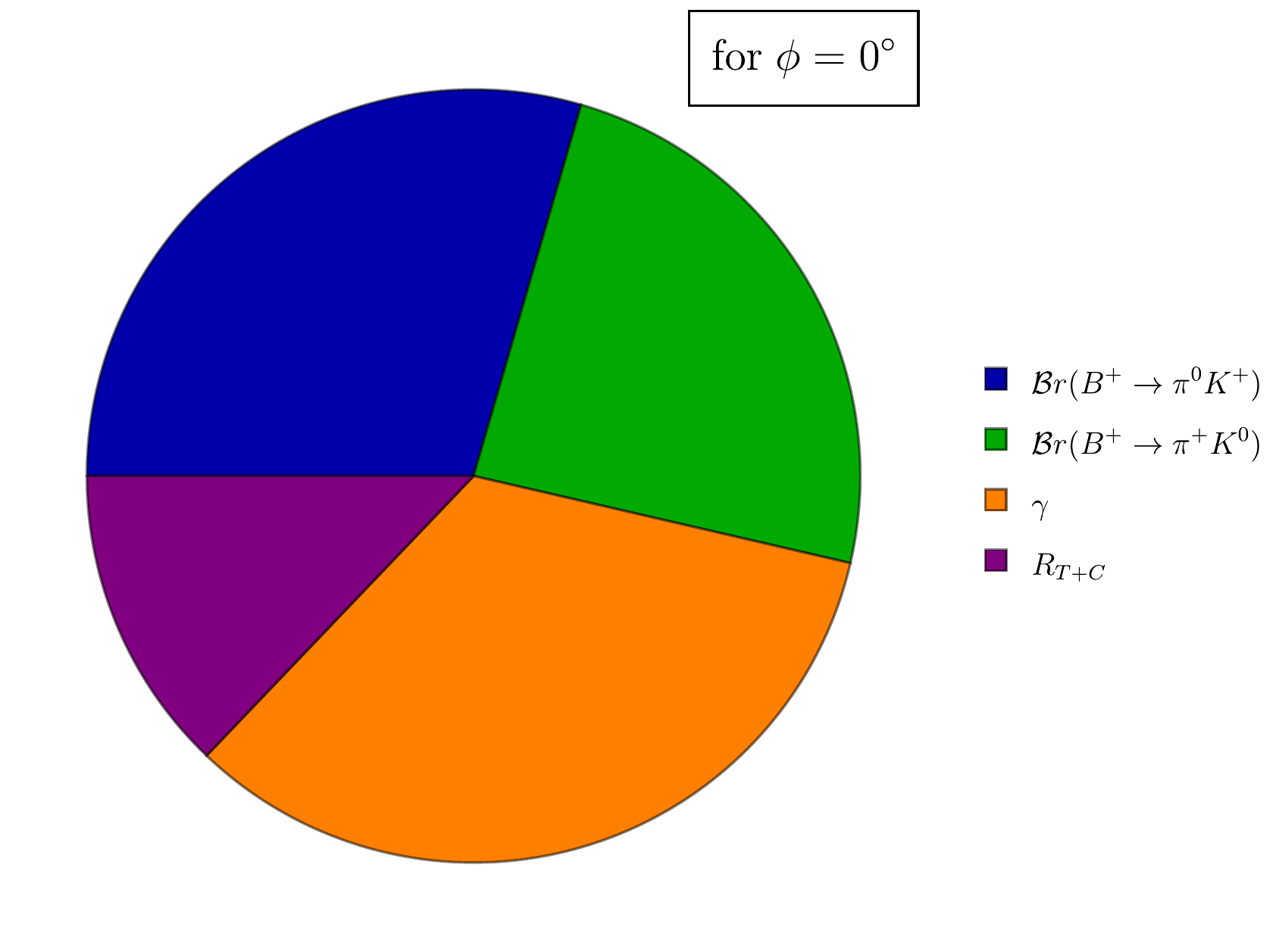}}
	\caption{Error budget for the isospin contour in the $\phi$--$q$ plane that is consistent with the $R_{\rm c}$ constraint in Eq.~\eqref{eq:consRc}: (a) impact of the various parameters when varying them individually within their $1\,\sigma$ ranges;
	(b) pie chart to illustrate the relative contributions of the parameters to the total uncertainty of $q$ for $\phi=0^\circ$.
}\label{fig:errortri-piechart}
\end{figure}

In Fig.~\ref{fig:errortri}, we discuss the uncertainties of the various input parameters, focusing on the contour in the 
$\phi$--$q$ plane in Fig.~\ref{fig:chargedreduced} that is in agreement with the $R_{\rm c}$ constraint. When adding the 
individual errors in quadrature, we obtain the uncertainty band in Fig.~\ref{fig:chargedreduced}. In Fig.~\ref{fig:piechart}, we illustrate the error budget as a pie chart. We observe that $\gamma$ 
and the branching ratios play the major roles, while $R_{T+C}$ has a slightly smaller impact on the error budget.

In analogy to the discussion of the charged $B\to \pi K$ system given above, we may also use the neutral $B\to\pi K$ decays 
and their isospin amplitude relations to determine contours in the $\phi$--$q$ plane. The key difference is that the measurement of the 
mixing-induced CP asymmetry of $B^0_d\to\pi^0 K_{\rm S}$ allows us to determine the angle $\phi_{00}$ in a clean way
through Eq.~(\ref{s2b}), thereby fixing the relative orientation of the neutral $B\to\pi K$ isospin triangle and its CP conjugate.
In contrast to using $\phi_{\rm c}$ in Eq.~(\ref{eq:phicval}) for the charged $B\to\pi K$ decays, this determination is theoretically 
clean (although also $\phi_{\rm c}$ is only affected by a small theoretical uncertainty). The charged and neutral $B\to\pi K$ decays
should result in constraints in the $\phi$--$q$ plane that are consistent with each other.

\boldmath 
\subsection{Utilizing mixing-induced CP violation in $B^0_d\to \pi^0K_{\rm S}$}
\unboldmath\label{sec:CSEWPs}
In order to not just constrain $q$ and $\phi$ but to determine these parameters, further information is needed. It is provided by the
mixing-induced CP asymmetry $S_{\text{CP}}^{\pi^0 K_{\text{S}}}$, which allows the extraction of the phase $\phi_{00}$. If we
use the values of the hadronic parameters $r_{\rm c}$, $\delta_{\rm c}$ and $r$, $\delta$ as determined 
in Subsection~\ref{sec:dethad}, we may convert this observable into a contour in the $\phi$--$q$ plane with the help of the 
following expression:
\begin{equation}\label{eq:qfromS}
q = \frac{- B_c + \sqrt{B_c^2 - 4 A_c D_c}}{2 A_c} \ , 
\end{equation}
where 
\begin{align}
A_c {}& \equiv r_c^2 (-\tan\phi_{00} \cos 2\phi - \sin 2 \phi) \ , \\
B_c {}& \equiv 2 r_c \cos\delta_c (\tan\phi_{00} \cos\phi+\sin\phi) - \frac{4}{3} \hat{c}_+ A_c \nonumber \\ - {}& (2r_c^2 -2 r_c r \cos(\delta_c-\delta)) (-\tan\phi_{00} \cos (\gamma+\phi) - \sin(\gamma+\phi)) \\
D_c {}& \equiv -\tan\phi_{00}  - (2 r_c \cos\delta_c - 2r\cos\delta)(\tan\phi_{00}  \cos\gamma+\sin\gamma) \nonumber\\
{}& + (r_c^2 +r^2 -2 r_c r \cos(\delta_c-\delta))(-\tan\phi_{00} \cos 2\gamma - \sin 2 \gamma) \nonumber\\
{}& + \frac{4}{3} \tilde{a}_C \;q r_c (-\tan\phi_{00}  \cos\phi-\sin\phi) + \frac{4}{9} q^2 (\tilde{a}_S^2 + \tilde{a}_C^2) A_c \nonumber\\
{}& + \frac{4}{3} (-\tan\phi_{00} \cos (\gamma+ \phi) - \sin (\gamma+ \phi))( r_c^2 \hat{c}_+ - r_c r (\tilde{a}_C \cos\delta + \tilde{a}_S \sin\delta )) 
\end{align}
with
\begin{align}
\hat{c}_+ \equiv \tilde{a}_C \;q \cos\delta_c + \tilde{a}_S \; q \sin\delta_c \ .
\end{align}
As discussed in Subsection~\ref{ssec:CSEWP}, we can determine the colour-suppressed EW penguin parameters $\tilde{a}_C$ 
and $\tilde{a}_S$ in Eqs.~\eqref{eq:actil} and \eqref{eq:astil} from experimental data using $R$ and $A_{\text{CP}}^{\pi^-K^+}$,
allowing us to take also these contributions into account.

\begin{figure}[t]
	\centering
	\includegraphics[width = 0.5\linewidth]{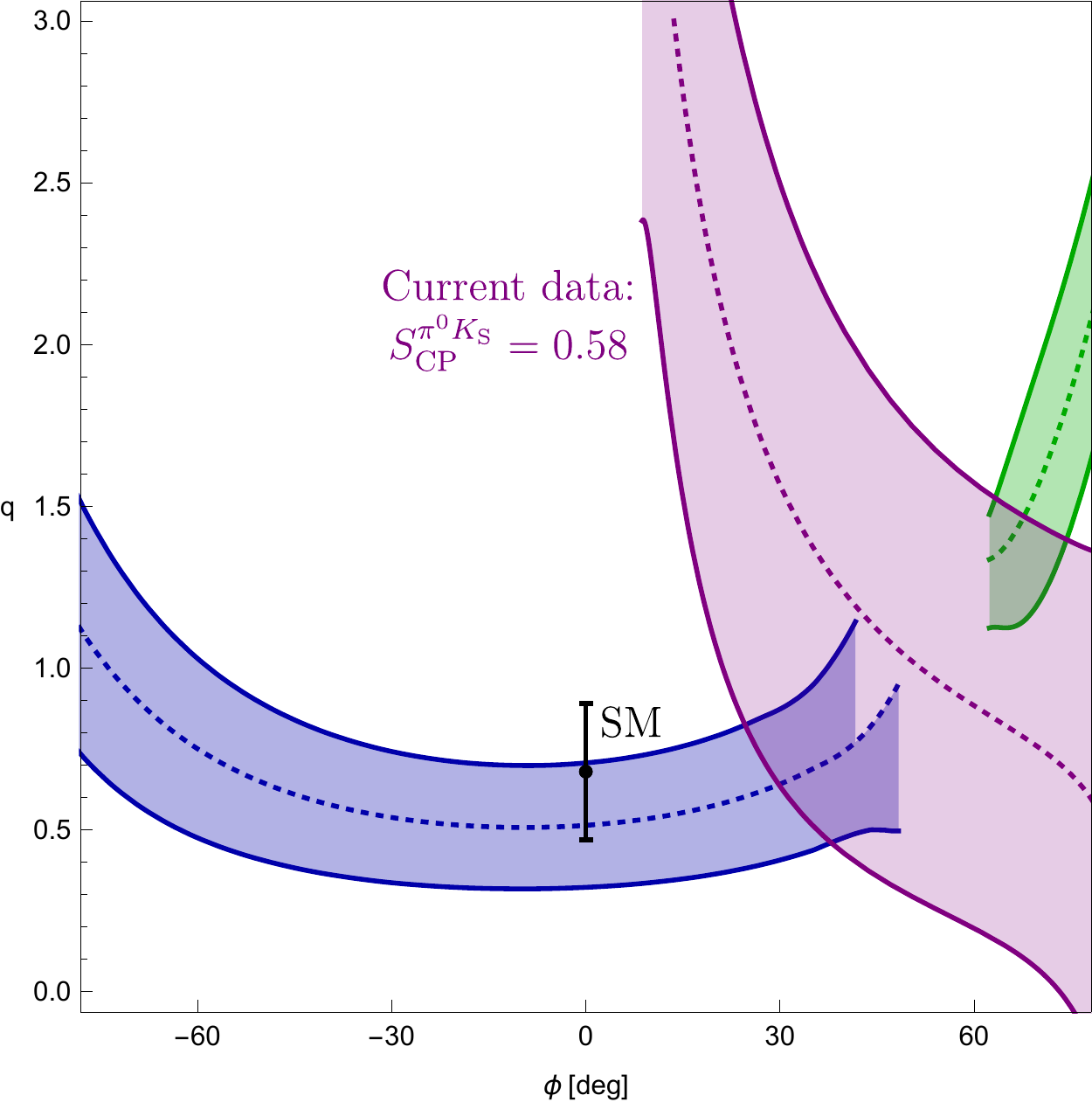}
	\caption{Constraints on the EW penguin parameters $q$ and $\phi$ from current data. The blue and green contours follow from the isospin analysis and are in agreement with the constraint from $R_{\rm c}$.}\label{fig:currentdata}
\end{figure}

Using the current measurement of $S_{\rm CP}^{\pi^0 K_{\rm S}}$ in Table~\ref{tab:CPs} gives 
\begin{equation}
\phi_{00} = (7.7 \pm 12.1)^\circ ,
\end{equation}
which should be compared with the SM prediction in Eq.~(\ref{eq:tanphi00-SM}). 
From Eq.~(\ref{eq:qfromS}) we then obtain the purple contour in Fig.~\ref{fig:currentdata}, which includes contributions from colour-suppressed EW penguin topologies. We show also the contours from the isospin analysis that agree with the $R_{\rm c}$ constraint, and the SM point from Eq.~(\ref{eq:qdef}).

\begin{table}
  \centering
    \begin{tabular}{l | c  |c |c |c }
    Scenario & $S_{\rm CP}^{\pi^0 K_{\rm S}}$ &  $A_{\rm CP}^{\pi^0 K_{\rm S}}$&  $\phi_{00}$   \\
    \hline \hline
1 & $0.67  \pm 0.042$ &$-0.07 \pm 0.042$ & $(0.9\pm 3.3)^\circ$ \\
2 & $0.33 \pm 0.042 $  & $-0.06 \pm 0.042$ &$(23.9\pm 2.6)^\circ$ \\ 
3 & $0.91 \pm 0.042$ &$-0.07  \pm 0.042$ &$(-23.0 \pm 6.0 )^\circ$ \\
    \end{tabular}
    \caption{Scenarios for future measurements of $S_{\rm CP}^{\pi^0 K_{\rm S}}$.}
    \label{tab:scen}
\end{table}

\begin{figure}[t!]
	\centering
	\subfloat[Scenario 1]{\label{fig:confinal1}\includegraphics[width = 0.33\linewidth]{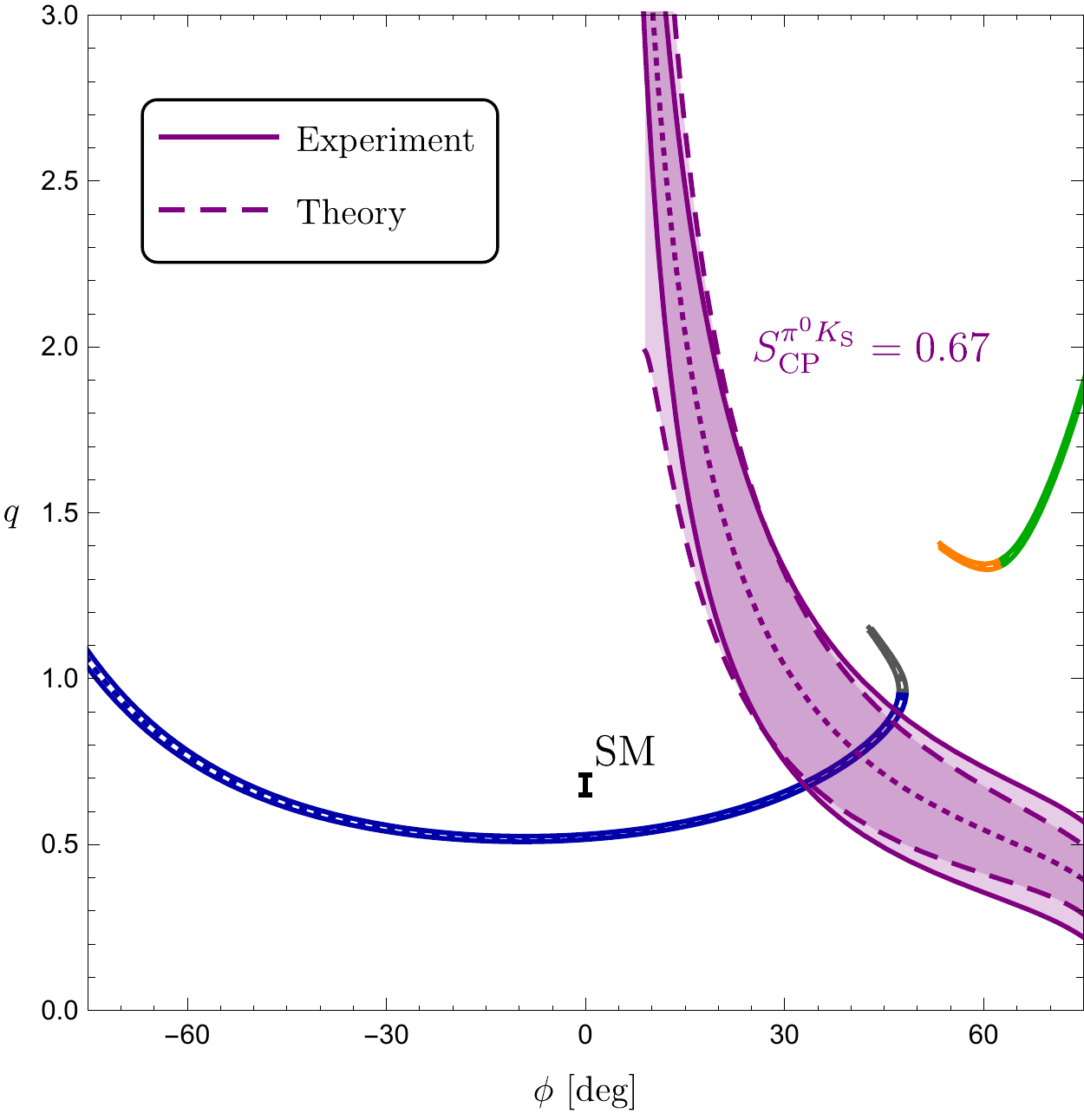}}
		\subfloat[Scenario 2]{\label{fig:confinal2}\includegraphics[width = 0.33\linewidth]{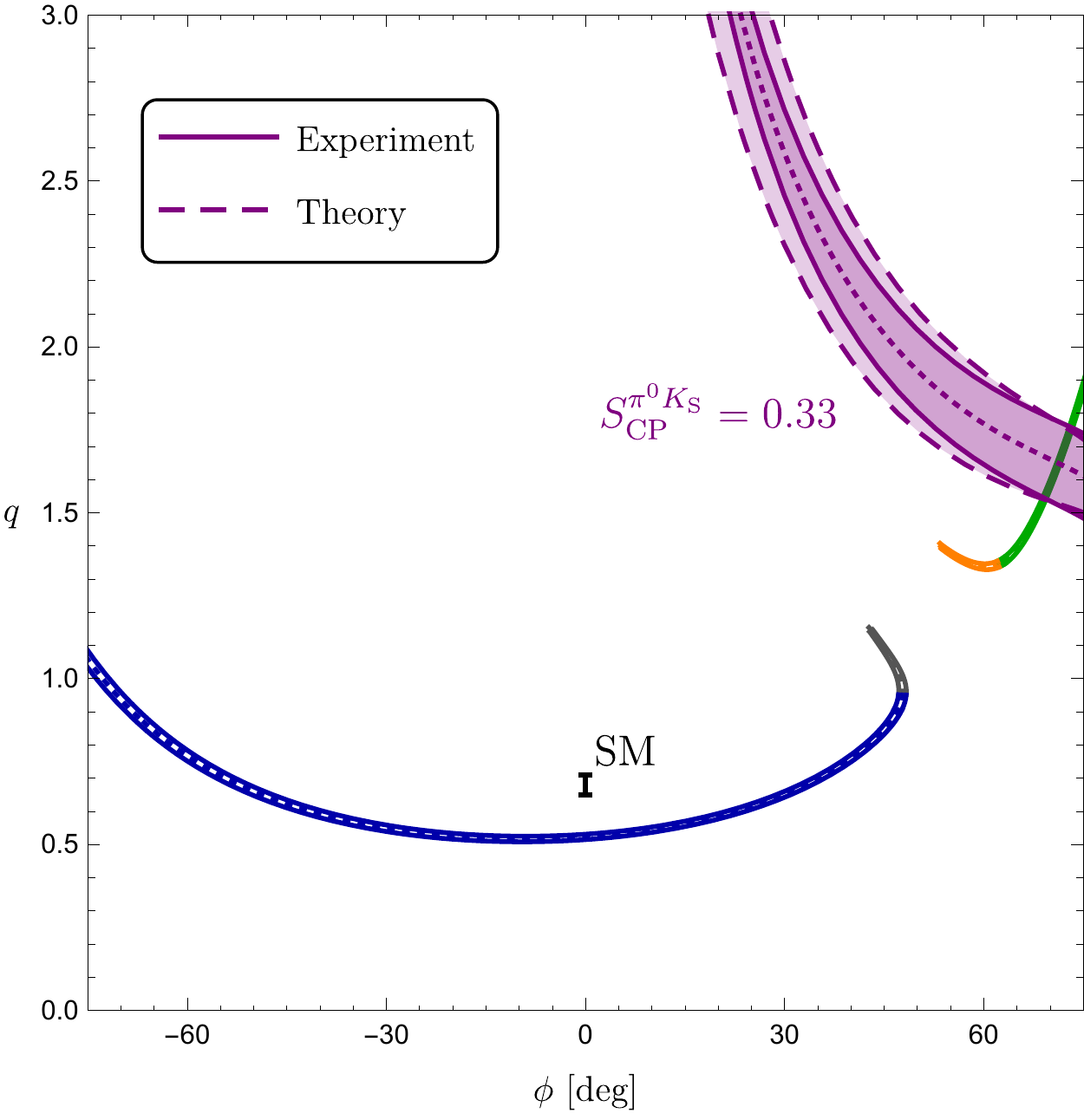}}
	\subfloat[Scenario 3]{\label{fig:confinal3} \includegraphics[width=0.33\textwidth]{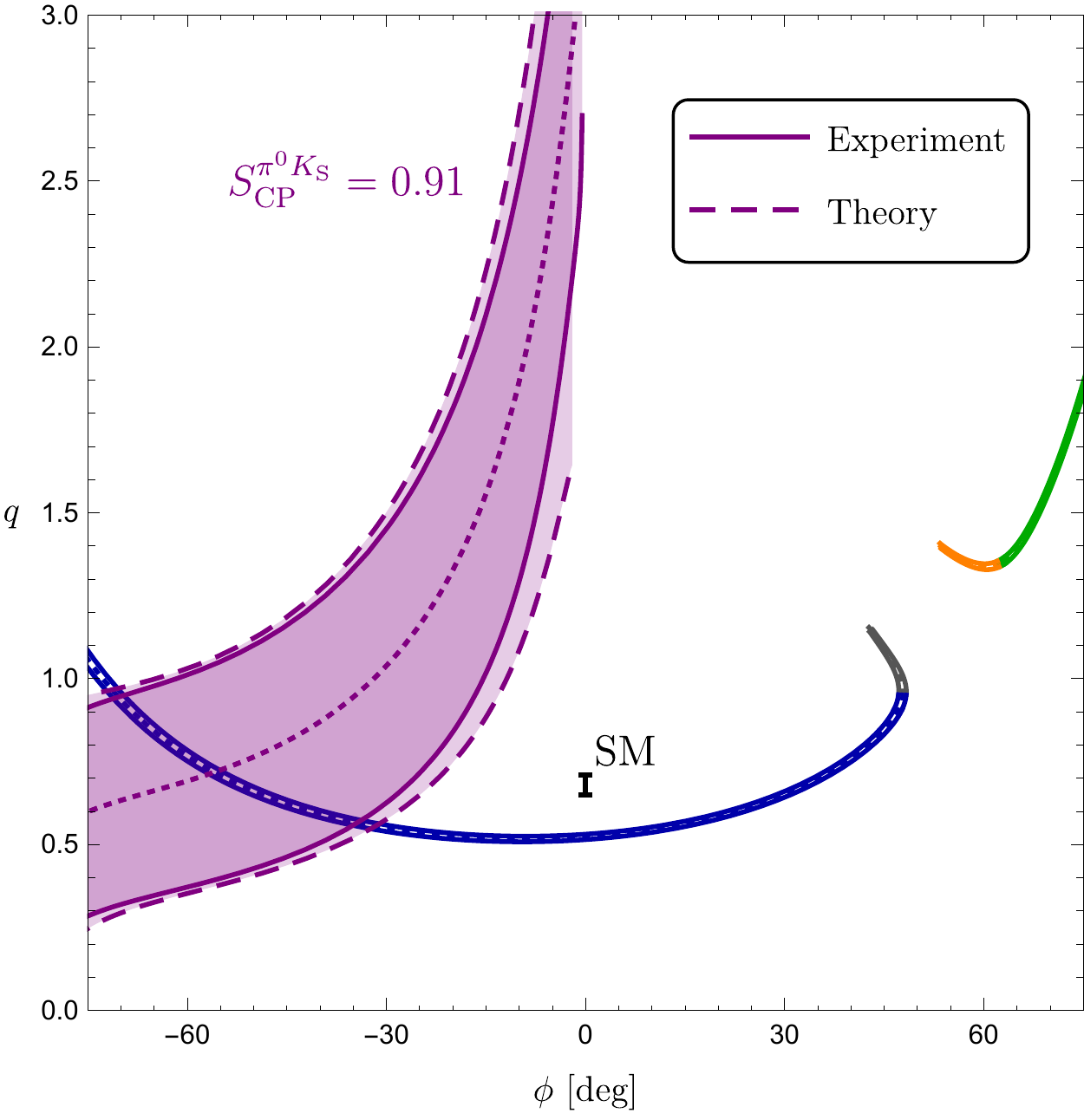}}
	\caption{Illustration of the future scenarios specified in Table~\ref{tab:scen}. For the constraints following from 
	measurements of $S^{\pi^0 K_{\rm S}}_{\text{CP}}$, the experimental and theory uncertainties are given separately.  }
	\label{fig:confinal}
\end{figure}

In order to demonstrate the future application of our strategy, we consider three scenarios for measurements of 
$S_{\text{CP}}^{\pi^0 K_{\text{S}}}$ as summarized in Table~\ref{tab:scen}. In the corresponding numerical analyses, 
we include effects of colour-suppressed EW penguin topologies for completeness. We assume that  
$S_{\text{CP}}^{\pi^0 K_{\text{S}}}$ has the same uncertainty as $A_{\text{CP}}^{\pi^0 K_{\text{S}}}$ and 
use the corresponding value anticipated for Belle II in Ref.~\cite{Belle-II}; unfortunately, the mixing-induced CP asymmetry was not
considered in this reference. In Fig.~\ref{fig:confinal}, we show the constraints in the $\phi$--$q$ plane resulting from $S_{\text{CP}}^{\pi^0 K_{\text{S}}}$ and the isospin determination separately for the three scenarios. We also give the SM point corresponding to the value of $R_q$ in Eq.~\eqref{eq:theoryupRq}. For the constraints  following from $S_{\text{CP}}^{\pi^0 K_{\text{S}}}$, we take into account the experimental uncertainties on $A_{\text{CP}}^{\pi^0 K_{\text S}}$ and $S_{\text{CP}}^{\pi^0 K_{\text S}}$ as given in Table~\ref{tab:scen}. In addition, we take into account the theoretical $SU(3)$ uncertainties for the hadronic parameters that are required to determine $q$ from Eq.~(\ref{eq:qfromS}). We show these experimental and theoretical uncertainties seperately in Fig.~\ref{fig:confinal}. In addition, for the isospin triangle constraints we only show the contours that remain after taking into account constraints from $\phi_{0+}$ and $R_{\rm c}$. For the uncertainty, we only consider the uncertainty on $R_{T+C}$ as in Eq.~\eqref{eq:theoryuprtc}. The theory uncertainty (dashed line) matches the future experimental uncertainty (solid line), which is very promising.

\begin{figure}[t]
	\centering
	\includegraphics[width = 0.33\linewidth]{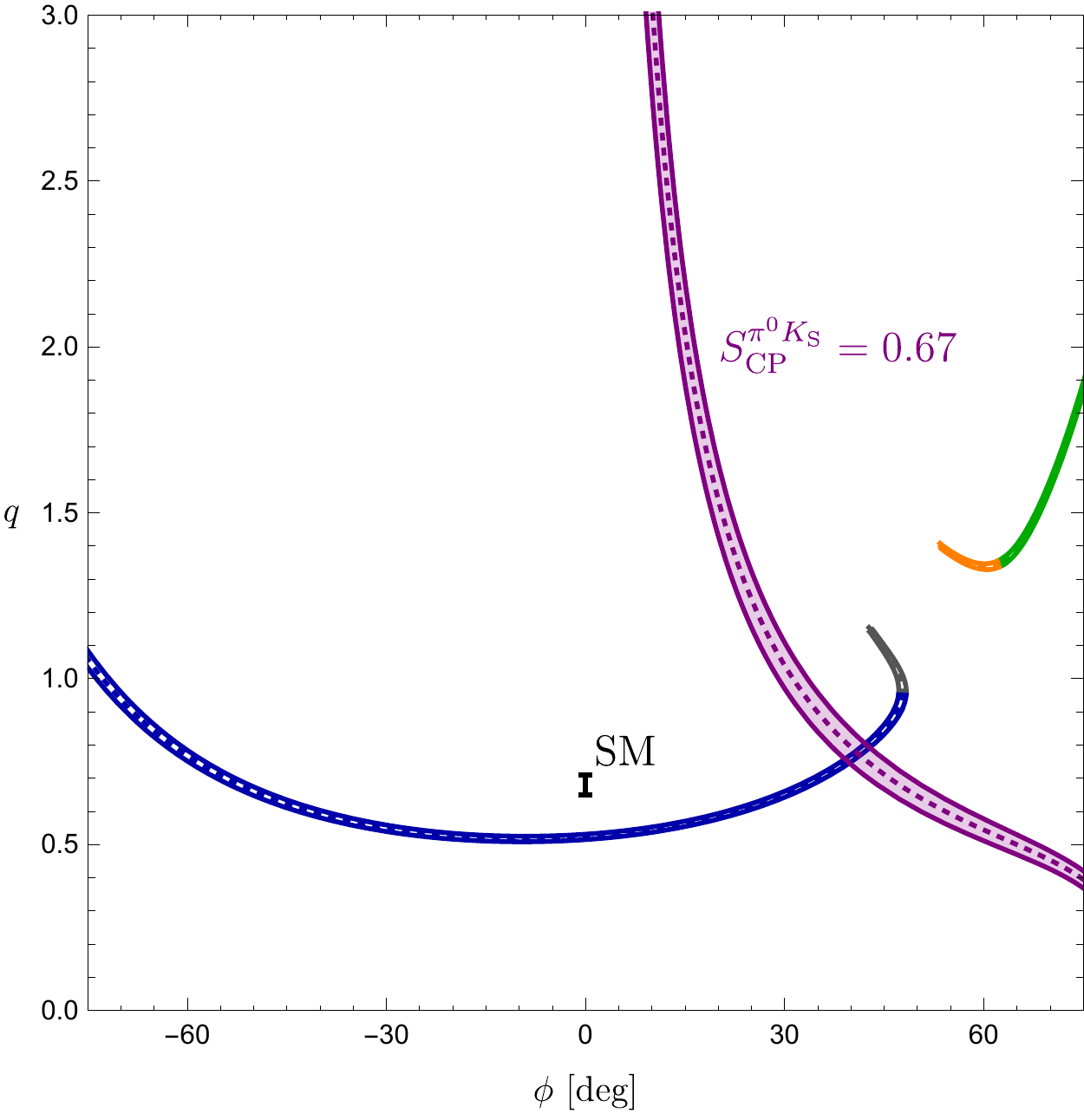}
	\caption{Scenario 1 taking only the expected future theory uncertainties into account. }\label{fig:futuretheory}
\end{figure}

Progress on theory and the interplay with experiment may lead to an even sharper picture for the hadronic parameters (see Section~\ref{sec:dethad} and Ref.~\cite{BIG}). As an illustration, we assume a scenario where the $SU(3)$-breaking corrections can be reduced by a factor of four with respect to the current situation. Taking only these uncertainties into account, we obtain the constraints in Fig.~\ref{fig:futuretheory}. These considerations show the exciting potential of the 
new strategy, going even beyond the next generation of $B$-decay experiments.

\begin{figure}   
	\centering
	\subfloat[]{\includegraphics[width=0.49\textwidth]{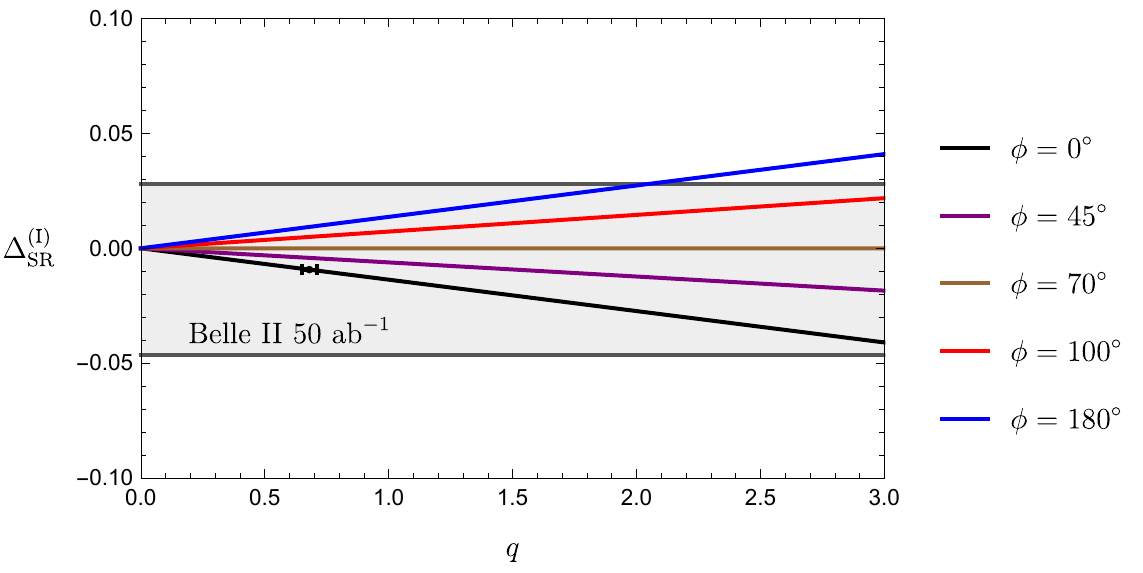} }
	\subfloat[]{\includegraphics[width=0.49\textwidth]{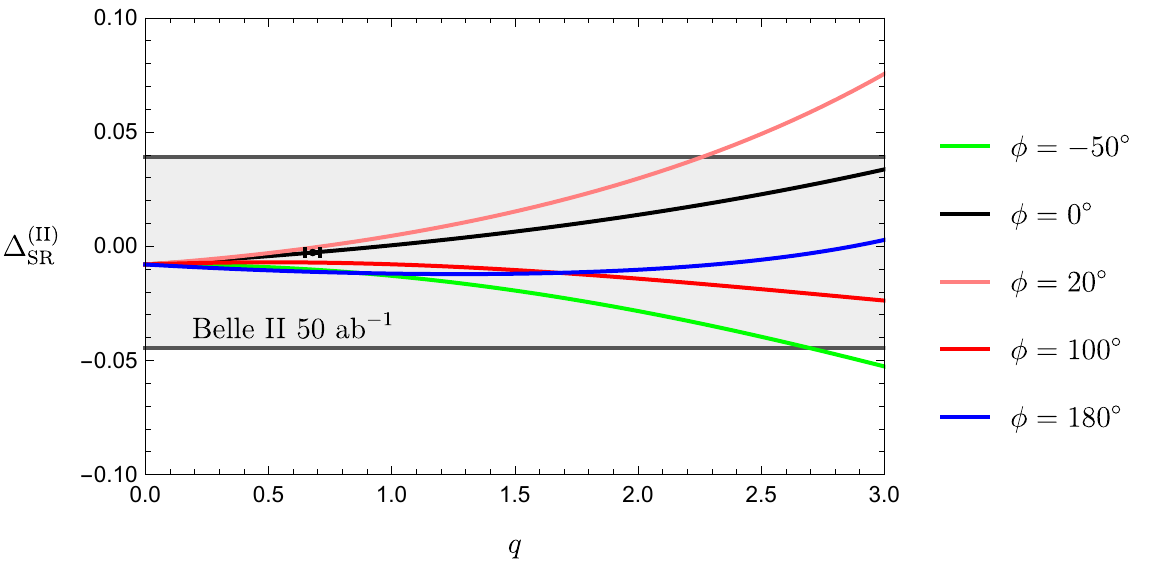}} 
	\caption{The sum rules introduced in Eqs.~(\ref{eq:sum-rule-I}) and (\ref{eq:sum-rule-II})  as functions of $q$ for different values of the CP-violating phase $\phi$. The grey horizontal band illustrates the ultimate experimental precision at Belle II, assuming an uncertainty of $\pm 0.042$ for $A_{\rm CP}^{\pi^0 K_{\rm S}}$ and perfect measurements of the other observables entering the sum rules.}
	\label{fig:SumRules}
\end{figure}

Finally, it is interesting to return to the sum rules discussed in Subsection~\ref{sec:SR}. The question arises whether they would be
significantly affected by the NP scenarios discussed above. In Fig.~\ref{fig:SumRules}, we show both sum rules as functions
of $q$ for several values of $\phi$, using the hadronic parameters in Table~\ref{tab:sumpara}. Here the outer curves correspond to the maximum values that the sum rules can take. The behaviour of $\Delta_{\rm SR}^{({\rm I})}$ can be easily derived from Eq.~(\ref{eq:SR-theo-1}), i.e.\ it is linear in $q$ with a slope proportional to $\sin(\gamma-\phi)$. On the other hand, $\Delta_{\rm SR}^{({\rm II})}$ also depends on $q^2$ as can be seen from Eq.~(\ref{eq:sum-rule-II-theo}). The grey horizontal bands show the sensitivity of the sum rules at Belle II, assuming an uncertainty for $A_{\rm CP}^{\pi^0 K_{\rm S}}$ of $\pm 0.042$ and perfect measurements of the other observables entering Eqs.~(\ref{eq:sum-rule-I}) and (\ref{eq:sum-rule-II}).  The black data point corresponds to the SM values of $q$ and $\phi$ using $R_q$ in Eq.~(\ref{eq:theoryupRq}). Consequently, we observe that the experimental resolution would not be sufficient to reveal the NP effects in the EW penguin sector with the sum rules, 
in contrast to the new method presented above.

\section{Conclusions} \label{sec:Conclusions}
 Employing information on the UT angle $\gamma$ and the
 $B^0_d$--$\bar B^0_d$ mixing phase $\phi_d$, we use the currently available data for $B\to\pi\pi$ decays to determine 
 hadronic parameters which characterize these modes and describe the interplay between various tree-diagram-like and 
 penguin topologies. We find agreement with previous studies although our results have higher precision. An important 
 new element in this endeavour is given by measurements of direct CP violation in $B^0_d\to\pi^0\pi^0$, allowing us to resolve a 
 twofold ambiguity. 
 The determination of the hadronic $B\to\pi\pi$ parameters relies only on the isospin symmetry and is hence theoretically clean.
 Consequently, the corresponding results represent reference values for the comparison with QCD calculations. 
 EW penguin topologies play a negligible role in the $B\to\pi\pi$ system for the current experimental uncertainties 
 but could be included in the future through more sophisticated analyses. 
 
Utilizing the $SU(3)$ flavour symmetry, we convert the hadronic $B\to\pi\pi$ parameters into their counterparts
in the $B\to\pi K$ system. We test also the $SU(3)$ flavour symmetry and obtain an impressive global picture which does not 
indicate any anomalously large non-factorizable $SU(3)$-breaking corrections. Correspondingly, we do not find indications of an
enhancement of colour-suppressed EW penguin topologies when analysing the data. 
The cleanest SM prediction of the $B\to\pi K$ observables is a correlation between the direct and mixing-induced CP asymmetries of the $B^0_d\to \pi^0K_{\rm S}$ decay. 
As we discussed in detail, it follows from an isospin relation between the neutral $B\to\pi K$ decay amplitudes and uses the 
$SU(3)$ flavour symmetry only to fix the magnitude of the $\hat{T}'+\hat{C}'$ amplitude. In comparison with a previous study,
a tension of the mixing-induced CP violation in $B^0_d\to \pi^0K_{\rm S}$ has become more pronounced 
due to a sharper determination of $\gamma$. Moreover, we have 
considered the angle $\phi_\pm$ as a new constraint, which also shows tension with respect to the SM. These discrepancies 
emerging from the current data suggest that either the values of the measured observables will change in the future or indicate NP effects with new sources of CP violation. In the former case, a reduction of the central value of the branching ratio of
$B^0_d\to\pi^0K^0$ by about $2.5\,\sigma$ with an increase of the mixing-induced CP asymmetry by about $1 \,\sigma$ would
give a situation in agreement with the SM. In the latter case, EW penguin topologies offer an attractive avenue for new particles 
to enter the $B\to\pi K$ modes. 
 
 In view of this intriguing $B\to\pi K$ puzzle and to test the corresponding sector of the SM, the EW penguin parameters $q$ 
 and $\phi$ are in the spotlight. We have presented a new strategy to determine these quantities from the data for the neutral and 
 charged $B\to\pi K$ decays, employing again the corresponding isospin relations. Applying this method to the current data, we 
 already obtain surprisingly stringent constraints in the $\phi$--$q$ plane. They are consistent with the SM but leave also a lot of space for possible NP effects. In order to actually pin down $\phi$ and $q$ further information is needed, which is provided 
 by the mixing-induced CP asymmetry of the $B^0_d\to \pi^0K_{\rm S}$ decay. Considering a variety of future scenarios, we 
 have illustrated this determination and have shown that the theory uncertainties can match the expected experimental precision in the era of 
 Belle II and the LHCb upgrade.  Following these lines, we may determine $(q,\phi)$ and reveal the dynamics 
 of the $B\to\pi K$ system with unprecedented accuracy. The resulting picture will either confirm once again the SM or 
 may eventually establish new flavour structures with possible new sources of CP violation.


\section*{Acknowledgements}
This research has been supported by the Netherlands Organisation for Scientific Research (NWO) and by the Deutsche Forschungsgemeinschaft (DFG), research unit FOR 1873 (QFET).


%
%
%

\begin{thebibliography}{99}

\bibitem{BG}A.~J.~Buras and J.~Girrbach,
  Rept.\ Prog.\ Phys.\  {\bf 77} (2014) 086201
  [arXiv:1306.3775 [hep-ph]].

\bibitem{cab}N.~Cabibbo,
  Phys.\ Rev.\ Lett.\  {\bf 10} (1963) 531.

\bibitem{KM}M.~Kobayashi and T.~Maskawa,
  Prog.\ Theor.\ Phys.\  {\bf 49} (1973) 652.
  
\bibitem{Belle-II} T.~Abe {\it et al.} [Belle-II Collaboration],
  arXiv:1011.0352 [physics.ins-det]; T.~Aushev {\it et al.},
  arXiv:1002.5012 [hep-ex]. 
  
\bibitem{LHCb}R.~Aaij {\it et al.} [LHCb Collaboration],
  Eur.\ Phys.\ J.\ C {\bf 73} (2013) 2373
  [arXiv:1208.3355 [hep-ex]].


\bibitem{NQ}Y.~Nir and H.~R.~Quinn,
  Phys.\ Rev.\ Lett.\  {\bf 67} (1991) 541.

\bibitem{GHLR}M.~Gronau, O.~F.~Hernandez, D.~London and J.~L.~Rosner,
  Phys.\ Rev.\ D {\bf 52} (1995) 6374
  [hep-ph/9504327].

\bibitem{GRL}M.~Gronau, J.~L.~Rosner and D.~London,
  Phys.\ Rev.\ Lett.\  {\bf 73} (1994) 21
  [hep-ph/9404282].


  \bibitem{RF-96} R.~Fleischer,
  Int.\ J.\ Mod.\ Phys.\ A {\bf 12} (1997) 2459
  [hep-ph/9612446].


\bibitem{BF-98}A.~J.~Buras and R.~Fleischer,
  Eur.\ Phys.\ J.\ C {\bf 11} (1999) 93
  [hep-ph/9810260].

  \bibitem{Neu-98} M.~Neubert,
  JHEP {\bf 9902} (1999) 014
  [hep-ph/9812396].

 \bibitem{BeNe} M.~Beneke and M.~Neubert,
  Nucl.\ Phys.\ B {\bf 675} (2003) 333
  [hep-ph/0308039].
    
\bibitem{FRS} R.~Fleischer, S.~Recksiegel and F.~Schwab,
  Eur.\ Phys.\ J.\ C {\bf 51} (2007) 55
  [hep-ph/0702275 [HEP-PH]].
         
\bibitem{groro} M.~Gronau and J.~L.~Rosner,
  Phys.\ Lett.\ B {\bf 666} (2008) 467
  [arXiv:0807.3080 [hep-ph]].

\bibitem{BGV}C.~Bobeth, M.~Gorbahn and S.~Vickers,
  Eur.\ Phys.\ J.\ C {\bf 75} (2015)  340
  [arXiv:1409.3252 [hep-ph]].
  

\bibitem{BFRS-2}
 A.~J.~Buras, R.~Fleischer, S.~Recksiegel and F.~Schwab,
  Nucl.\ Phys.\ B {\bf 697} (2004) 133
  [hep-ph/0402112].

 
 \bibitem{BFRS-1} A.~J.~Buras, R.~Fleischer, S.~Recksiegel and F.~Schwab,
  Phys.\ Rev.\ Lett.\  {\bf 92} (2004) 101804
  [hep-ph/0312259].
  \bibitem{BEJLLW-1}V.~Barger, L.~Everett, J.~Jiang, P.~Langacker, T.~Liu and C.~Wagner,
  Phys.\ Rev.\ D {\bf 80} (2009) 055008
  [arXiv:0902.4507 [hep-ph]].
  
\bibitem{BEJLLW-2} V.~Barger, L.~L.~Everett, J.~Jiang, P.~Langacker, T.~Liu and C.~E.~M.~Wagner,
  JHEP {\bf 0912} (2009) 048
  [arXiv:0906.3745 [hep-ph]].
  
 \bibitem{RF-FPCP} R.~Fleischer,
  PoS FPCP {\bf 2015} (2015) 002
  [arXiv:1509.00601 [hep-ph]].
 
\bibitem{BDLRR}  N.~B.~Beaudry, A.~Datta, D.~London, A.~Rashed and J.~S.~Roux,
  JHEP {\bf 1801} (2018) 074
  [arXiv:1709.07142 [hep-ph]].
     
    
\bibitem{ANSS}W.~Altmannshofer, C.~Niehoff, P.~Stangl and D.~M.~Straub,
  Eur.\ Phys.\ J.\ C {\bf 77} (2017)  377
  [arXiv:1703.09189 [hep-ph]].


 \bibitem{RF-95}R.~Fleischer,
  Phys.\ Lett.\ B {\bf 365} (1996) 399
  [hep-ph/9509204].
  
  
 
\bibitem{FJPZ}
  R.~Fleischer, S.~J\"ager, D.~Pirjol and J.~Zupan,
  Phys.\ Rev.\ D {\bf 78} (2008) 111501
  [arXiv:0806.2900 [hep-ph]].
  
\bibitem{FJV18} 
   R.~Fleischer, R.~Jaarsma and K.~K.~Vos,
  arXiv:1712.02323 [hep-ph].
 
 \bibitem{Moriond-18}R.~Fleischer, R.~Jaarsma, E.~Malami and K.~K.~Vos,
  arXiv:1805.06705 [hep-ph]; talk given at Rencontres de Moriond 2018, QCD and High Energy Interactions, La Thuile, Italy, 17--24 March 2018, to appear in the Proceedings.
  
\bibitem{GL}M.~Gronau and D.~London,
  Phys.\ Rev.\ Lett.\  {\bf 65} (1990) 3381.
  
\bibitem{PDG}
  C.~Patrignani {\it et al.} [Particle Data Group],
  Chin.\ Phys.\ C {\bf 40} (2016)  100001.
   
\bibitem{Amh14}
 Y.~Amhis {\it et al.} [Heavy Flavor Averaging Group (HFAG)],
  arXiv:1412.7515 [hep-ex].
  for updates, see \href{http://www.slac.stanford.edu/xorg/hfag/}{\tt http://www.slac.stanford.edu/xorg/hfag/}.
    \bibitem{Lee13}
  J.~P.~Lees {\it et al.} [BaBar Collaboration],
  Phys.\ Rev.\ D {\bf 87} (2013)  052009
  [arXiv:1206.3525 [hep-ex]].

\bibitem{Belle17} 
  T.~Julius {\it et al.} [Belle Collaboration],
  Phys.\ Rev.\ D {\bf 96} (2017)  032007
  [arXiv:1705.02083 [hep-ex]].
  
\bibitem{Aaij:2018tfw}
  R.~Aaij {\it et al.} [LHCb Collaboration],
  arXiv:1805.06759 [hep-ex].
    
   
\bibitem{Wol83}
  L.~Wolfenstein,
  Phys.\ Rev.\ Lett.\  {\bf 51} (1983) 1945.
    
\bibitem{BLO} A.~J.~Buras, M.~E.~Lautenbacher and G.~Ostermaier,
  Phys.\ Rev.\ D {\bf 50} (1994) 3433
  [hep-ph/9403384].
  
     \bibitem{Cha15} 
 J.~Charles {\it et al.},
  Phys.\ Rev.\ D {\bf 91} (2015)  073007
  [arXiv:1501.05013 [hep-ph]];
  for updates, see  \href{http://ckmfitter.in2p3.fr}{\tt http://ckmfitter.in2p3.fr}.


\bibitem{gw}
  M.~Gronau and D.~Wyler,
  Phys.\ Lett.\ B {\bf 265} (1991) 172.
  


\bibitem{ADS}D.~Atwood, I.~Dunietz and A.~Soni,
  Phys.\ Rev.\ Lett.\  {\bf 78} (1997) 3257
  [hep-ph/9612433];
  Phys.\ Rev.\ D {\bf 63} (2001) 036005
  [hep-ph/0008090].
  
\bibitem{FR-gam}R.~Fleischer and S.~Ricciardi,
  proceedings of the 6th International Workshop on the CKM Unitarity Triangle (CKM 2010)
  [arXiv:1104.4029 [hep-ph]].
  
  
 \bibitem{UTfit}A.~Bevan {\it et al.},
  arXiv:1411.7233 [hep-ph];
  for updates, see http://www.utfit.org.  
  
  
\bibitem{BIG}
  R.~Fleischer, R.~Jaarsma and K.~K.~Vos,
  JHEP {\bf 1703} (2017) 055
  [arXiv:1612.07342 [hep-ph]].

 
\bibitem{FJV-S}R.~Fleischer, R.~Jaarsma and K.~K.~Vos,
  Phys.\ Rev.\ D {\bf 94} (2016) no.11,  113014
  [arXiv:1608.00901 [hep-ph]].



    
\bibitem{Gro98}
M.~Gronau, D.~Pirjol and T.~M.~Yan,
  Phys.\ Rev.\ D {\bf 60} (1999) 034021
   Erratum: [Phys.\ Rev.\ D {\bf 69} (2004) 119901]
  [hep-ph/9810482].
  
\bibitem{Fle02} R.~Fleischer,
  Phys.\ Rept.\  {\bf 370} (2002) 537
  [hep-ph/0207108].
  
\bibitem{DeBrFl} K.~De Bruyn and R.~Fleischer,
  JHEP {\bf 1503} (2015) 145
  [arXiv:1412.6834 [hep-ph]].
     \bibitem{BBNS}M.~Beneke, G.~Buchalla, M.~Neubert and C.~T.~Sachrajda,
  Nucl.\ Phys.\ B {\bf 606} (2001) 245
  [hep-ph/0104110].
  
\bibitem{NR}M.~Neubert and J.~L.~Rosner,
  Phys.\ Rev.\ Lett.\  {\bf 81} (1998) 5076
  [hep-ph/9809311];
  Phys.\ Lett.\ B {\bf 441} (1998) 403
  [hep-ph/9808493].
\bibitem{Aaij:2016elb} 
  R.~Aaij {\it et al.} [LHCb Collaboration],
  Phys.\ Rev.\ Lett.\  {\bf 118} (2017) 081801
  [arXiv:1610.08288 [hep-ex]].

 
 \bibitem{Rosner:2015wva}
  J.~L.~Rosner, S.~Stone and R.~S.~Van de Water,
  [arXiv:1509.02220 [hep-ph]].


\bibitem{Khodjamirian:2003xk}
  A.~Khodjamirian, T.~Mannel and M.~Melcher,
  Phys.\ Rev.\ D {\bf 68} (2003) 114007
  [hep-ph/0308297].
  

\bibitem{FM}R.~Fleischer and T.~Mannel,
  Phys.\ Rev.\ D {\bf 57}, 2752 (1998)
  [hep-ph/9704423].
  
\bibitem{gro} M.~Gronau,
  Phys.\ Lett.\ B {\bf 627} (2005) 82
  [hep-ph/0508047].
      
\bibitem{GR} M.~Gronau and J.~L.~Rosner,
  Phys.\ Rev.\ D {\bf 74} (2006) 057503
  [hep-ph/0608040].
  
 \bibitem{Aub08}
 B.~Aubert {\it et al.} [BaBar Collaboration],
  Phys.\ Rev.\ D {\bf 79} (2009) 052003
  [arXiv:0809.1174 [hep-ex]].
    
\bibitem{Fuj08}
 M.~Fujikawa {\it et al.} [Belle Collaboration],
  Phys.\ Rev.\ D {\bf 81} (2010) 011101
  [arXiv:0809.4366 [hep-ex]].
    

    

\bibitem{FFM}  S.~Faller, R.~Fleischer and T.~Mannel,
  Phys.\ Rev.\ D {\bf 79} (2009) 014005
  [arXiv:0810.4248 [hep-ph]].
      
  
\end{thebibliography}
\end{document}